\documentclass[a4paper,fleqn,usenatbib]{mnras}

\usepackage{newtxtext,newtxmath}
\usepackage[T1]{fontenc}
\usepackage{ae,aecompl}

\usepackage{graphicx}	
\usepackage{amsmath}	
\usepackage{adjustbox}
\usepackage{xspace}
\usepackage{natbib}
\usepackage{siunitx}
\setcitestyle{notesep={}}
\usepackage{cleveref}
\crefname{figure}{Figure}{Figures}
\crefname{section}{Section}{Sections}
\crefname{table}{Table}{Tables}

\newcommand{\Msol}{M$_{\odot}$\xspace}
\newcommand{\squotes}[1]{\lq {#1}\rq\xspace}
\newcommand{\dquotes}[1]{\lq\lq {#1}\rq\rq\xspace}
\newcommand{\eagle}{{\sc eagle}\xspace}

\newcommand{\galform}{{\sc galform}\xspace}
\newcommand{\borg}{{\sc borg}\xspace}
\newcommand{\swift}{{\sc swift}\xspace}
\newcommand{\csd}[1]{#1$_{\mathrm{SD}}$\xspace}
\newcommand{\cob}[1]{#1$_{\mathrm{*}}$\xspace}
\newcommand{\sibeliusdark}{{\sc sibelius-dark}\xspace}
\newcommand{\sibelius}{{\sc sibelius}\xspace}

\DeclareFixedFont{\ttb}{T1}{txtt}{bx}{n}{8} 
\DeclareFixedFont{\ttm}{T1}{txtt}{m}{n}{8}  

\usepackage{color}
\definecolor{deepblue}{rgb}{0,0,0.5}
\definecolor{deepred}{rgb}{0.6,0,0}
\definecolor{deepgreen}{rgb}{0,0.5,0}
\definecolor{grey}{rgb}{0.4,0.4,0.4}

\usepackage{listings}

\newcommand\sqlstyle{\lstset{
language=SQL,
showspaces=false,
basicstyle=\ttfamily,
numbers=left,
numberstyle=\tiny,
commentstyle=\color{grey},
frame=single
}}

\title[\sibeliusdark]{SIBELIUS-DARK: a galaxy catalogue of the Local Volume from a constrained realisation simulation}

\author[S. McAlpine et al.]{Stuart McAlpine,$^{1}$\thanks{E-mail: stuart.mcalpine@helsinki.fi}
John C. Helly,$^{2}$
Matthieu Schaller,$^{3,4}$
Till Sawala,$^{1}$
Guilhem Lavaux,$^{5}$
\newauthor
Jens Jasche,$^{6}$
Carlos S. Frenk,$^{2}$
Adrian~Jenkins,$^{2}$
John R. Lucey$^{7}$
and Peter H. Johansson$^{1}$
\\
$^{1}$Department of Physics, Gustaf H\"allstr\"omin katu 2, University of Helsinki, Finland\\
$^{2}$Institute for Computational Cosmology, Durham University, South Road, Durham DH1 3LE, United Kingdom \\
$^{3}$Lorentz Institute for Theoretical Physics, Leiden University, PO Box 9506, NL-2300 RA Leiden, The Netherlands\\
$^{4}$Leiden Observatory, Leiden University, PO Box 9513, NL-2300 RA Leiden, The Netherlands\\
$^{5}$CNRS \& Sorbonne Universit\'e, UMR7095, Institut d'Astrophysique de Paris, 75014 Paris, France\\
$^{6}$The Oskar Klein Centre, Department of Physics, Stockholm University, Albanova University Center, 106 91 Stockholm, Sweden \\
$^{7}$Centre for Extragalactic Astronomy, University of Durham, Durham DH1 3LE, UK
}

\date{Accepted XXX. Received YYY; in original form ZZZ}

\pubyear{2021}

\begin{document}
\label{firstpage}
\pagerange{\pageref{firstpage}--\pageref{lastpage}}
\maketitle

\begin{abstract}
We present \sibeliusdark, a constrained realisation simulation of the local volume to a distance of 200~Mpc from the Milky Way. \sibeliusdark is the first study of the \dquotes{\textit{Simulations Beyond The Local Universe}} (\sibelius) project, which has the goal of embedding a model Local Group-like system within the correct cosmic environment. The simulation is dark-matter-only, with the galaxy population calculated using the semi-analytic model of galaxy formation, \galform. We demonstrate that the large-scale structure that emerges from the \sibelius constrained initial conditions matches well the observational data. The inferred galaxy population of \sibeliusdark also match well the observational data, both statistically for the whole volume and on an object-by-object basis for the most massive clusters. For example, the $K$-band number counts across the whole sky, and when divided between the northern and southern Galactic hemispheres, are well reproduced by \sibeliusdark. We find that the local volume is somewhat unusual in the wider context of $\Lambda$CDM: it contains an abnormally high number of supermassive clusters, as well as an overall large-scale underdensity at the level of $\approx 5$\% relative to the cosmic mean. However, whilst rare, the extent of these peculiarities does not significantly challenge the $\Lambda$CDM model. \sibeliusdark is the most comprehensive constrained realisation simulation of the local volume to date, and with this paper we publicly release the halo and galaxy catalogues at $z=0$, which we hope will be useful to the wider astronomy community.
\end{abstract}


\section{Introduction}
\label{sect:introduction}

Over the past few decades cosmological computer simulations have
become an increasingly effective tool for advancing our understanding
of structure and galaxy evolution in the Universe \citep[see][ for
comprehensive reviews]{Frenk2012, Vogelsberger2020}. The Lambda
cold-dark-matter model ($\Lambda$CDM), frequently referred to as the
standard model, is the leading paradigm to describe the nature of the
cosmos. Structure formation proceeds from primordial density
fluctuations in a \squotes{bottom up} manner, with low-mass structures
(referred to as haloes) collapsing first and larger structures forming
later \citep{Davis1985}. The traditional goal of $\Lambda$CDM cosmological
simulations, such as the recent
Horizon-AGN \citep{Dubois2014}, Magneticum \citep{Hirschmann2014},
EAGLE \citep{Schaye2015} and IllustrisTNG \citep{Pillepich2018}
simulations, has been to produce a \squotes{random} representative patch of the Universe that
can be statistically compared to the one we observe, in terms of the
properties of the large-scale structure, clustering statistics, the
galaxy abundance and diversity, etc.  However, whilst these simulated
universes might reflect the observed Universe statistically, they do
not contain the specific objects (such as the Local Group, or the Virgo and Coma clusters), embedded within the correct
large-scale structure, that we actually observe. There remains an
underlying tension between an observational dataset, which is subject
to cosmic variance, and the ensemble mean predictions from
cosmological simulations.

How then can one determine the nature of specific objects within our
Universe? Is it possible to deduce unambiguously the evolutionary
pathways that led them to the point at which we observe them? One
method that broadly attempts to answer this question involves a brute
force approach: scouring large random $\Lambda$CDM cosmological
simulations for model analogues that are as similar as possible to the
particular object in question. For example, to examine the nature of
the Milky Way and the Local Group, our most robust observational
environment to study small-scale astrophysics, studies such as the
ELVIS \citep{GarrisonKimmel2014,GarrisonKimmel2019} and APOSTLE
\citep{Sawala2016} projects have simulated, in exquisite detail, halo
pairs of Local Group analogues extracted from large random
$\Lambda$CDM cosmological simulations. Such studies have made
meaningful advancements on our understanding of galaxy formation
physics, particularly in relation to the so-called small-scale
tensions between N-body (i.e. dark-matter-only) simulations of the
$\Lambda$CDM model and  observations \citep[see][ for a
review]{Bullock2017}, such as the core/cusp problem
\citep{Moore1994,Flores1994}, the missing satellites problem
\citep{Moore1999,Klypin1999} and the \squotes{too big to fail} problem
\citep{Boylan-Kolchin2011}. Solutions to these problems based on 
cosmological hydrodynamics simulations of Local Group analogues have
been proposed by \cite{Sawala2016}.

However, whilst one can infer from studies using this brute force approach as to the prevalence of
particular types of objects within the context of $\Lambda$CDM, for
example, of a Local Group-like system, one cannot adequately ascertain
the nature of \emph{the} Local Group, as the non-linear formation of
structure within our Universe allows an almost infinite number of
evolutionary pathways towards the same end point. To truly probe the
nature of our Local Group from cosmological simulations, or indeed of
any particular object that we observe, it must be simulated within the
correct cosmological environment, rather than a random one. This is the goal of
\squotes{constrained realisation} simulations.

A constrained realisation simulation has a focused objective: 
to deliberately construct a set of $\Lambda$CDM initial conditions that
will evolve into the particular large-scale structure distribution of
the observed local volume. Thus, the full phase-space distribution of
the observed individual clusters, filaments and voids in the local
volume will each be reproduced by their model analogues, at the
correct location, within the simulation. Simulations of this nature
can go beyond asking questions such as how prevalent particular observed structures are
in the context of $\Lambda$CDM, to predicting more accurately the
particular formation pathways of the structures in the local volume
with which we are so familiar, such as the Virgo cluster, the Coma
cluster, and of course the Local Group.

The initial conditions for a constrained realisation simulation can be
derived from two related approaches. In the first, the initial density
field is inferred from a dataset of galaxy redshifts or radial
peculiar velocities using the ideas pioneered by
\cite{Bertschinger1987} and \cite{Hoffman1991}. Examples of such
simulations include those by \cite{Mathis2002}, the
\dquotes{\textit{Constrained Local UniversE Simulations}} \citep[{\sc
  clues},][]{Yepes2014,Carlesi2016} project, the \dquotes{\textit{Constrained LOcal \& Nesting Environment Simulations}} \citep[{\sc clones},][]{Sorce2021} project, the \dquotes{\textit{ELUCID}
  simulation}\citep{Wang2016} and the \dquotes{\textit{High-resolution
    Environmental Simulations of The Immediate Area}} \citep[{\sc
  hestia},][]{Libeskind2020} project. In the second approach, the initial
conditions are derived using Bayesian inference through physical
forward modelling with a Hamiltonian Monte Carlo sampling approach,
such as in the simulations by \cite{Wang2014}, or the
\dquotes{\textit{Bayesian Origin Reconstruction from Galaxies}}
(\borg) project \citep{Jasche2013}. Each technique can only constrain
the density field \emph{above} non-linear scales. Thus, the
small-scale properties of the initial conditions remain
random. However, for random realisations of the small-scale
features, massive objects formed within constrained realisation
simulations, for example the Virgo cluster, retain consistent properties
at the 10--20\% level \citep{Sorce2016b}, and similarly the Coma cluster \citep{Jasche2019}, indicating that their formation is
largely dictated by much larger scales. Thus, the scatter among random
realisations that are constrained at the linear scale is substantially
smaller, by a factor of 2--3 on scales of 5~$h^{-1}$~Mpc, 
than that found for random simulations \citep{Sorce2016a}, which
allows us to zero-in on the most plausible evolutionary pathways of
particular objects within the local volume.

\begin{figure*} 
\includegraphics[width=\textwidth]{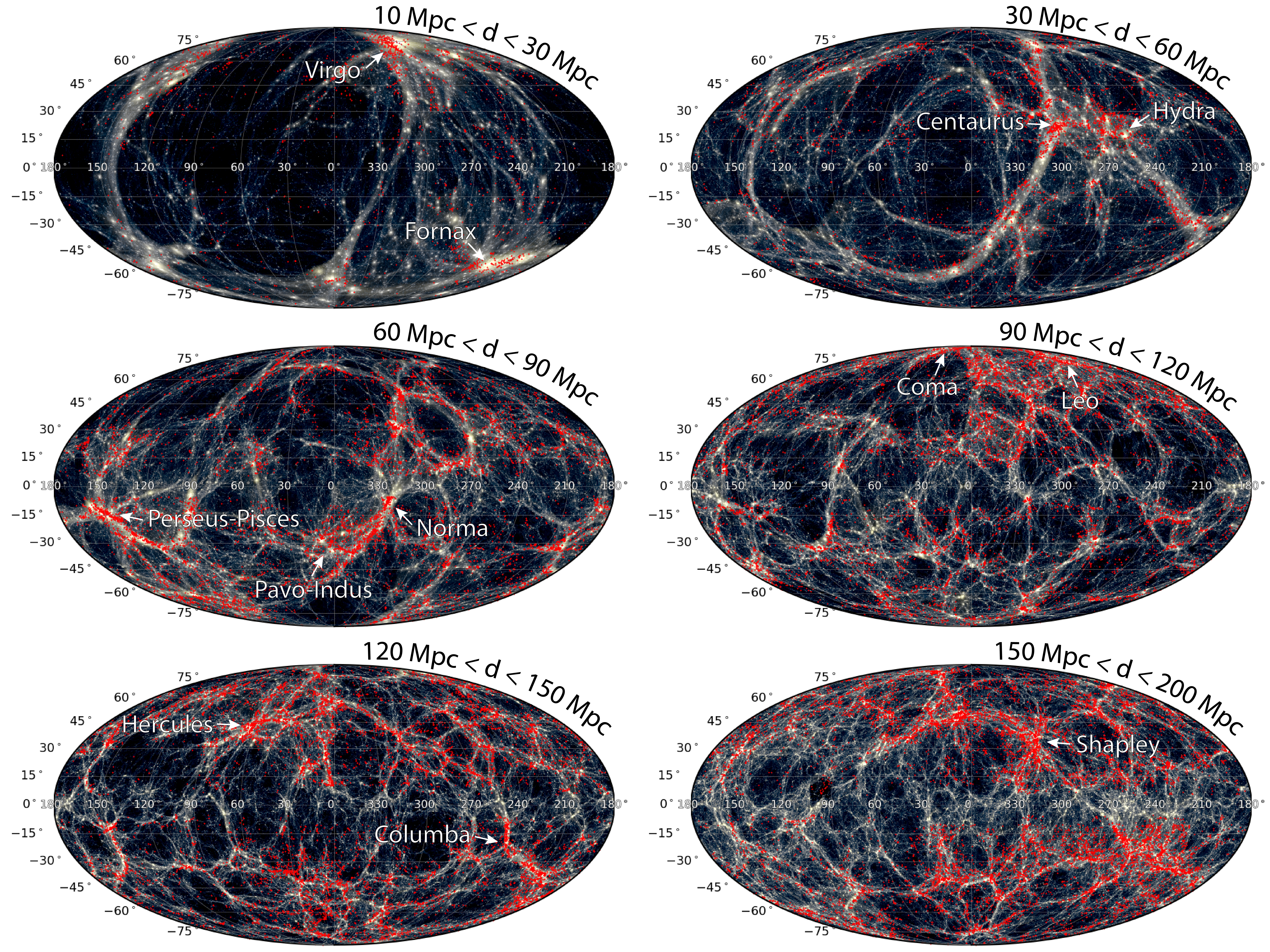}

\caption{The dark matter distribution of the entire \sibeliusdark volume ($d_{\mathrm{MW}} \leq 200$~Mpc), viewed in six spherical shells centred on the Milky Way (blue/green). Each slice is presented as an all-sky-map in a Mollweide projection using the Galactic coordinate system. Overplotted as red points are the galaxies from the 2M++ galaxy sample \citep{Lavaux2011}, demonstrating just how well the non-linear structure of the local volume has been encapsulated in the \borg reconstruction. The locations of twelve famous clusters/concentrations are also highlighted.}

\label{fig:sky_maps}

\end{figure*}

Here we present \sibeliusdark, the first of the
\dquotes{\textit{Simulations Beyond The Local Universe}} (\sibelius)
project \citep{Sawala2021b}. The \sibelius project has the goal of
connecting the Local Group to the local environment by embedding a
Local Group-like analogue within the correct large-scale structure
produced by the \borg algorithm. It
encompasses, at high resolution, the constrained large-scale structure
out to a distance 200~Mpc from the Milky Way, and so includes
well-known 
clusters such as \textit{Virgo}, \textit{Coma} and
\textit{Perseus}. Additionally, at the centre of \sibeliusdark there
is a Local Group-like halo pair with the correct dynamics. The
simulation is dark-matter-only. The galaxy population, which matches
well many observational datasets both statistically and, in
particular, 
massive clusters, is calculated using the semi-analytic model of
galaxy formation, \galform
\citep{Lacey2016}. Often, previous works using constrained initial conditions have focused on the evolution and properties of individual objects within the local volume, such as the Local Group \citep[e.g.,][]{Libeskind2011,Carlesi2020} or the Virgo cluster \citep[e.g.,][]{Sorce2021}, or have simulated the local volume as a whole, yet to much smaller radii \citep[$\lesssim 8000$~km/s, e.g.,][]{Mathis2002,Klypin2003}. Here, the combination of such a large constrained region at high resolution that is self-consistently connected to the evolution Local Group makes \sibeliusdark the most
comprehensive constrained realisation simulation to date.

The layout of this paper is as follows. In \cref{sect:method} we
present an overview of the method for generating the \sibelius
constrained initial conditions, describing how a Local Group-like
object is embedded within the large-scale structure produced by the
\borg algorithm. We then demonstrate how well the \sibeliusdark halo
and galaxy population match the data on a statistical level in
\cref{sect:galaxies}, and on a cluster-by-cluster level in
\cref{sect:clusters_and_groups}. In \cref{sect:virgo_and_coma} we
further investigate how well the \sibeliusdark analogues of the Virgo
and Coma clusters match the data, and make predictions for the
location and observability of their \squotes{splashback radius} in
\cref{sect:splashback_radius}. The nature of the Local Group analogue
at the centre of the volume is explored in
\cref{sect:local_group}. Finally, we discuss our results and conclude
in \cref{sect:conclusion}. In \cref{sect:public_data_release} we
present the details of how to access the \sibeliusdark data at $z=0$,
which we make public with the publication of this paper.    

\section{Method}
\label{sect:method}

\subsection{Generating phase information to construct a constrained realisation of the local volume and the Local Group}

The aim of the \sibelius project is to construct Lambda-Cold-Dark-Matter ($\Lambda$CDM) initial conditions for a simulation that will evolve into the observed density and velocity fields of our local volume (i.e., to a distance $d_{\mathrm{MW}} \lesssim 200$~Mpc from our Milky Way), with a correctly placed and suitably representative Local Group analogue at its centre. Typically, representative cosmological simulations assume periodic boundary conditions. However, as the local volume is clearly non-periodic, the constrained phase information that describes the local volume is instead embedded within a larger periodic parent volume, $L=1$~Gpc on a side. Therefore \sibeliusdark, and all subsequent \sibelius simulations, are performed using the "zoom-in" technique, whereby only a region of interest is resimulated at high resolution, with the remainder of the volume being populated by low-resolution elements. We note that the initial conditions for \sibeliusdark, as with all zoom-in resimulations, are designed to remain \squotes{uncontaminated} through the course of the simulation, i.e., a sufficiently large initial volume is populated with high-resolution particles such that no low-resolution particles enter the high-resolution region of interest (in our case a sphere with radius 200~comoving Mpc from the simulated Milky Way position at $z=0$) at any time.

In the \sibelius setup, the fiducial observer is the centre of the parent volume ($[x,y,z] = [500, 500, 500]$~Mpc). The \squotes{constrained} phase information propagates out to a radius of $200$~Mpc from this point ($\approx 5$\% of the total volume of the parent box), with the remainder of the volume being filled with random, or otherwords unconstrained, phase information. The initial conditions are designed such that there is no sharp boundary in the phase information at the edge of the constrained region, and the cumulative phase information through the entirety of the volume remains statistically consistent with $\Lambda$CDM. 

The phase information that generates the initial density field is constructed in two distinct steps: the linear and mildly non-linear modes that govern the formation of the large-scale structure are produced using the \dquotes{\textit{Bayesian Origin Reconstruction from Galaxies}} (\borg) algorithm \citep{Jasche2013,Jasche2019}. The modes that govern the formation of systems at the size of the Local Group are largely dictated on scales below those included in the \borg constraints \citep{Sawala2021a}, and thus are included after, via a random shuffling of the smaller scale modes \citep{Sawala2021b}. We briefly summarise these two steps below.  

\subsubsection{The \borg algorithm}
\label{sect:constraints}

\begin{table*}

\caption{The \squotes{Loose} and \squotes{Strict} criteria a halo pair must satisfy to be classified as a Local Group, used for generating the \sibeliusdark initial conditions \citep[see \cref{sect:embedding_lg} and][]{Sawala2021b}. From left to right, the total halo mass of the MW + M31, their mass ratio, relative distance, radial velocity, tangential velocity and angular separation from the observed location of M31 on the sky. The final two columns show how many candidates (with and without orientation constraint) satisfy each criteria from the 60,000 exploration runs performed in \citet{Sawala2021b}. }

\begin{tabular}{lrrrrrrcc} \hline

Criteria & $M_{\mathrm{tot}}$ & $\frac{M_{\mathrm{MW}}}{M_{\mathrm{M31}}}$ & d & $v_{\mathrm{r}}$ & $v_{\mathrm{t}}$ & $\delta$ & $N[M_{\mathrm{tot}},\frac{M_{\mathrm{MW}}}{M_{\mathrm{M31}}},d,v_{\mathrm{r}},v_{\mathrm{t}}]$ & $N[M_{\mathrm{tot}},\frac{M_{\mathrm{MW}}}{M_{\mathrm{M31}}},d,v_{\mathrm{r}},v_{\mathrm{t}},\delta]$ \\
 & [$10^{12}$~\Msol] & & [Mpc] & [km/s] & [km/s] & [$^\circ$] & & \\

\hline\hline

\dquotes{Loose} & $1.2 \rightarrow 6$ & $0.4 \rightarrow 5$ & $0.5 \rightarrow 1.5$ & $-200 \rightarrow 0$ & $<150$ & $< 45$ & 6385 & 2309 \\

\dquotes{Strict} & $2 \rightarrow 4$ & $1 \rightarrow 2$ & $0.74 \rightarrow 0.8$ & $-109 \rightarrow -99$ & $<40$ & $< 15$ & 1 & 0 \\

\hline
\end{tabular}
\label{table:criteria}
\end{table*}

\borg is a fully-probabilistic inference algorithm designed to reproduce the three-dimensional cosmic matter distribution from local galaxy surveys at linear and mildly-non-linear scales. It incorporates a physical model for gravitational structure formation within the inference process itself, connecting the initial density field to the evolved large scale structure via a particle mesh, turning the task of analysing the present non-linear galaxy distribution into a statistical initial conditions problem. The result is a highly non-trivial Bayesian inverse problem \citep{Kitaura2008, Ensslin2009}, requiring the exploration of a very high-dimensional and non-linear space of possible solutions to the initial conditions problem from incomplete observations (which are flux-limited  and missing at some locations, e.g.  in the Zone of Avoidance). Samples of the posterior distribution are obtained via an efficient implementation of the Hamiltonian Markov Chain Monte Carlo (MCMC) method. The algorithm self-consistently accounts for observational uncertainty such as noise, survey geometry, selection effects and luminosity-dependent galaxy biases \citep[full details of the process are described in][]{Jasche2013}.

The limiting mode/scale the \borg algorithm can constrain is dependent on the resolution of the particle mesh used for the gravitational structure formation simulation. Here, the analysis consists of 256$^{3}$ grid nodes within a cubic Cartesian domain of side length $L = 1000$~Mpc, resulting in a total of $\approx 1.6 \times 10^{7}$ inference parameters for the Bayesian analysis, corresponding to the amplitudes of the primordial density fluctuation at the respective grid nodes. The inference procedure therefore yields constrained realisations with a resolution of $\approx 3.91$~Mpc \citep[for full details see][]{Jasche2019}. Although coarse, this resolution is sufficiently adequate to resolve galaxy clusters, concentrations and voids of the local volume. 

The input to the \borg algorithm is the observed three dimensional density field, inferred in this instance from the 2M++ galaxy sample \citep{Lavaux2011}\footnote{The 2M++ galaxy sample is based on photometry from the Two-Micron-All-Sky Extended Source Catalog \citep[2MASS-XSC,][]{Skrutskie2006} and redshifts from the 2MASS redshift survey \citep{Erdogdu2006}, the 6dF galaxy redshift survey Data Release Three \citep[6dFGRS,][]{Jones2009} and the
Sloan Digital Sky Survey Data Release Seven \citep[SDSS-DR7,][]
{Abazajian2009}.}.  However, being Bayesian, \borg cannot tell us the \squotes{true} initial density field for this dataset, but instead the most probable initial density fields given the observational constraints. For \sibeliusdark we settled on iteration 9350 of the MCMC chain \citep{Sawala2021b}, as, of the most probable realisations of the entire constrained volume, it best reproduced the masses and positions of the nearby Virgo and Fornax clusters relative to the Milky Way. An alternative choice of realisation would result in a very similar constrained region as a whole, but will contain differences in the overall dark matter distribution and the masses of individual dark matter haloes. Future work will investigate the level of these differences on the halo and galaxy populations.

To demonstrate how well the non-linear structure of the local volume has been encapsulated via the \borg reconstruction, \cref{fig:sky_maps} shows the dark matter distribution of the \sibeliusdark volume in six spherical shells centred on the Milky Way. These are presented as all sky maps in a Mollweide projection in the galactic coordinate system, and cover the entirety of the constrained region ($d_{\mathrm{MW}} < 200$ Mpc). Highlighted are some familiar structures: from our nearest cluster neighbours, Virgo and Fornax, to the concentrations of Hydra, Centaurus and Norma, thought to make up the \dquotes{Great Attractor}, as well as the dominant wall of Perseus-Pisces, the Coma supercluster and the Hercules superclusters. In addition to the concentrations of structure, prominent underdense voids are also visible. Overplotted in red are the galaxies from the 2M++ galaxy sample \citep{Lavaux2011}, which map extremely well to the underlying dark matter density field \citep[see also][ for more analysis of the \borg constraints]{Jasche2019}.

\begin{figure} \includegraphics[width=\columnwidth]{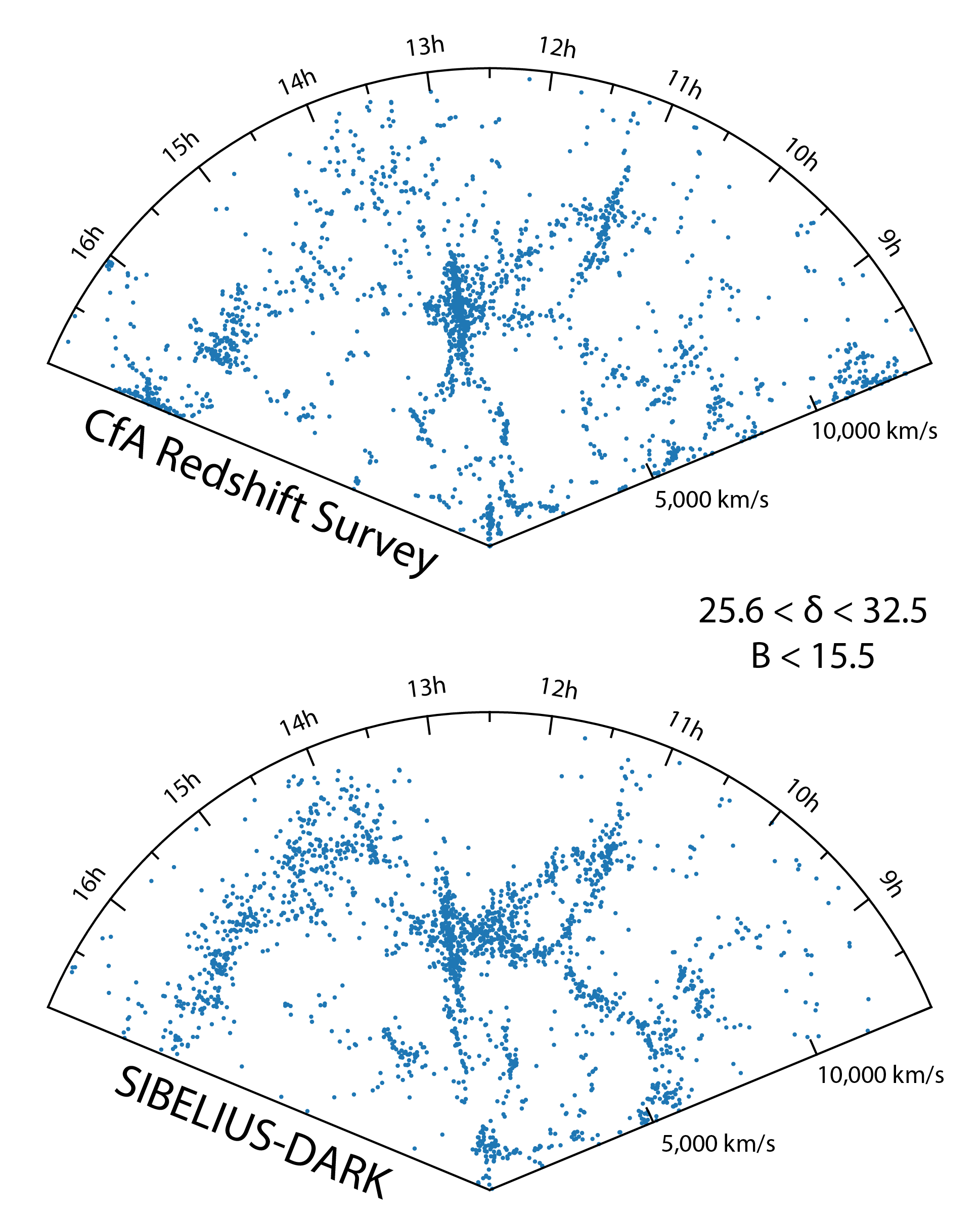}

\caption{The top panel shows the distribution of galaxies in redshift space brighter than $B < 15.5$ from a famous slice of the CfA redshift survey \citep[8h $<$ RA $<$ 17h \& $25.6^\circ <$ DEC $< 32.5^\circ$ \& $0$~km/s $< v_{\mathrm{r}} < 14,500$~km/s,][]{Huchra1983}. Below is the same slice from the inferred \sibeliusdark galaxy population, demonstrating how well the {\sc borg} constraints have captured the unique structure of the local volume.}
\label{fig:cfa}

\end{figure}

Taking this one step further, we also include \cref{fig:cfa}, which compares the inferred galaxy distribution of \sibeliusdark in redshift space to a famous slice of the CfA redshift survey \citep{Huchra1983}, demonstrating again how well the {\sc borg} constraints have captured the unique structure of our local volume.

\subsubsection{Embedding a Local Group within the \borg constraints}
\label{sect:embedding_lg}

As mentioned above, the constrained phase information produced by the \borg algorithm is limited to scales larger than the mesh size of $\approx 3.91$~Mpc, a scale greater than those important for the formation of $\sim 10^{12}$~\Msol haloes \citep[$\lambda_{\mathrm{cut}} \approx 1.62$~Mpc,][]{Sawala2021a}. Therefore the smaller-scale phase information at the location of the Local Group currently remains random. In order to embed a Local Group-like system at the centre of the \sibelius volume, we must therefore supply additional phase information. To do this, we leave the larger-scale phases set by the \borg constraints fixed, and then perform an exploration of the smaller-scale phases within the 16~cMpc cubic region surrounding the Local Group location (i.e., the centre of the box). That is, we randomise the phases below 3.2~cMpc within the central 16~cMpc cubic region\footnote{The supplementary random phases are generated using the publicly available {\sc panphasia} Gaussian white noise field \citep{Jenkins2013}.} until a system with accurate Local Group-like properties forms at the correct location. This results in multiple realisations that share the same large-scale phase information of the \borg constraints, but differ substantially in their halo populations at lower masses \citep[a method similar to the original CLUES project,][]{Yepes2014}.

\citet{Sawala2021b} performed 60,000 low-resolution simulations randomising the phases below 3.2~cMpc within the central 16~cMpc cubic region of the same \borg realisation. These variations were then probed for a dark matter halo pair that is no more than 5~Mpc from the centre of the box, and satisfies at least a \squotes{loose} criteria for a Local Group- like system (see \cref{table:criteria}), i.e., 

\begin{enumerate}

    \item There is no third system more massive than the Milky Way within 2~Mpc of the Local Group centre.
    
    \item The combined mass of the Milky Way and M31 falls within the range (1.2--6) $\times 10^{12}$ \Msol, and the ratio of the Milky Way halo mass with respect to the M31 halo mass is between 0.4 and 5.
    
    \item The relative distance between the Milky Way and M31 is between 500 and 1500 kpc.
    
    \item The relative radial velocity between the Milky Way and M31 is between -200 and 0 km/s and the relative tangential velocity between the pair is less than 150 km/s.
    
    \item M31 is oriented at the right location on the sky when viewed from the Milky Way (to within an angular separation of 45$^{\circ}$).
    
\end{enumerate}

\noindent From the 60,000 simulations, 2309 contained a halo pair that satisfied this loose criteria (with the orientation on the sky being the most restrictive factor). When we refined this selection to a \dquotes{strict} criterion (see again \cref{table:criteria}), based on the most recent observational limits, we were reduced to no suitable candidates from the 60,000 runs. To avoid simply making more attempts in the hope of eventually finding a good Local Group, we instead took the nine most suitable candidates and resimulate each of them 1000 times, now only randomising the scales below 0.8~cMpc, which serves to slightly perturb their properties. Of these resimulations, three systems then fulfilled the strict criteria, one of which is used here for \sibeliusdark. The full details of this embedding process is presented in \citet{Sawala2021b}. 

Naturally, the embedment process described above does not guarantee that the Local Group, or more specifically the Milky Way (i.e., the desired observer), will go on to form at exactly the centre of the parent volume at redshift $z=0$ ([$x,y,z] = [500,500,500]$~Mpc), i.e., the location of the fiducial observer as defined by the \borg constraints. The \sibeliusdark Milky Way is located at coordinates $[x,y,z] = [499.343, 504.507, 497.311]$ at redshift $z=0$, indicating it has drifted slightly outwith this limit, at a distance 5.3~Mpc from the centre of the parent volume. Yet to remain more consistent with the observations, for this study the observer is set to be the location of the simulated Milky Way at $z=0$, and not the centre of the parent volume. This has very little impact on the results, however; for example the distance to the simulated Virgo cluster is changed by only $\approx 7$\% depending on the observing location selected, and the distance to the simulated Coma cluster by less than 1\%.

\subsection{Numerical setup}
\label{sect:numerical_setup}

\subsubsection{Initial conditions}
\label{sect:ics}

The $\Lambda$CDM initial conditions for \sibeliusdark were generated using first order (Zel'dovich) Lagrangian perturbation theory as set out by \citet{Jenkins2010}, calculated down to redshift $z=127$. The displacement field for the zoom region is computed using two concentric meshes each of size 15360$^{3}$, centred on the middle of the parent volume. The top level mesh covers the entire domain ($L=1000$~Mpc) and the second mesh just covers the high-resolution region ($L=500$~Mpc). The second mesh, while not strictly necessary to reach the particle Nyquist frequency in the zoom region, improves both the accuracy and the fidelity with which the Panphasia phase information is reproduced on the smallest length-scales. A more in depth discussion of the initial conditions generation for \sibelius can be found in \citet{Sawala2021b}.

Generating the initial conditions was performed using 183 compute nodes of the COSMA-7 DiRAC facility hosted by Durham University and required 91.5 TB of run-time memory. The $L = 1000$~Mpc volume is sampled by 131 billion ($N_{\mathrm{p}}^{3}=5078^{3}$) collisionless dark matter particles, less than $1$\% of which are low-resolution boundary particles, with a high-resolution particle mass of $1.15 \times 10^{7}$~\Msol. The comoving gravitational softening length is set as $\epsilon_{\mathrm{CM}} = 0.05 \times (L/N_{\mathrm{p}}) = 3.32$~ckpc and the maximum physical softening length is $\epsilon_{\mathrm{phys}} = 0.0022 \times (L/N_{\mathrm{p}}) = 1.48$~kpc, following \citet{Ludlow2020}. The mass resolution and gravitational softening lengths were chosen to match that of the upcoming {\sc eagle-xl} model \citep[the successor to the \eagle model,][]{Schaye2015}, to allow for an optimal comparison between \sibeliusdark and later hydrodynamical resimulations of the \sibelius volume.

The combination of a large volume and a comparatively high resolution will make \sibeliusdark a useful tool even when the constrained aspect of the simulation is not considered. For context, the resolution of \sibeliusdark is comparable to that of the \textit{Millennium-II} simulation \citep[within $\approx 20$\%,][]{BoylanKolchin2009} yet has $\approx 13$ times more volume; shares the same number of particles as the \textit{P-Millennium} simulation \citep{Baugh2019} at $\approx 13$ times higher resolution (yet has $\approx 15$ times less volume); and contains many massive clusters sampled by more particles than the majority of the {\sc phoenix} \citep{Gao2012} cluster sample (there are $\approx 235,000,000$ particles within $r200c$ of the \sibeliusdark Perseus cluster).     

The simulation adopts a flat $\Lambda$CDM cosmogony with parameters inferred from analysis of {\it Planck} data \citep{Planck2013}: $\Omega_\Lambda = $\,0.693, $\Omega_{\rm m} = $\,0.307, $\Omega_{\rm b} =$\,0.04825, $\sigma_8 =$\,0.8288, $n_{\rm s} = $\,0.9611 and $H_0 = $\,67.77\,km\,s$^{-1}$\,Mpc$^{-1}$.

\subsubsection{The \swift simulation code}
\label{sect:swift}

\begin{figure*} \includegraphics[width=\textwidth]{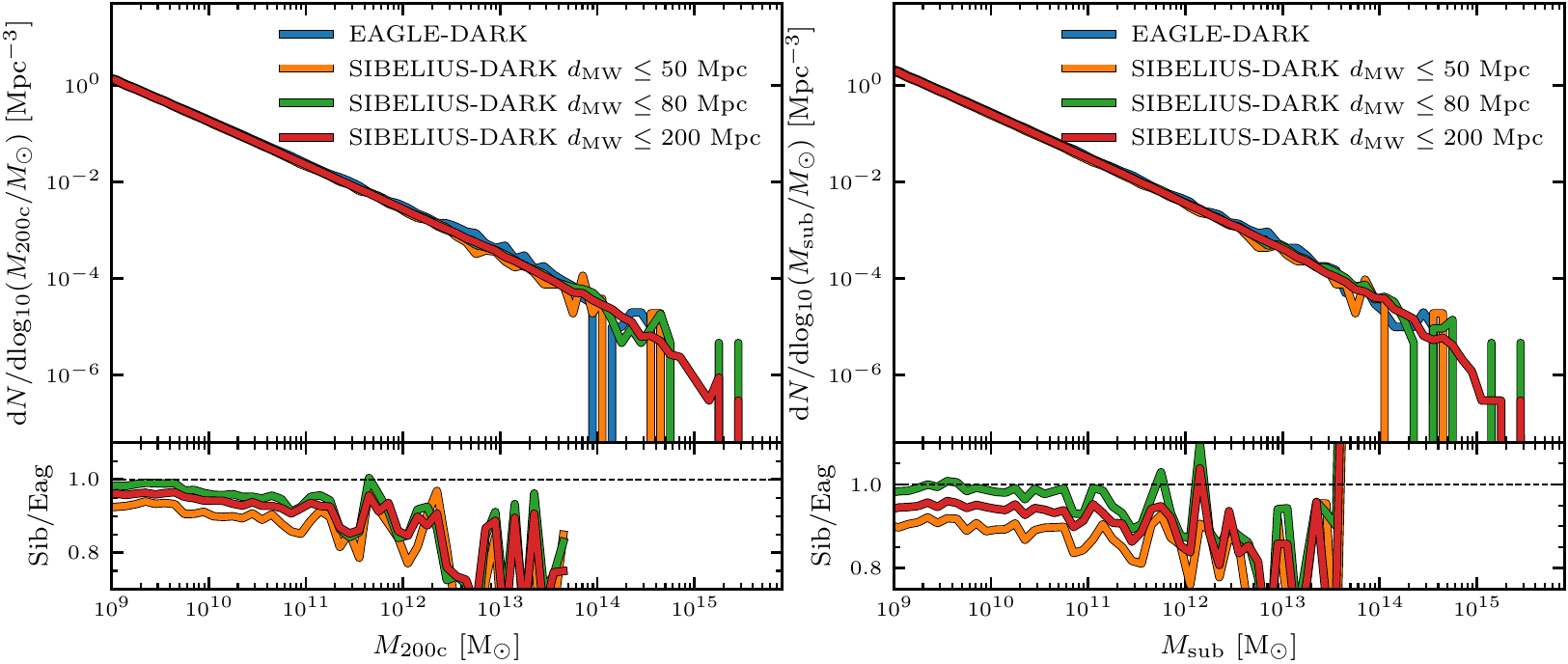}

\caption{The halo (left) and subhalo (right) mass functions within three spherical volumes centred on the Milky Way. For comparison, the halo and subhalo mass functions of the unconstrained (100~Mpc)$^{3}$ reference \eagle DMO volume ({\sc eagle-dark}), and the ratios to this volume (lower panels), are also shown. There are fewer haloes/subhaloes per unit volume in \sibeliusdark compared to {\sc eagle-dark} ($\approx 2$--$10$\% at $10^{10}$~\Msol). The level of difference within each spherical volume directly reflects the volumes average density relative to the mean density (which the {\sc eagle-dark} volume is at, see also \cref{fig:density_vs_rad}), suggesting that the local volume, particularly the innermost 50~Mpc, is  underdense relative to the cosmic mean.}

\label{fig:mass_function}

\end{figure*}

The simulation was performed using \swift \citep{Schaller2018} \footnote{Publicly available, including the exact version used for the \sibeliusdark simulations, at \url{www.swifstim.com}}, an open source coupled gravity, hydrodynamics, and galaxy formation code. \swift exploits task-based parallelism within compute nodes themselves, and interacts between compute nodes via \texttt{MPI} using non-blocking communications, resulting in excellent strong- and weak-scaling for cosmological calculations out to many tens of thousands of compute nodes \citep{Schaller2016}. The short- and long-range gravitational forces are computed using a 4$^{\mathrm{th}}$-order fast multipole method \citep[e.g.][]{Cheng1999} and particle-mesh method solved in Fourier space, respectively, with an imposed adaptive opening angle criterion similar to the one proposed by \cite{Dehnen2014}. For this project, \swift was run using only its $N$-body solver.

\sibeliusdark was run on 160 compute nodes (using 320 \texttt{MPI} ranks for a total of 4480 compute cores) of the COSMA-7 DiRAC facility\footnote{Each compute node hosts 2 Intel Xeon Gold 5120 CPU at 2.20GHz with 14 cores each and a total of 512GB of RAM. The nodes are connected via Mellanox EDR Infiniband switches in a 2:1 blocking configuration.} hosted by Durham University, for a total of 3.5 million CPU hours over 14,845 timesteps. We note, however, that the version of \swift used did not exploit a domain decomposition algorithm tailored for zoom simulations. Up-coming code improvements in this direction are expected to reduce the required CPU time for such a simulation by at least 25\%. 

\subsection{Post-processing}
\label{sect:post-processing}

Through the course of the \sibeliusdark simulation 200 \squotes{snapshots} were stored between redshifts $z=25$ and $z=0$, spaced linearly in the logarithm of the scale factor. The snapshots were post-processed to produce catalogues of Friend-Of-Friends (FOF) groups, using a linking length of 0.2 times the mean inter-particle separation, and to produce catalogues of bound subhaloes using a heavily modified\footnote{We added a \texttt{MPI} domain decomposition scheme suitable for zoom-in simulations, a feature not present in the public version of {\sc hbt+}.} version of the publicly available \squotes{Hierarchical Bound-Tracing} ({\sc hbt+}) algorithm \citep{Han2012,Han2018}. As a final step, the merger trees of these dark matter subhaloes were constructed using the {\sc dhalos} algorithm described by \citet{Jiang2014}. The combination of snapshots, dark matter subhalo catalogues and merger trees serve as input to semi-analytic models of galaxy formation to produce mock galaxy catalogues, described in the next section. Our high cadence between outputs is necessary to capture accurately the evolution of dark matter halos, as each snapshot is separated by less than the freefall time of the overdensities, and comfortably exceeds the recommended number of outputs required for use with semi-analytic models \citep[e.g.,][]{Benson2012}.

Due to their large size (5.3~TB each), all but 11 of the snapshots containing the complete particle data were deleted following the creation of the subhalo catalogues (the $z$ = 5, 3, 2, 1, 0.5, 0.09, 0.07, 0.05, 0.03, 0.02 \& 0 snapshots were kept).

Halo mass, denoted $M_{\mathrm{200c}}$ ({\tt m200crit} in the public database), is defined as the total mass enclosed within $r_{\mathrm{200c}}$, the radius at which the mean enclosed density is 200 times the critical density of the Universe (i.e., $200 \rho_{\mathrm{crit}}$). Dark matter haloes are populated by one (a central) or more (a central plus satellites) bound substructures (i.e., subhaloes). The total bound mass of the subhalo is denoted $M_{\mathrm{sub}}$ ({\tt Mbound} in the public database).

\subsubsection{The semi-analytic model \galform}
\label{sect:galform}

To infer how the galaxy population of the \sibeliusdark volume evolves, we use the semi-analytic model of galaxy formation, \galform \citep{Lacey2016}. Semi-analytic models, or SAMs, are a computationally efficient method to describe the physical processes giving rise to the formation and evolution of galaxies. They are built upon the output of dark matter-only (DMO) N-body simulations, which offers the possibility of exploring galaxy evolution down to the smallest scales within extremely large cosmological volumes, such as \sibeliusdark, at a fraction of the expense of a full hydrodynamical simulation. The downside of this method, compared to a full hydrodynamical simulation, is its simplicity. For example, SAMs are limited in their ability to explore the internal structure of galaxies, non-symmetric features, and intergalactic gas. Yet recent improvements to semi-analytic modelling do show reasonable agreement with hydrodynamical counterparts \citep{Guo2016,Hou2018,Hou2019}. Moreover, the semi-analytic approach has a distinct statistical advantage: being computationally so inexpensive, it is possible to explore the parameter space of models thoroughly, resulting in an accurate calibration against many observational datasets with a strong predictive power.

The latest \galform model of \citet{Lacey2016} includes a different initial mass function for quiescent star formation and starbursts, feedback from active galactic nuclei to suppress gas cooling in massive halos, and a new empirical star formation law in galaxy disks based on the molecular gas content. In addition, there is a more accurate treatment of dynamical friction acting on satellite galaxies and an updated stellar population model. The magnitudes of galaxies in \galform include the reprocessing of starlight by dust, leading to both dust extinction at ultraviolet to near-infrared wavelengths, and dust emission at far-infrared to sub-mm wavelengths. The absorption and emission is calculated self-consistently from the gas and metal contents of each galaxy and the predicted scale lengths of the disk and bulge components using a radiative transfer model \citep{Lacey2011,Cowley2015,Lacey2016}.

\bigbreak
The simulation data used for this study, and that made available for public release (see \cref{sect:public_data_release}), is not a lightcone, it is the halo and galaxy catalogue at $z=0$. Thus we have assumed a negligible evolution in the positions and properties of the \sibeliusdark galaxies between $z=0.045$ (the edge of the constrained region) and $z=0$. We define the redshift of a galaxy as $z=v_{\mathrm{r}} / c$, where $v_{\mathrm{r}}$ is the radial velocity (which includes the Hubble flow) and $c$ is the speed of light, and we define the apparent magnitude as $m = M + 5 \mathrm{log}_{\mathrm{10}}(d/10)$, where $d$ is the distance to the simulated Milky Way in pc and $M$ is the absolute magnitude.

\section{Results}
\label{sect:results}

\subsection{The halo and galaxy population of \sibeliusdark}
\label{sect:galaxies}

\begin{figure} \includegraphics[width=\columnwidth]{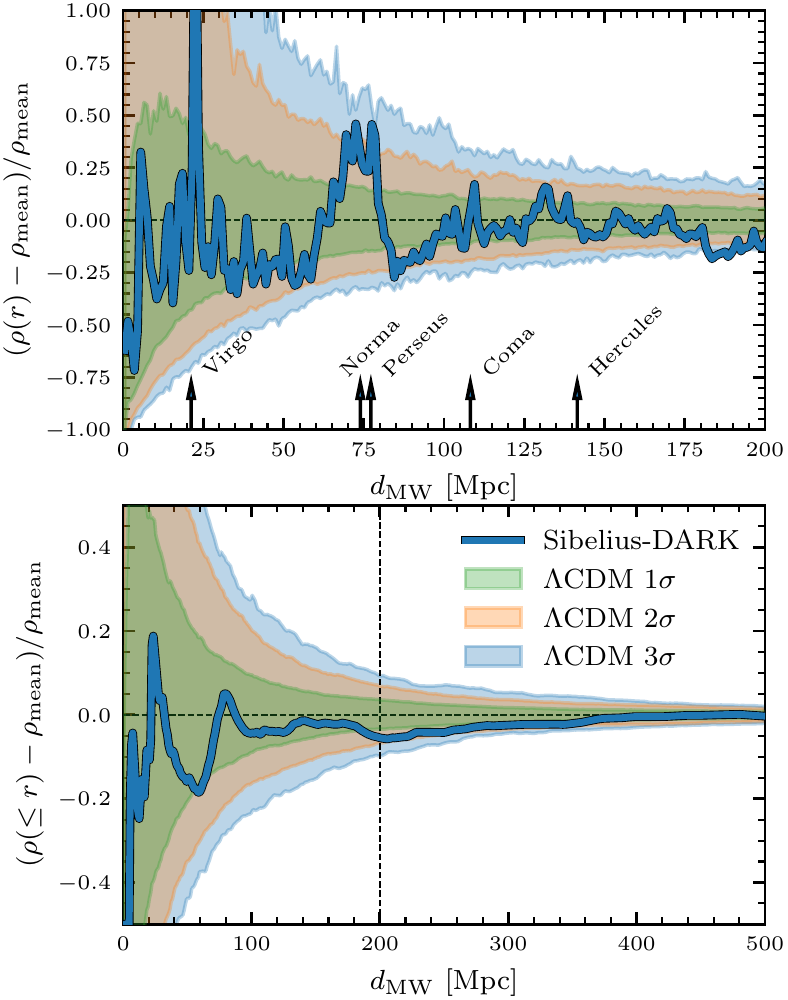}

\caption{Total matter density, relative to the mean cosmic density ($\Omega_{\mathrm{m}} \rho_{\mathrm{crit}}$), within shells (top) and increasingly larger spheres (bottom) centred on the simulated Milky Way. In the upper panel we show the location of 5 \sibeliusdark clusters. The vertical dashed line in the lower panel indicates the boundary of the constrained region, beyond this mark we enter the unconstrained parent volume. The shaded regions indicate the expected range of density fluctuations for $\Lambda$CDM, computed from 1000 random samplings of $R=500$~Mpc spheres within a unconstrained 3.2 Gpc DMO volume. There is a large variation in the average density depending on the volume considered, which explains the systematic shifts of the halo/subhalo mass function in \cref{fig:mass_function}. At the boundary of the constraints, $d_{\mathrm{MW}} = 200$~Mpc, our local volume is predicted to have an overall underdensity of $\approx 5$\%, a $\approx 2 \sigma$ deviation in $\Lambda$CDM. }

\label{fig:density_vs_rad}

\end{figure}

\begin{figure*} \includegraphics[width=\textwidth]{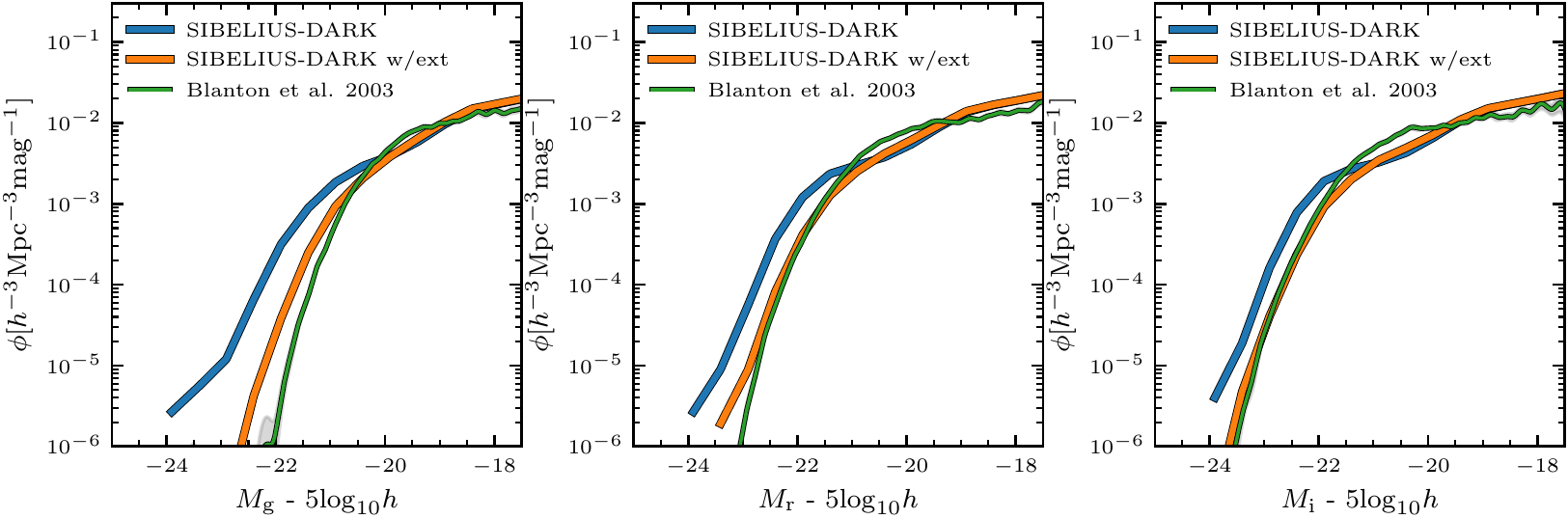}

\caption{The $g,r$ and $i$-band luminosity functions of \sibeliusdark galaxies at $z=0$, both including (orange) and excluding (blue) the effects of dust extinction. Observational data in green is from the SDSS survey \citep[errors as grey bands, but they are almost always smaller than the line width,][]{Blanton2003}.}

\label{fig:sdss_lum}

\end{figure*}

\begin{figure*} \includegraphics[width=\textwidth]{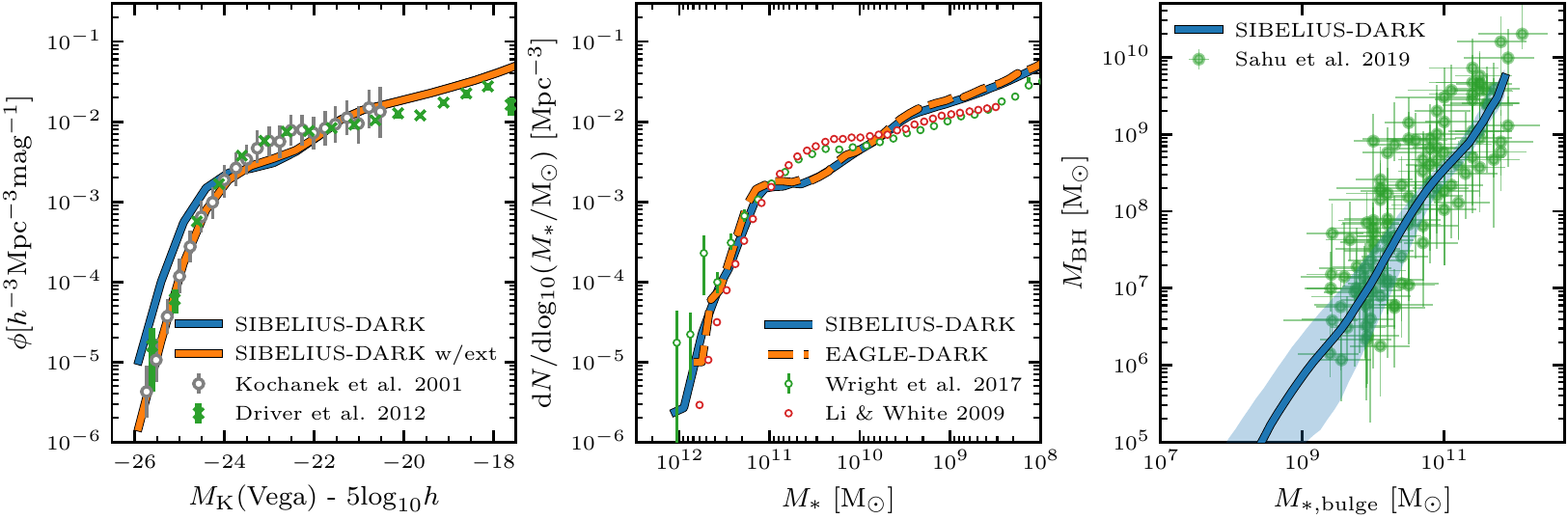}

\caption{\textit{Left}: the $K$-band luminosity function at $z=0$, both including (orange) and excluding (blue) the effects of dust extinction, compared to observational estimates from \citet{Kochanek2001} and \citet{Driver2012}. \textit{Middle}: the galaxy stellar mass function (GSMF) at $z=0$, compared to observational estimates from \citet{Li2009} and \citet{Wright2017}. Overplotted as an orange dashed line is the \galform GSMF from the {\sc eagle-dark} 100~Mpc unconstrained volume, which demonstrates that the ensemble properties of galaxies within \sibeliusdark are predicted to be very similar to those from unconstrained volumes. \textit{Right}: the central supermassive black hole mass--stellar bulge mass relation at $z=0$, compared to the observational estimates from \citet{Sahu2019}. The line is the median value and the shaded region highlights the 10$^{\mathrm{th}}$ to 90$^{\mathrm{th}}$ percentile range.}

\label{fig:K_band_lum}

\end{figure*}

We begin with a statistical investigation of the \sibeliusdark volume in its entirety, that is, a study of all galaxies out to a distance of $d_{\mathrm{MW}} \leq 200$~Mpc from the Milky Way. As a reminder, \sibeliusdark was performed as a DMO simulation; the properties of the galaxies populating those dark matter haloes are computed in post-processing using the semi-analytic model \galform (see \cref{sect:galform}).

At $z=0$ \sibeliusdark hosts 22,904,767 dark matter haloes with a mass in excess of $M_{\mathrm{200c}} \geq 10^{9}$~\Msol, which in turn host 33,220,267 bound dark matter subhaloes above the same mass threshold. Of these haloes, seven exceed $M_{\mathrm{200c}} \geq 1 \times 10^{15}$~\Msol (which is potentially an unusual amount, we discuss this more in \cref{sect:conclusion}), the most massive of which is located at approximately $\approx 77$~Mpc from the Milky Way, the Perseus cluster, with a mass of $M_{\mathrm{200c}} = 2.72 \times 10^{15}$~\Msol. The distribution of halo and subhalo masses within three spherical volumes centred on the Milky Way are shown in \cref{fig:mass_function}. For a comparison, the distribution of halo and subhalo masses from the DMO \eagle reference simulation \citep[{\sc eagle-dark},][]{Schaye2015,Crain2015,McAlpine2016}, a (100~Mpc)$^{3}$ unconstrained periodic volume, are also shown, with the ratio between the \sibeliusdark volumes and the {\sc eagle-dark} volume shown in the lower panels. We note that the {\sc eagle-dark} simulation was performed using the same cosmology and at the same resolution as \sibeliusdark. In addition, we re-post-processed the {\sc eagle-dark} outputs using the {\sc hbt+} structure finder to remain consistent with the \sibeliusdark catalogues.

Over the mass range we explore ($\geq 10^{9}$~\Msol), there are generally fewer haloes/subhaloes per unit volume within \sibeliusdark compared to the {\sc eagle-dark} volume. This deficit, in the range $\approx 2$--10\%, is considerably larger than one would expect from sample variance alone at these scales \citep[$\approx 1$\% at $M_{\mathrm{200c}} \approx 10^{10}$~\Msol;][]{Sawala2021a}. Here, \squotes{sample variance} refers to the expected level of scatter in the halo population depending on the initial Gaussian random field that went in to producing the initial conditions. However, it is important to realise that the level of sample variance measured from studies such as \citet{Sawala2021a} are inferred from periodic cubic volumes that are fixed to the mean density. As the inner regions of \sibeliusdark, and indeed our own local volume, are not guaranteed to reside at the mean density exactly, this creates an additional level of scatter when comparing to unconstrained simulations, such as \eagle, which we refer to as \squotes{cosmic variance}.

The level of cosmic variance within \sibeliusdark is most clearly demonstrated in \cref{fig:density_vs_rad}, showing the density of matter, relative to the mean density, within shells (top) and increasingly larger spheres (bottom) centred on the simulated Milky Way. There are substantial variations in the average shell density depending on the volume being considered, rapidly changing between underdense and overdense regions correlating with the presence (and absence) of massive structures. Any one volume of the Universe is unique, but to give context we show the one, two and three $\sigma$ ranges of density fluctuations which one would expect in the confines of $\Lambda$CDM. These ranges are computed by performing the equivalent calculation from a sampling of 1000 randomly located $R=500$~Mpc spheres within a 3.2~Gpc DMO simulation. With this context, the upper panel of \cref{fig:density_vs_rad} reveals three regions within the \sibeliusdark volume that stand out as \squotes{unusual}: it is rare to have something as massive as the Virgo cluster so nearby, the density of the shells that collectively contain the Norma cluster and Perseus-Pisces superstructure is unusually high (and the voids that they create on either side are unusually underdense), and the shells towards the outer regions of the local volume are anonymously underdense (which could potentially be linked to the evacuated regions before the Shapley concentration). When considered more generally, the spherical volumes in the lower panel reveal that the local volume is largely \squotes{normal}, remaining within $\Lambda$CDM's 1$\sigma$ range. Yet we note that the spherical volumes typically always verge on the side of an underdensity, particularly at $d_{\mathrm{MW}} \approx 50$~Mpc and at the outskirts of the local volume ($d_{\mathrm{MW}} \approx 175-200$~Mpc) where an overall underdensity of $\approx 5$\% is found, a 2$\sigma$ deviation in $\Lambda$CDM \citep[an underdensity that is consistently found across all the \borg realisations, see Figure 10 of ][]{Jasche2019}. It is not until a distance of $d_{\mathrm{MW}} \approx 400$~Mpc that the volume recovers from this underdensity and stabilises to the mean density, which is well into the unconstrained region of the parent volume\footnote{We note that, like with any simulated periodic cosmological volume, when the entirety of the (1~Gpc)$^{3}$ parent volume is considered, \sibeliusdark is forced to be at the mean density by construction.}. We would therefore require constrained initial conditions that go beyond our current limit of $d_{\mathrm{MW}} = 200$~Mpc in order to see the true extent of this underdensity.

 Translating this back to the halo and subhalo mass functions in \cref{fig:mass_function}, it is then clear why, and to what level, there are systematic differences in the number of haloes relative to the unconstrained {\sc eagle-dark} volume, which as a reminder is fixed to the mean density. Overall, this suggests that our local volume, particularly our most immediate neighbourhood ($d_{\mathrm{MW}} < 50$~Mpc), is underdense relative to the cosmic mean. We discuss this more in \cref{sect:local_hole}.

\subsubsection{The galaxy population}
\label{sect:luminosity function}

Turning now to the model galaxy population, \cref{fig:sdss_lum} shows the luminosity function of galaxies at $z=0$ in the three primary bands of the \textit{Sloan Digital Sky Survey} (SDSS); $g,r$ and $i$. Here we include the model prediction both including and excluding the effects of dust extinction, and, to compare, data from the SDSS survey itself \citep{Blanton2003}. There is a general good agreement between the model prediction and the data, with only the brightest $g$-band galaxies being slightly overproduced by the \galform model.  

\cref{fig:K_band_lum} shows, from left to right, the predicted $K$-band luminosity function compared to observational estimates from \citet{Kochanek2001} and \citet{Driver2012}, the galaxy stellar mass function (GSMF) compared to observational estimates from \citet{Li2009} and \citet{Wright2017} and the central supermassive black hole mass--stellar bulge mass relation compared to the observational estimates from \citet{Sahu2019}. The behaviour of the $K$-band magnitudes and the GSMF is similar between the model and the data, differing mostly in the normalisation at the position of the \squotes{knee}, the area of the GSMF most sensitive to the particular implementation of stellar and AGN feedback \citep[e.g.,][]{Bower2012}. This underprediction of intermediate mass galaxies ($M_* \sim 10^{10}$~\Msol) seen in the model population relative to observational estimates is not unique to \sibeliusdark, and is predicted also by the \galform semi-analytic model performed on unconstrained volumes. For example, we overplot the equivalent \galform GSMF using the same 100~Mpc {\sc eagle-dark} volume that was used for \cref{fig:mass_function} \citep[see also the results from][]{Lacey2016, Guo2016}, which additionally demonstrates that the \emph{ensemble} properties of galaxies from the \sibeliusdark volume are predicted to be very similar to those within unconstrained volumes. It is likely that the discrepancy at the knee of the GSMF, at least in part, is due to how the stellar mass is estimated. Here the model stellar masses are directly measured, yet observationally they are inferred from SED fitting, the results of which depend on the stellar population synthesis model used, on assumptions about galaxy star formation histories and metallicity distributions, on the model for dust attenuation, and on the assumed initial mass function. When \galform stellar masses are estimated using observational techniques, the normalisation around the knee is boosted, bringing the prediction much closer to the observational estimates \citep[e.g.,][]{Mitchell2013,Lacey2016}. Finally, the relation between the mass of the central supermassive black hole and the mass of the stellar bulge matches extremely well to the recent observational data of \citet{Sahu2019}. 

\subsubsection{Galaxy number counts}
\label{sect:number_counts}

\begin{figure} \includegraphics[width=\columnwidth]{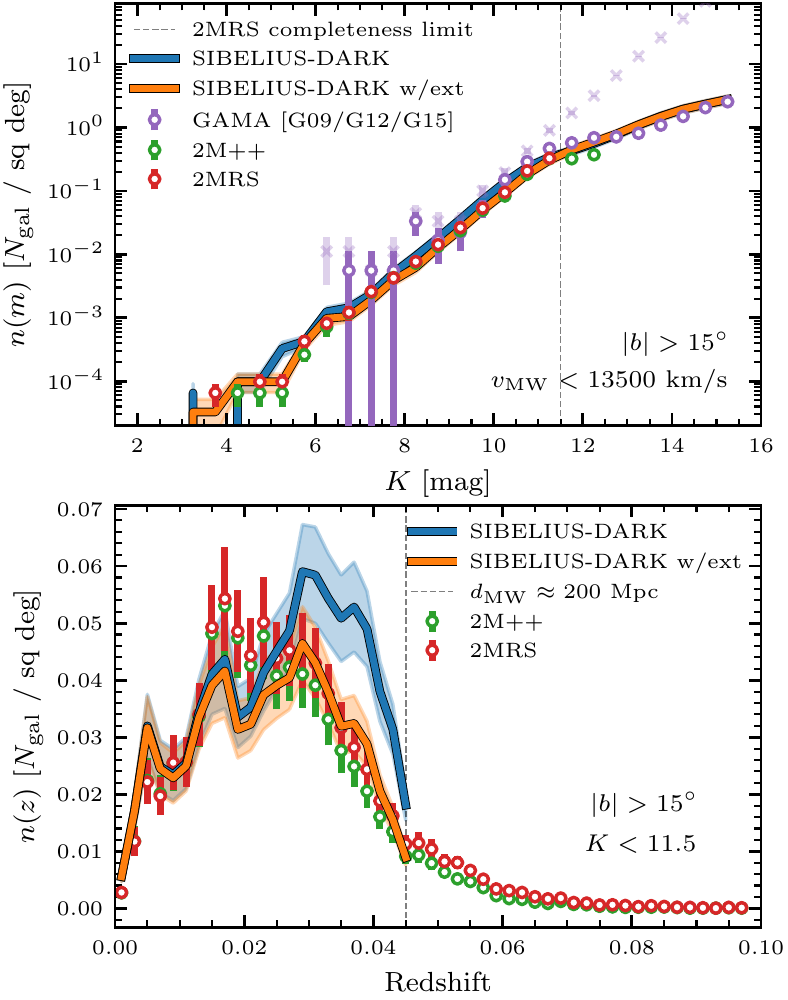}

\caption{\textit{Top}: $K$-band number counts ($n(m)$) of \sibeliusdark galaxies both with (w/) and without dust extinction compared to the 2M++ and 2MRS galaxy samples and the GAMA survey. The galaxies from each dataset are limited to those with $v_{\mathrm{MW}} < 13500$ km/s ($d_{\mathrm{MW}} \lesssim 200$~Mpc). We find excellent agreement between the model galaxies from the simulation and the observational data. Beyond $K \approx 11.5$ the power-law slope shallows due to the limited volume we consider (this is demonstrated by the opaque crosses, which are the same GAMA data with no volume cut applied). \textit{Bottom}: redshift distribution ($n(z)$) of galaxies with $K < 11.5$. Again, there is a good agreement between the simulation and the observations, with only a potential slight underabundance ($\approx 20$\%) of galaxies within \sibeliusdark around $z=0.02$. Errorbars and shaded regions in both panels are \squotes{field-field} errors over $n=12$ equal-area sub-fields covering the entire sky (see text).}

\label{fig:K_band_number_counts}

\end{figure}

\begin{figure} \includegraphics[width=\columnwidth]{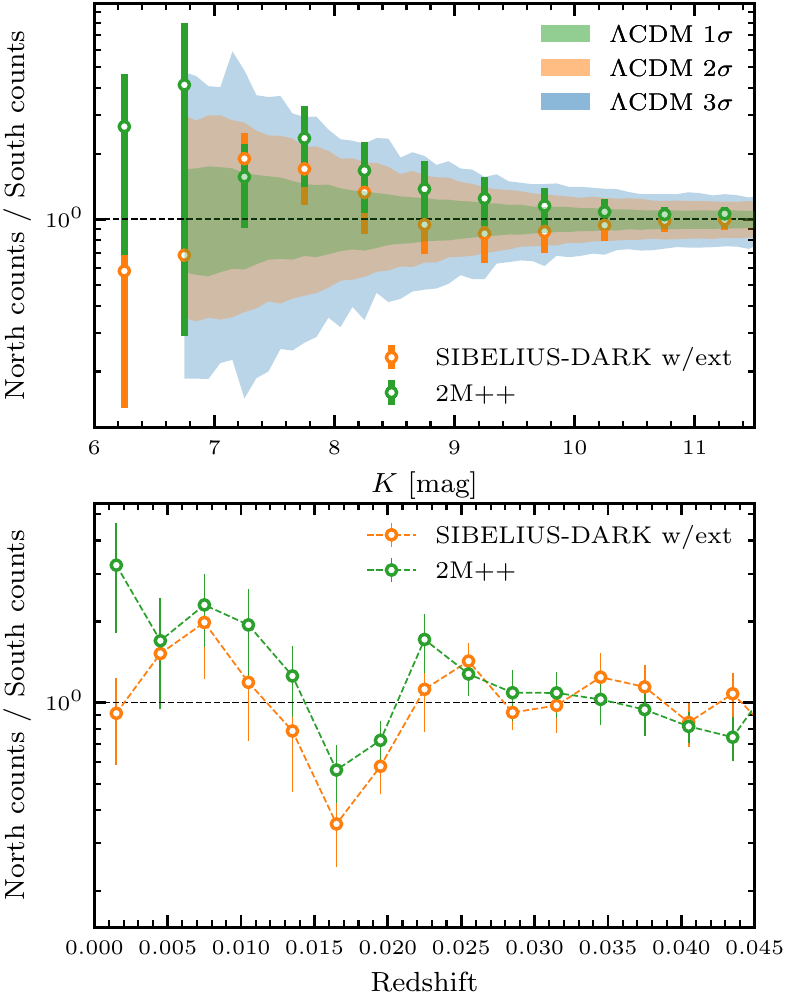}

\caption{Similar to \cref{fig:K_band_number_counts}, now investigating the ratio of counts between galaxies in the Galactic north ($b > 15^\circ$) and the Galactic south ($b < -15^\circ$). \textit{Top}: for both the simulation and the data we find faint galaxies ($K \gtrsim 9$) have approximately equal numbers between the two hemispheres, whereas brighter galaxies ($K \lesssim 9$) are more abundant in the north relative to the south (by a factor of $\approx 2$--3). For the brightest galaxies ($K<7$) however, 2M++ galaxies remain northern dominated, whereas \sibeliusdark galaxies are more equally split. By comparing to 1000 random samplings of the \textit{Millennium} simulation, we find these behaviours are not unusual within a $\Lambda$CDM context, generally falling within the 1--2$\sigma$ range. \textit{Bottom}: the north/south ratio now considered as a function of redshift, which is particularly sensitive to the distribution of large scale structure. For example, there are more galaxies in the southern hemisphere at $z \approx 0.017$ due to the Perseus-Pisces superstructure. The generally good agreement between \sibeliusdark and the data indicates a similar distribution of the large-scale structure.}

\label{fig:north_south_counts}

\end{figure}
When considering the global characteristics of a model galaxy population, such as the stellar mass function, cosmic variance is often mitigated by calculating an ensemble average over many possible distributions of the large scale structure. One can then investigate how these average characteristics compare to our local observations in order to gain theoretical understanding. The power of \sibeliusdark, as a constrained simulation, is that we can now directly compare to the particular large-scale structure distribution of \emph{our} local volume, subject to the same biases in the same directions of the sky. This allows us to mimic observational surveys more directly, not only by applying the limitations of the instrument, but also by focusing on the same particular region of the sky that is surveyed. Here we demonstrate a direct comparison to the 2M++ and 2MRS all-sky galaxy samples \citep{Lavaux2011,Huchra2012}, and the SDSS survey, Data Release Twelve \citep[SDSS-DR12,][]{Ahumada2020}.

The analysis in this section investigates galaxy number counts, both as a function of magnitude ($n(m)$) and as a function of redshift ($n(z)$). We estimate the variance ($\sigma$) for the galaxy counts using the \squotes{field-field} error technique, whereby we compute the area-density of galaxies over $n$ equal-area sub-fields ($\rho_i$) relative to the area-density of galaxies over the entire field we are considering ($\bar{\rho}$), i.e.,

\begin{equation}
    \sigma^2 = \frac{1}{n(n-1)} \sum^{n}_{i=1} (\rho_i - \bar{\rho})^2,
\end{equation}

\begin{figure} \includegraphics[width=\columnwidth]{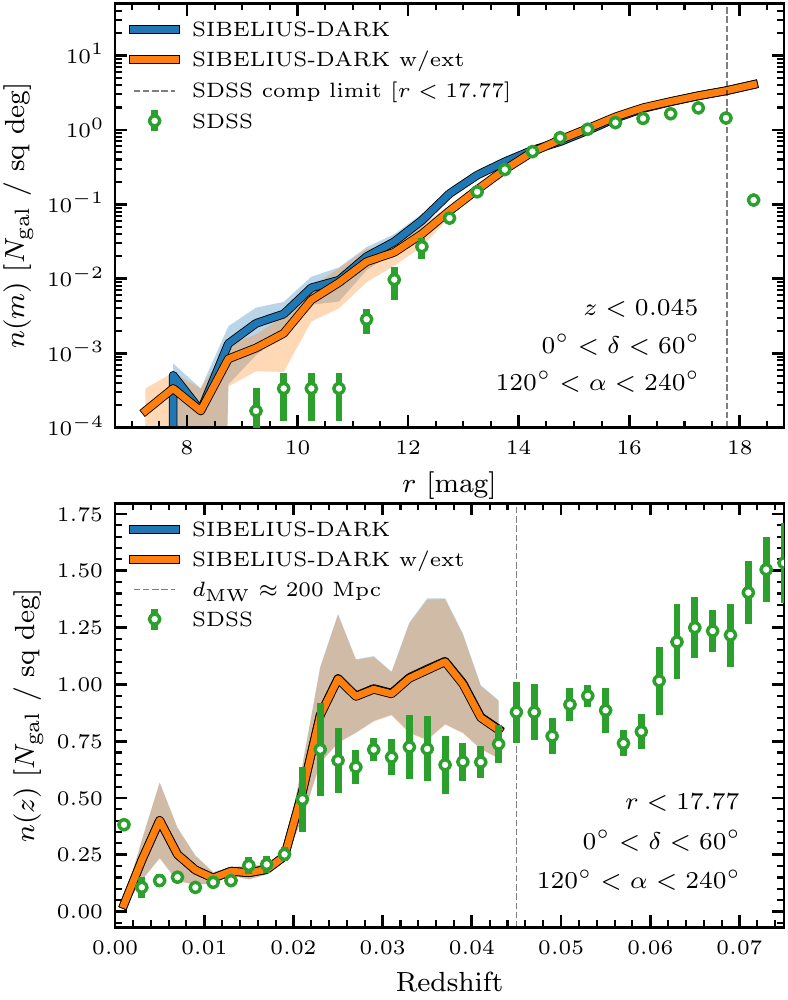}

\caption{The layout of this figure is the same as \cref{fig:K_band_number_counts}, now showing the $n(m)$ and $n(z)$ distributions in the $r$-band for galaxies in the North Galactic Cap region (bounds detailed in figure) of the SDSS survey. In the upper panel only galaxies with $z <0.045$ ($d_{\mathrm{MW}} \lesssim 200$~Mpc) are considered. In the lower panel only galaxies within the spectroscopic completeness limit for SDSS are considered ($r < 17.77$). Errors are \squotes{field-field} errors over $n=6$ equal-area slices of right ascension within the North Galactic Cap (see text). Note in the lower panel the two \sibeliusdark lines overlap.}

\label{fig:r_band_number_counts}

\end{figure}

\noindent \citep[see Section 2.3 of ][ for more details]{Wong2021}. For \cref{fig:K_band_number_counts} we define $n=12$ equal area sub-fields covering the whole sky (excluding $|b| < 15^\circ$), for \cref{fig:north_south_counts} we estimate the field-field errors for the galaxy counts in the Galactic north and the Galactic south from the six sub-fields in the relevant hemisphere, and for \cref{fig:r_band_number_counts} we divide the North Galactic Cap region into $n=6$ equal area slices of right ascension. 

The upper panel of \cref{fig:K_band_number_counts} shows the $K$-band number counts of \sibeliusdark galaxies compared to the 2M++ and 2MRS galaxy samples, the all-sky samples from which the {\sc borg} constraints were derived. Given that the galaxies in both 2M++ and 2MRS go slightly deeper than $d_{\mathrm{MW}} = 200$~Mpc (the boundary of the \sibeliusdark constraints), we limit all model and observed galaxies to those with radial velocities less than $v_\mathrm{MW} < 13500$ km/s. In addition, we only consider galaxies at galactic latitudes $|b| > 15^\circ$ to minimise the effects of obscuration from the galactic disk in the data. We show the counts of \sibeliusdark galaxies both with (w/) and without dust extinction to demonstrate the widest possible range in the semi-analytic model prediction. The $n(m)$ counts reveal a familiar power-law behaviour, with the model galaxies from \sibeliusdark being in excellent agreement, both for the slope and the normalisation, with the galaxies in the 2M++ and 2MRS samples. Approximately just beyond the completeness limit of the data ($K=11.5$) the \sibeliusdark galaxies start to taper off, and continue with a shallower slope towards the faint end. To clarify why this is, we include galaxies from three complete fields of the GAMA survey \citep[G09, G12 \& G15,][]{Baldry2018}, which go much deeper in the $K$-band than 2MRS and 2M++, but cover a much smaller area of the sky. We again find excellent agreement with the \sibeliusdark prediction for the GAMA galaxies, both brighter and fainter than $K=11.5$. The change in slope towards fainter galaxies is due to the limited volume that we are considering ($d_\mathrm{MW} < 200$~Mpc), as the distant bright galaxies are simply not present to bolster the numbers of the nearby intrinsically faint galaxies. Indeed, if we were not to enforce the distance cut of $v_\mathrm{MW} < 13500$~km/s to the GAMA data, the power-law slope continues uninterrupted beyond $K=11.5$ (shown as opaque crosses).

Next we investigate how the model \sibeliusdark galaxies and those from the 2M++ and 2MRS samples are distributed in redshift, which we show in the lower panel of \cref{fig:K_band_number_counts}. Here we only consider the galaxies within the completeness limit of the observational data ($K < 11.5$), and now no longer consider the galaxies from the GAMA survey (as they are not all sky). Both the counts of the observed galaxies and those from \sibeliusdark rise and fall within a radius of 200~Mpc, coming to a peak between redshifts $z=0.02$ and $z=0.03$. There is good general agreement between the trend of the simulation and the trend of the observations, with the 2M++ and 2MRS galaxies often overlapping with the curves of \sibeliusdark. However, there is potentially up to a $\approx 20$\% deficit of \sibeliusdark galaxies around $z=0.02$ compared to the observations, which is approximately at the distance of the Perseus-Pisces and Norma superstructures.     

Previously, in \cref{fig:K_band_number_counts}, we considered the galaxy distribution across the entire sky (excluding only the galactic plane, $|b| > 15^\circ$). Yet we can also narrow our focus to particular sub-regions of the sky, in a test of homogeneity. \cref{fig:north_south_counts} shows, again for the $K$-band, the ratio of the number of galaxies in the Galactic north ($b>15^\circ$) compared to those in the Galactic south ($b< -15^\circ$), both for the 2M++ data and for \sibeliusdark. For both datasets, the behaviour is generally similar over a wide range of magnitudes; galaxies fainter than $K \gtrsim 9$ have approximately equal numbers between the northern and southern hemisphere, whereas brighter galaxies in the range $7 \gtrsim K \gtrsim 9$ are more abundant in the north relative to the south (up to a factor of $\approx 2$ for \sibeliusdark and up to a factor of $\approx 2.5$ for 2M++). It is the brightest galaxies in the 2M++ sample that show the largest difference between the two hemispheres, with 58 of the 79 galaxies brighter than $K < 7$ found in the Galactic north, whereas there are more \sibeliusdark galaxies in the Galactic south in this regime. In future work it will be interesting to see how sensitive the number of bright galaxies between the northern and southern hemispheres is to the particular realisation of the \borg constraints. 

To test the likelihood of such a north--south disparity for bright galaxies is in the context of $\Lambda$CDM, we use the \textit{Millennium} simulation (MS-W7)\footnote{We use specifically the WMAP7 \textit{Millennium} simulation, MS-W7, as it was performed using the same \galform semi-analytic model as \sibeliusdark \citep{Guo2013}. We note that the resolution of MS-W7 is $\approx 100$ times lower than that of \sibeliusdark, and a slightly different cosmology was used (WMAP7).}, a publicly available $\Lambda$CDM simulation with a volume $\approx 15$ times greater than \sibeliusdark. We perform 1000 random samplings of $R=200$~Mpc spheres within the MS-W7 simulation and compare the counts in the $K$-band between two arbitrary hemispheres (excluding declinations within 15$^\circ$ to remain consistent with our data). The one, two and three $\sigma$ ranges for the counts are shown as shaded regions on \cref{fig:north_south_counts}. We find that a factor of $\approx 2$ prevalence for bright galaxies in the northern hemisphere is entirely consistent with a random $\Lambda$CDM realisation, falling within the $1$-$2\sigma$ ranges.

For completeness, the lower panel of \cref{fig:north_south_counts} shows again the ratio of the counts in the Galactic north relative to the Galactic south, but now distributed by redshift. The north/south divide here is particularly sensitive to the layout of structures within the volume. For example, there are more galaxies in the north when the distribution is dominated by the Hydra and Centaurus clusters ($z \approx 0.005$). This then changes to an increased number of galaxies in the south as we pass through the Perseus and Norma clusters ($z \approx 0.017$). The ratio briefly reverts to northern dominated as we pass by Coma and Leo ($z \approx 0.025$), and then averages closer to unity as the distance increases and the volumes becomes larger ($z \gtrsim 0.030$). As was seen in the panel above, the galaxies from the 2M++ sample are often slightly more northern dominated than the galaxies from \sibeliusdark, which is most evident for nearby galaxies where the volume is the smallest ($z \lesssim 0.005$). Overall, the pattern of behaviour between \sibeliusdark and the observations is encouragingly consistent, indicating a similar distribution of the large-scale structure.

It is extremely encouraging how well the \sibeliusdark $n(m)$ and $n(z)$ distributions in the $K$-band match to the 2M++ and 2MRS observations, the same all-sky datasets the \borg constraints were derived from. In \cref{fig:r_band_number_counts} we conduct a similar investigation, now comparing to data from the SDSS survey, which was not used for constructing the constraints. Here we are now considering galaxies in the $r$-band, again with the upper panel showing the $n(m)$ distribution and the lower panel showing the $n(z)$ distribution. We select galaxies from the largest contiguous region of the SDSS survey, the North Galactic Cap ($0^\circ < \delta < 60^\circ$ \& $120^\circ < \alpha < 240^\circ$), and, again, only consider galaxies within $d_{\mathrm{MW}} < 200$~Mpc ($z \lesssim 0.045$). 

The counts from \sibeliusdark show a dual power-law behaviour, transitioning at magnitude $r \approx 15$. Compared to the galaxies from SDSS, \sibeliusdark contains 5--10 times more bright galaxies ($r \lesssim 12$). This is not entirely unexpected, as the spectroscopic target selection in SDSS removes the brightest galaxies from the sample as not to saturate the spectroscopic CCDs, or contaminate the spectra of adjacent fibers \citep{Strauss2002}. In fact, this result could serve as a prediction for filling in the missing galaxies at the bright end. At the low mass end, \sibeliusdark also predicts more galaxies near the completeness limit. However this may simply be due to the fact that the completeness limit is not a hard cut, or a particular feature relating to the limited volume we are considering ($z \leq 0.045$). Indeed, we already see the telltale signs of incompleteness in the SDSS data before the $r = 17.77$ limit in this region. The $n(z)$ distribution of \sibeliusdark galaxies in the lower panel also matches well the behaviour of the SDSS data, sharing the same jump when passing through the Coma and Leo clusters at $z \approx 0.02$. Reflecting the differences in the panel above, \sibeliusdark has up to $\approx 25$\% more galaxies at $z \approx 0.025$--0.045 relative to the SDSS data.

\subsection{Clusters within the local volume}
\label{sect:clusters_and_groups}

\begin{figure*} 
\includegraphics[width=\textwidth]{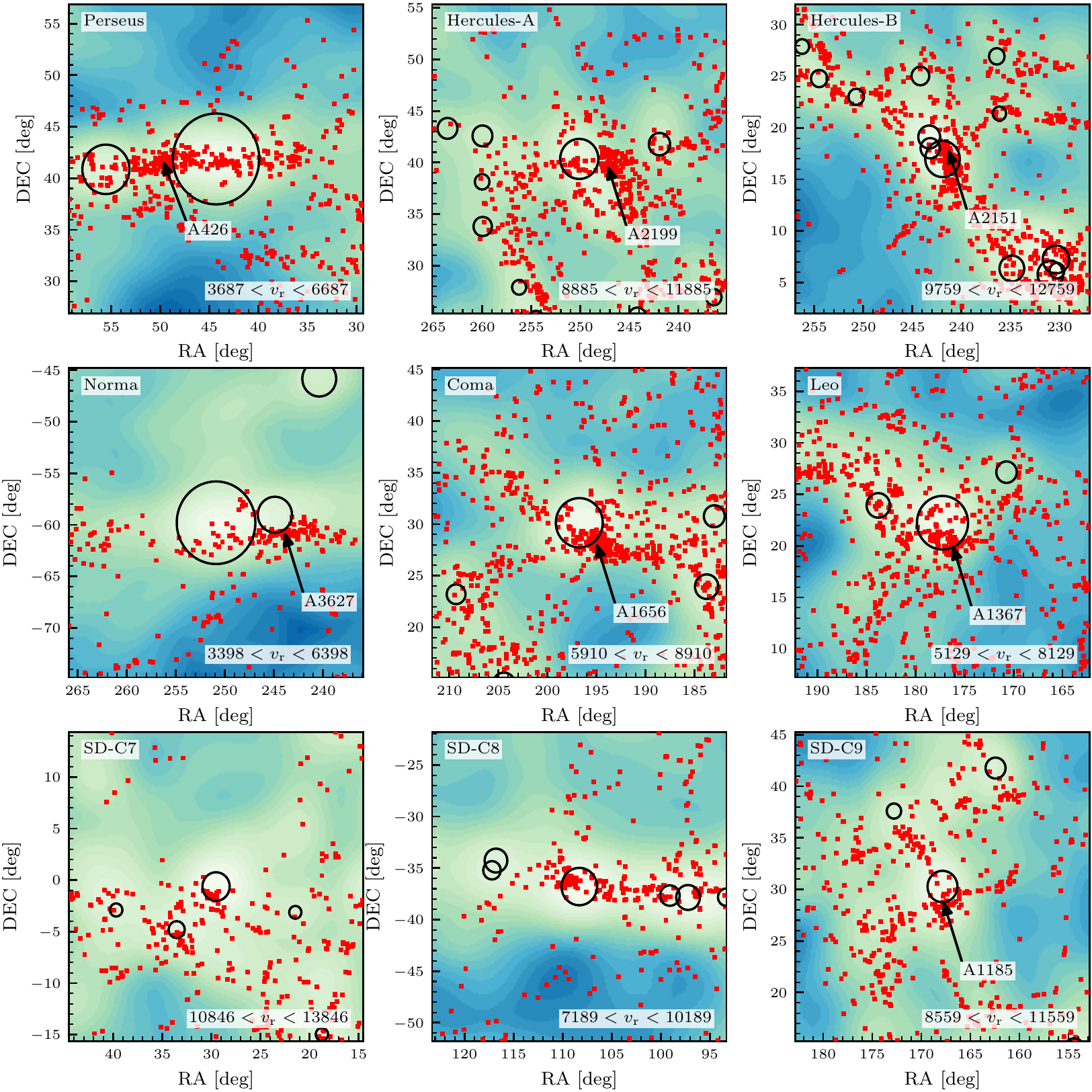}

\caption{The large-scale structure surrounding the nine most massive haloes in the \sibeliusdark volume. Each panel covers an RA and DEC of $\pm 15$ degrees and $v_{\mathrm{r}} \pm 1500$ km/s surrounding the particular \sibeliusdark halo. The contours in the background show the \sibeliusdark galaxy density, increasing in logarithmic density from blue to green. Any \sibeliusdark haloes more massive than $M_{\mathrm{200c}} \geq 10^{14}$~\Msol are shown as black circles, with a radius of $r_{\mathrm{200c}}$. Overplotted in red are galaxies from the 2M++ catalogue and, annotated with arrows, the location of the richest Abell cluster \citep{Abell1958,Abell1989} in the region. There is no Abell cluster in the vicinity of SC-C7 and SC-C8, however there is a concentration of 2M++ galaxies at the location of the haloes. The properties of the nine clusters are listed in \cref{table:cluster_masses} and the properties of their most massive galaxies are listed in \cref{table:cluster_bcgs}.}

\label{fig:cluster_locations_1}

\end{figure*}

\begin{figure*} 
\includegraphics[width=\textwidth]{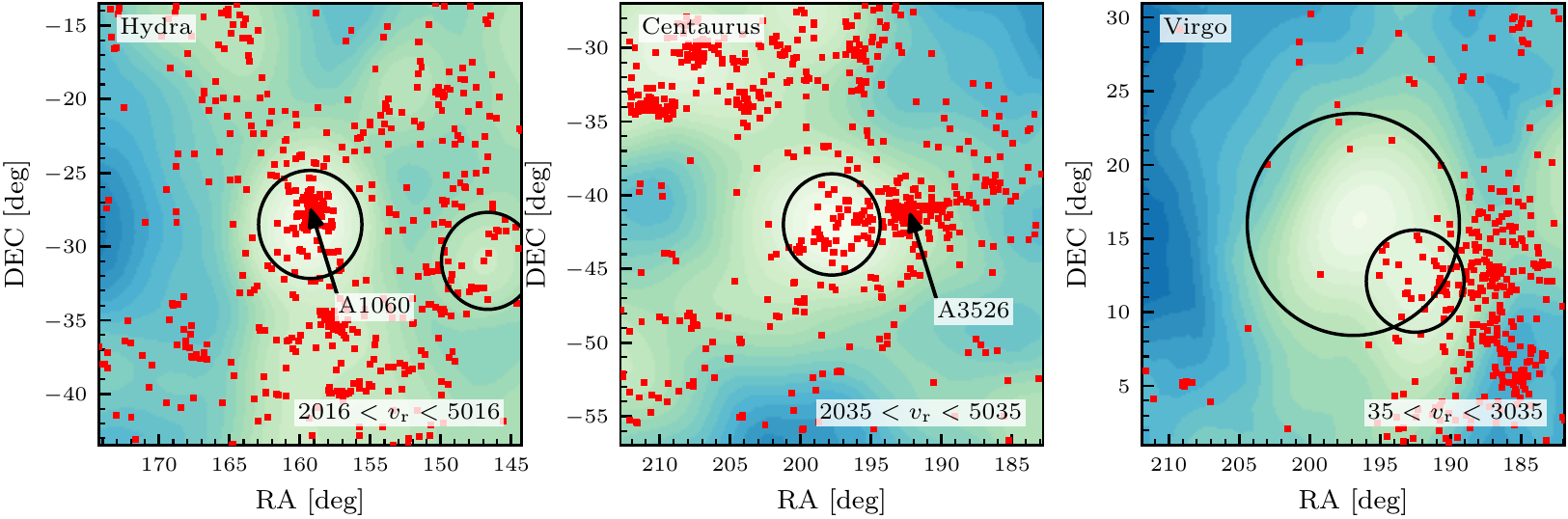}

\caption{As \cref{fig:cluster_locations_1}, now for the \textit{Centaurus}, \textit{Hydra} and \textit{Virgo} clusters.}

\label{fig:cluster_locations_2}

\end{figure*}

We now focus attention on particular structures within the \sibeliusdark volume, comparing directly the properties of a selection of observed clusters to their \sibeliusdark counterparts. As a reminder, the constrained phases that generate the initial conditions of the simulation are designed statistically to reproduce the density field of the local volume \emph{en masse}, providing no guarantee that any one specific object will be oriented at the exact position on the sky or located at the exact distance that we observe. To that end, we must identify the \sibeliusdark analogues initially by eye, a process we outline here for twelve massive clusters in the simulation, including: \textit{Perseus}, \textit{Hercules}, \textit{Norma}, \textit{Coma}, \textit{Leo}, \textit{Centaurus}, \textit{Hydra} and \textit{Virgo}. For users of the public database who wish to compare to other specific structures in the \sibeliusdark volume, we recommend a similar association approach as done below. The halo properties of the twelve clusters are listed in the appendix in \cref{table:cluster_masses}, including their associated cluster from the Abell catalogue \citep{Abell1958,Abell1989}. Here, we directly compare only the halo properties of these clusters to observations, leaving a direct comparison of the Brightest Cluster Galaxies (BCGs) to future work. However we list the properties of the most massive model galaxies in each cluster in the appendix in \cref{table:cluster_bcgs}.

To avoid confusion when describing our results, in this section we refer to the model clusters from \sibeliusdark using a \squotes{SD} suffix (e.g., \csd{Virgo}). Similarly, when referring to a cluster from the real Universe, we refer to it either via its Abell catalogue number (e.g., A1060), or suffixed with a \squotes{*} (e.g., \cob{Virgo}). All halo masses quoted below, both from the simulation and the observations, are $M_{\mathrm{200c}}$.  

\cref{fig:cluster_locations_1} shows the density field of galaxies surrounding the nine most massive haloes within the \sibeliusdark volume. All galaxies within an RA and DEC of $\pm 15$ degrees and $v_{\mathrm{r}} \pm 1500$ km/s of the \sibeliusdark cluster centre are shown. Within the region, any \sibeliusdark halo more massive than $M_{\mathrm{200c}} \geq 10^{14}$~\Msol is shown as a black circle, with a radius of $r_{\mathrm{200c}}$. Overplotted in red are galaxies from the 2M++ galaxy sample that cover the same volume, and we also indicate the location of the richest cluster from the Abell catalogue in the region. We then associate the most massive \sibeliusdark halo that is closest to the Abell cluster location as the observational analogue. For seven of the nine haloes we are able unambiguously to associate an observed Abell cluster to their model halos. In two cases, \csd{SD-C7} and \csd{SD-C8}, there is no Abell cluster in the region; yet there is a clear concentration of 2M++ galaxies in each case. We repeat this process for three famous lower mass clusters, \textit{Hydra}, \textit{Centaurus}, and \textit{Virgo} in \cref{fig:cluster_locations_2}. Below, we examine how these \sibeliusdark clusters compare to their observed counterparts. For many of the observed halo mass estimates we refer to those within the recent compilation of \citet{Stopyra2021} and references therein.

\textbf{Perseus}: the \cob{Perseus-Pisces} Supercluster is one of the most massive structures in the Local Universe, a long, dense wall of galaxies with a length of almost 100~Mpc (see \cref{fig:sky_maps}). At one end of this wall lies the supercluster's most dominant member, the \cob{Perseus} cluster (A426), one of the most massive clusters observed within the local volume, and with a mass of $M_{\mathrm{200c}} = 2.72 \times 10^{15}$~\Msol, the most massive halo within \sibeliusdark. There is considerable disagreement in the literature as to the mass of the \cob{Perseus} cluster, ranging from a lower value of $\approx 9 \times 10^{14}$~\Msol using X-ray data \citep{Simionescu2011}, to an upper value of $\approx 3 \times 10^{15}$~\Msol from dynamical estimates \citep{Meusinger2020}. This puts \csd{Perseus} within the observed range, albeit towards the upper limit.

\textbf{Hercules-A \& Hercules-B}: the \cob{Hercules} Superclusters are a pairing of two superclusters connected over a large area of the northern sky: a northern supercluster, to whose richest member, A2199, we will refer to as \csd{Hercules-A}, and a southern supercluster, whose richest member is the \cob{Hercules} cluster itself (A2151), to which we will refer as \csd{Hercules-B}. These two clusters, \csd{Hercules-A} and \csd{Hercules-B}, constitute the 2$^{\mathrm{nd}}$ and 3$^{\mathrm{rd}}$ most massive haloes in the \sibeliusdark volume, with masses $M_{\mathrm{200c}} = 1.89 \times 10^{15}$~\Msol and $M_{\mathrm{200c}} = 1.78 \times 10^{15}$~\Msol, respectively. Observationally, the mass estimates for \cob{Hercules-A} are again uncertain, spanning almost a decade in mass from $\approx 2 \times 10^{14}$~\Msol to $\approx 2 \times 10^{15}$~\Msol between X-ray \citep{Piffaretti2011}, dynamical \citep{Kopylova2013,Lopes2018}, SZ \citep{Planck2016b} and weak lensing \citep{Kubo2009} estimators. The estimates for the mass of \cob{Hercules-A} are likely complicated by the ongoing merger with its partner cluster, A2197 \citep{Kremp2002}. For \cob{Hercules-B}, the estimated mass range from the observations of (0.79-- 1.89) $\times 10^{15}$\Msol \citep{Lopes2018} is similar to \cob{Hercules-A}. As with the supercluster of \cob{Hercules-A}, the supercluster of \cob{Hercules-B} is likely also dynamically bound and collapsing \citep{Kremp2002,Kopylova2013}, which could significantly disrupt the resulting mass estimate. Compared to the \sibeliusdark haloes, both \csd{Hercules-A} and \csd{Hercules-B} lie within the observed mass range, both towards the upper ends of the estimated ranges.

\textbf{Norma}: the \cob{Norma} cluster lies extremely close to the galaxy's \squotes{zone-of-avoidance}, where extinction is severe, and our understanding of the surrounding local large-scale structure remains incomplete. \cob{Norma} is perhaps most famous for its connection to the \dquotes{Great Attractor}, a region of the sky (shared by the Hydra, Centaurus, Norma and Shapley clusters) towards which many local galaxies, including the Local Group, are streaming \emph{en masse} against the direction of the Hubble flow \citep{Lynden-Bell1988}. Due to its location, the mass of the \cob{Norma} cluster, and its nature, remain poorly constrained, but is estimated to be in the range $\approx 3 \times 10^{14}$~\Msol to $\approx 2 \times 10^{15}$~\Msol, with the upper end coming from dynamical estimates \citep{Woudt2008}. In \sibeliusdark, \csd{Norma} is the 4$^{\mathrm{th}}$ most massive halo, with a mass of $M_{\mathrm{200c}} = 1.72 \times 10^{15}$~\Msol, close to the observed dynamical mass estimates of the cluster.

\textbf{Coma \& Leo}: the \cob{Coma} Supercluster is one of the most famous and well-studied structures within the Local Universe. Located at a distance of approximately 100 Mpc, it consists of two primary members, the \cob{Coma} Cluster (A1656) and the \cob{Leo} Cluster (A1367), which are connected to one another via a rich network of filaments, in which a large number of groups and galaxies are embedded \citep[e.g.,][]{Williams1981,Rines2001,Seth2020}. \csd{Coma} is the 5$^{\mathrm{th}}$ most massive halo within the \sibeliusdark volume, with a mass of $M_{\mathrm{200c}} = 1.27 \times 10^{15}$~\Msol, which lies within the range of observational estimates $\approx 3 \times 10^{14}$~\Msol to $\approx 2 \times 10^{15}$~\Msol \citep{Piffaretti2011,Planck2016b,Kubo2009}. \csd{Leo}, the second primary member of the \cob{Coma} Supercluster, is the 6$^{\mathrm{th}}$ most massive halo in \sibeliusdark directly behind \csd{Coma}, with a mass of $M_{\mathrm{200c}} = 1.17 \times 10^{15}$~\Msol. This makes \csd{Leo} noticeably more massive than the observational estimates \citep[with an upper limit of $\approx 4 \times 10^{14}$~\Msol from dynamical estimates,][]{Rines2003}. However, we note that \cob{Leo} is observed to be a young, dynamically active system, with multiple ongoing mergers in the inner regions \citep{Cortese2004}, which could affect the mass estimates.

\textbf{SD-C7 \& SD-C8 \& SD-C9}: the 7$^{\mathrm{th}}$, 8$^{\mathrm{th}}$ and 9$^{\mathrm{th}}$ most massive haloes within \sibeliusdark have no commonly named counterpart in the data. Thus, we simply refer to them by their mass rank. Only \csd{SD-C9} has a counterpart in the Abell catalogue of clusters, A1185. Yet we note that \csd{SD-C7} and \csd{SD-C8} are clearly associated with a grouping of galaxies in the same location in the 2M++ catalogue.   

\textbf{Hydra \& Centaurus}: \cob{Hydra} and \cob{Centaurus} are a famous cluster pair in the southern sky, located in the foreground of \cob{Norma}. The two clusters, A1060 and A3526, are less massive than the nine clusters discussed above, yet are thought to be major contributors to the structure of the Great Attractor \citep[e.g.,][]{Raychaudhury1989}. In \sibeliusdark, \csd{Hydra} is the 32$^{\mathrm{nd}}$ most massive halo in the volume at $M_{\mathrm{200c}} = 4.9 \times 10^{14}$~\Msol, and \csd{Centaurus} is the 44$^{\mathrm{th}}$ most massive, at $M_{\mathrm{200c}} = 4.1 \times 10^{14}$~\Msol. These masses are both well in line with observational estimates for the two clusters \citep[$\sim 10^{14}$~\Msol, e.g.,][]{Richtler2011}. 

\textbf{Virgo}: the \cob{Virgo} cluster, at a distance of $\approx 16$ Mpc, is our most massive neighbour, a dynamically young and unrelaxed cluster with a relatively large  number of  substructures \citep{Binggeli1987}. Our \csd{Virgo} Cluster, shown in the right panel of \cref{fig:cluster_locations_2}, is located close to the observed location on the sky, yet is slightly too distant ($\approx 21$~Mpc). From the right panel of \cref{fig:cluster_locations_2}, we see two haloes more massive than $10^{14}$~\Msol in the vicinity which, at first glance, would suggest an ongoing merger; however the smaller of the two haloes is approximately 15~Mpc behind \csd{Virgo}, and is not itself a member of the cluster. \csd{Virgo} is the 56$^{\mathrm{th}}$ most massive halo in \sibeliusdark, at $M_{\mathrm{200c}} = 3.5 \times 10^{14}$~\Msol, which is slightly below recent mass estimates for the cluster \citep[$\approx 5$--$7 \times 10^{14}$~\Msol,][]{Shaya2017,Kashibadze2020}.   
  
In this section we have investigated twelve clusters on the \sibeliusdark sky to see how well the constraints match the observations for particular objects. For each model cluster, we are encouraged that there is always a concentration of observed galaxies in close proximity, and the halo masses predicted by the simulation are often within the bounds of observational estimates. For all but two of these model haloes, \csd{SD-C7} and \csd{SD-C8}, we are able to associate a catalogued Abell cluster directly with them. This gives us confidence that these structures are reasonable analogues of the observed Universe, allowing us to make predictions as to their particular structure and formation. In the next section we focus more closely on the particular structure of the \csd{Virgo} and \csd{Coma} clusters.   

\subsection{A closer look at the Virgo and Coma clusters}
\label{sect:virgo_and_coma}

\begin{figure} \includegraphics[width=\columnwidth]{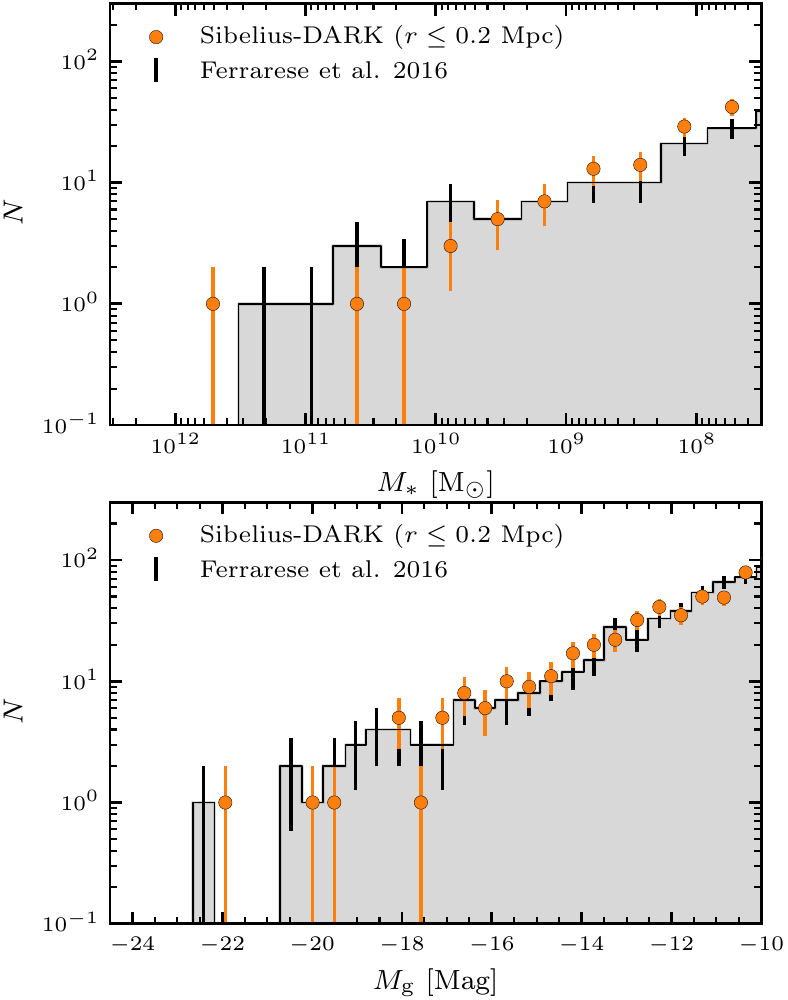}

\caption{The galaxies within the core ($r=0.2$~Mpc) of the \sibeliusdark Virgo cluster (orange points); distributed by stellar mass (upper panel) and $g$-band absolute magnitude (lower panel). We compare against observational data from the \textit{Next Generation Virgo Cluster Survey} \citep[grey histograms and black errorbars,][]{Ferrarese2016}. Poisson errors are shown in both cases. There is a good agreement between the member population of the model \sibeliusdark Virgo cluster and the data.}

\label{fig:virgo_lf}

\end{figure}

In the previous section we established that the large-scale structure surrounding the most massive objects in the \sibeliusdark volume matches well the observational data. Here we take the comparison one step further, by investigating the member galaxy populations of two of the local volume's most famous occupants; Virgo and Coma. The analysis in \cref{sect:virgo_cluster,sect:coma_cluster} are deliberately brief, as we reserve a more in depth comparison to the properties and evolution of particular structures to future work. Our goal here is to reassure the reader that \sibeliusdark produces reasonable analogues to observed clusters, giving confidence to then go on and use these analogues to make predictions as to a plausible nature and evolution. One such example prediction is presented in \cref{sect:splashback_radius}, where we investigate the location and observability of the \squotes{splashback radius} for the two clusters. We remind the reader however, that as \sibeliusdark is built upon a single \borg realisation, one designed to reproduce the local volume \textit{en masse}, we caution against over-interpreting the exact nature of individual structures.  

\subsubsection{Observing the Virgo cluster}
\label{sect:virgo_cluster}

The Virgo cluster remains to this day one of the most well studied objects within our Universe; at relatively close proximity \citep[$\approx 16$~Mpc,][]{Blakeslee2009}, with such a large mass \citep[$\approx 5 \times 10^{14}$~\Msol,][]{Kashibadze2020}, Virgo is a prime target for studying cluster environments. Indeed, the Virgo cluster has long been a testbed for the luminosity function in cluster environments, with results often being revised as deeper and more complete data becomes available, particularly for the prediction of the faint-end slope \citep[$\alpha \approx -1 \rightarrow -2$, e.g.,][]{Sandage1985,Binggeli1985,Impey1998,Rines2008}.    

The \textit{Next Generation Virgo Cluster Survey} \citep[NGVCS,][]{Ferrarese2012} is the deepest and most complete optical survey of the Virgo cluster's core region (the innermost $\approx 3.7$~deg$^{2}$), down to a point-source depth of $g \approx 25.7$ mag. In \cref{fig:virgo_lf} we show the distributions of stellar mass and absolute $g$-band magnitude for the galaxies in the NGVCS survey \citep{Ferrarese2016}. As the data have already been corrected to remove foreground and background interlopers, we compare against the distributions of \sibeliusdark galaxies within a $r=0.2$~Mpc spherical aperture centred on the \sibeliusdark M87 analogue, which approximately equates to the surveyed area. The match between the \sibeliusdark Virgo analogue and the data is extremely encouraging; the magnitude gap of $\approx 1$ between M87 and the next brightest member is reproduced, the overall number of bright galaxies in the cluster is consistent, and the slope and normalisation of the faint end agree well. Considering also the stellar mass in the upper panel, we again find a good agreement between the two datasets.       
 
At the centre of the Virgo cluster lies M87, its central elliptical galaxy. The \sibeliusdark analogue for M87 has a stellar bulge mass of $M_{\mathrm{*,bulge}} = 3.6 \times 10^{11}$~\Msol and hosts a supermassive black hole of mass $M_{\mathrm{BH}} = 3.2 \times 10^{9}$~\Msol. Compared to the compilation data from \citet{Sahu2019} (plotted in \cref{fig:K_band_lum}), with observed masses of $M_{\mathrm{*,bulge}} = (1.7$--$5.6) \times 10^{11}$~\Msol and $M_{\mathrm{BH}} = (5.2$--$6.3) \times 10^{9}$~\Msol, this puts the \sibeliusdark analogue for M87 well within a factor of 2 for both properties.

\subsubsection{Observing the Coma cluster}
\label{sect:coma_cluster}

\begin{figure} \includegraphics[width=\columnwidth]{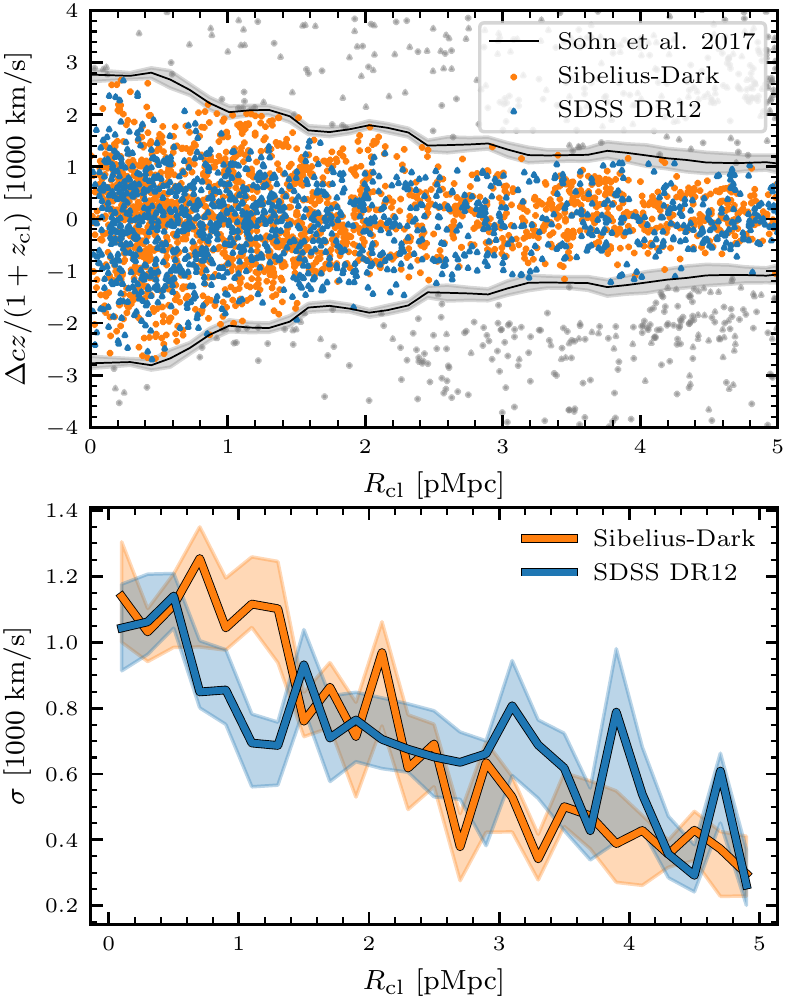}

\caption{\textit{Top}: rest-frame cluster-centric radial velocity versus projected cluster-centric distance for SDSS (blue) and \sibeliusdark (orange) galaxies surrounding their respective Coma clusters. Member galaxies are those that fall within the caustic outlines (grey), computed by \citet{Sohn2017} using SDSS data. \textit{Bottom}: velocity dispersion of member galaxies within each projected radial bin. Errors are from bootstrap resampling.}

\label{fig:coma_vr}

\end{figure}

The Coma cluster is more massive ($> 10^{15}$~\Msol) and more distant ($\approx 100$~Mpc) than Virgo, which makes identifying cluster members much more challenging, and requires robust spectroscopic redshifts to reduce contamination by foreground and background objects. Applying the caustic technique is a popular observational method to identify cluster members using spectroscopic samples \citep{Diaferio1997}.

In the upper panel of \cref{fig:coma_vr} we examine the observed galaxies from SDSS DR12 data that surround the Coma cluster, showing the rest-frame cluster-centric radial velocity versus projected cluster-centric distance. The values for the radial velocity and position of the observed Coma cluster (i.e., the reference frame) are taken from \citet{Sohn2017}, and only galaxies within the spectroscopic completeness limit of SDSS, $r < 17.77$, are considered. Cluster members are identified as those that fall within the outlined caustic, which has been computed by \citet{Sohn2017} using the same SDSS data. Also shown are the \sibeliusdark Coma member galaxies, with both properties now computed from the reference frame of the \sibeliusdark Coma BCG (detailed in \cref{table:cluster_bcgs}). We use the same caustic outline to identify the member galaxies of the \sibeliusdark Coma cluster.

By eye, in the upper panel of \cref{fig:coma_vr}, there is overall good agreement between the model and observed populations. To examine the comparison more closely, the bottom panel of \cref{fig:coma_vr} shows the one-sigma dispersion of velocities within each projected radial bin. The errors here are computed by bootstrap resampling the velocities within each projected radial bin, showing the 10$^{\mathrm{th}}$ to 90$^{\mathrm{th}}$ percentile range of the resulting dispersion. The trends for the model and observed Coma cluster are encouragingly similar; the dispersion of velocities decreases with increasing distance from the cluster centre, as one would expect, from $\approx 1100$~km/s at the cluster core, to $\approx 650$~km/s at 2.25~projected Mpc \citep[the approximate location of the observed virial radius for Coma,][]{Sohn2017}, with the only disparity coming at $\approx 1$~projected Mpc, where the \sibeliusdark Coma cluster has a higher dispersion.

At the centre of the Coma cluster lies NGC4889, its brightest galaxy. The \sibeliusdark analogue for NGC4889 has a stellar bulge mass of $M_{\mathrm{*,bulge}} = 8.6 \times 10^{11}$~\Msol and hosts a supermassive black hole of mass $M_{\mathrm{BH}} = 5.3 \times 10^{9}$~\Msol. Compared to the compilation data from \citet{Sahu2019} (plotted in \cref{fig:K_band_lum}), with observed masses of $M_{\mathrm{*,bulge}} = (0.9$--$1.7) \times 10^{12}$~\Msol and $M_{\mathrm{BH}} = (0.6$--$3.7) \times 10^{10}$~\Msol, this puts the \sibeliusdark analogue right at the lower mass estimate for both the galaxy and the central black hole. However, it is worth noting that NGC4889 hosts one of the most massive black holes ever discovered in the nearby Universe, lying far above the predicted black hole mass for the mass of the stellar bulge \citep{McConnell2011}, which would be challenging for any model of galaxy formation to reproduce exactly.

The results from \cref{fig:virgo_lf,fig:coma_vr} have given us confidence that the clusters in \sibeliusdark are reasonable analogues of the observational counterparts, which we can then go on and use to make predictions.

\subsubsection{A theoretical prediction for the \dquotes{Splashback Radius} of Virgo and Coma}
\label{sect:splashback_radius}

\begin{figure} \includegraphics[width=\columnwidth]{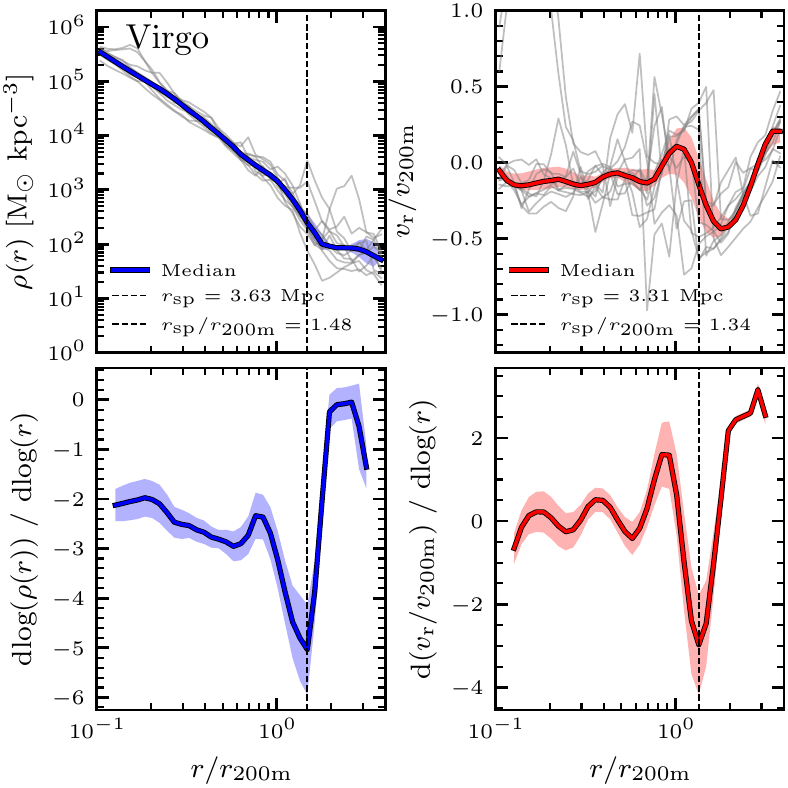}

\caption{\textit{Top}: the density (left) and radial velocity (right) profiles of the dark matter particles measured from the Virgo clusters potential centre. Individual grey lines show the profiles measured within 10 arbitrary bins of azimuthal angle, with the solid coloured lines representing the median density/velocity within each radial bin between the 10 sightlines. The error on the median, shown as a shaded region, is computed via bootstrap resampling. \textit{Bottom}: the logarithmic slopes of the above profiles. Here we can clearly identify the characteristic change in gradient at the outer edges of the halo relating to the \dquotes{splashback radius}, $r_{\mathrm{sp}}$, measured to be $3.63 \pm 0.34$~Mpc and $3.31 \pm 0.31$~Mpc from the density and velocity profiles, respectively. The quoted errors for $r_{\mathrm{sp}}$ are the bin width.}

\label{fig:caustics_virgo}

\end{figure}

\begin{figure} \includegraphics[width=\columnwidth]{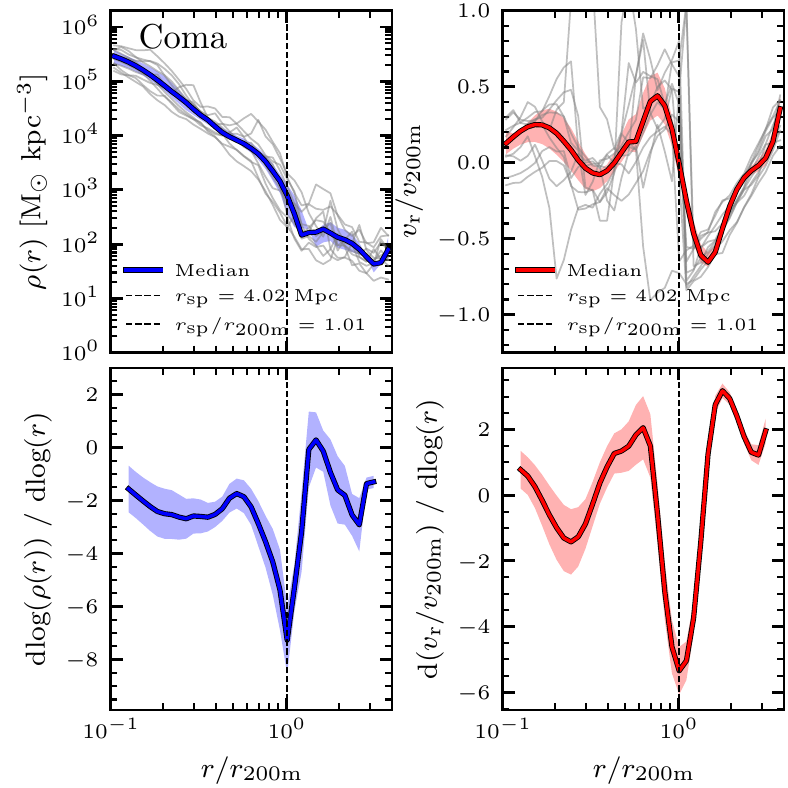}

\caption{As \cref{fig:caustics_virgo}, now for the Coma cluster. The \dquotes{splashback radius} for the Coma cluster is measured to be $4.02 \pm 0.38$~Mpc from both the density and velocity profiles.}

\label{fig:caustics_coma}

\end{figure}

In recent years an increasing amount of attention has been focused on the \dquotes{splashback radius} of dark matter haloes, a caustic formed from the \squotes{bunching up} of mass elements that have just reached the apocenter of their first orbits \citep[e.g,][]{Adhikari2014}. The physical signature of this effect is seen as a sudden drop in the outer regions of the density profile \citep[e.g.,][]{Diemer2014}, creating a divergence from the universal analytic profiles discussed in the literature, such as the Navarro-Frenk-White \citep[NFW,][]{Navarro1996,Navarro1997} and Einasto profiles \citep{Navarro2010}. It is suggested that the splashback radius provides a more intuitive metric to define the size of a dark matter halo, and avoids the shortcomings of more arbitrary definitions of halo extent which are coupled to the expansion of the Universe, such as those based on a chosen density contrast \citep[e.g.,][]{Diemer2013}. To recover the splashback radius one requires an adequate tracer of the underlying density profile within the halo, from either the dark matter or stellar distribution, or the subhalo population. The splashback radius can additionally be recovered in velocity space from the same tracers. We note that whilst \sibeliusdark is a DMO simulation, it has been shown that the inclusion of hydrodynamics has almost no effect on the location of the splashback radius \citep[e.g.,][]{Contigiani2020, ONeil2020}.

\cref{fig:caustics_virgo,fig:caustics_coma} show, respectively, Virgo and Coma's clustercentric dark matter density and radial velocity profiles. As is common in this field, we scale the radii by $r_{\mathrm{200m}}$, i.e., the radius at which the haloes enclosed mass reaches 200 times the \emph{mean} density ($200\Omega_m \rho_{\mathrm{crit}}$), and the radial velocities by $v_{\mathrm{200m}} = \sqrt{G M_{\mathrm{200m}} / r_{\mathrm{200m}}}$, i.e., the circular velocity at $r_{\mathrm{200m}}$. We compute the profiles in 40 equally spaced logarithmic bins of $r/r_{\mathrm{200m}}$ in the range $\in [-1, 0.6]$, and in ten azimuthal bins of 36 degrees centred on the clusters potential minimum. The profiles presented are the median density/velocity in each radial bin between the multiple sightlines, a process designed to minimise the underlying bias from interloping massive substructures within the halo in a particular direction\footnote{We acknowledge that if the accretion onto the halo is skewed by the presence of a few dense filaments, then the splashback radius may itself be a function of direction, a subtlety that this smoothing process would wash out \citep{Contigiani2020}. However we retain the spherical assumption here for simplicity.} \citep[e.g.,][]{Mansfield2017,Deason2020}. We estimate the error on the median value by bootstrap resampling the multiple sightlines, indicated by the shaded regions, and we then apply a fourth-order smoothing algorithm \citep{Savitzky1964} to the median profiles in order to better recover the features. Finally, we compute the gradient of the median lines in the upper panels using linear regression, fitting the slope of the median lines in intervals of five bins using SciPy's {\sc curve\_fit} function \citep{SciPy2020}. This gradient is then shown in the lower panels. The splashback radius, $r_{\mathrm{sp}}$, is identified as the sudden dip in these gradients, seen at approximately the virial radius.

For both the Virgo and Coma clusters, the density and velocity profiles in \cref{fig:caustics_virgo,fig:caustics_coma} each show a clear primary caustic in the lower panels at $\approx r_{\mathrm{200m}}$. It is worth noting that both haloes additionally show tentative evidence for a secondary caustic within $r_{\mathrm{200m}}$, which is theoretically linked to the second apocenter passage of material within the halo \citep[a feature which becomes more prominent for haloes at lower accretion rates, e.g.,][]{Adhikari2014}. However, given that here we are only considering the profiles of individual haloes, whose results can be particularly sensitive to recent/ongoing mergers or asymmetrical accretion, we choose not to overinterpret the trends, and only concentrate on the primary caustic at the splashback radius.

The splashback radius of Virgo is $r_{\mathrm{sp}} = 3.63 \pm 0.34$~Mpc and that of Coma is $4.02 \pm 0.38$~Mpc measured from the density profile and $r_{\mathrm{sp}} = 3.31 \pm 0.31$~Mpc and $4.02 \pm 0.38$~Mpc measured from the velocity profile. Or, when scaled by $r_{\mathrm{200m}}$, they are equivalently $r_{\mathrm{sp}} / r_{\mathrm{200m}} = 1.48 \pm 0.14$ and $1.01 \pm 0.09$ measured from the density profile and $r_{\mathrm{sp}} / r_{\mathrm{200m}} = 1.34 \pm 0.12$ and $1.01 \pm 0.09$ measured from the velocity profile for Virgo and Coma respectively. The quoted errors are the bin width. Many factors have an impact on the location of the splashback radius for a given halo; its mass, the cosmology, the redshift, and perhaps most importantly, the halo accretion rate,

\begin{figure} \includegraphics[width=\columnwidth]{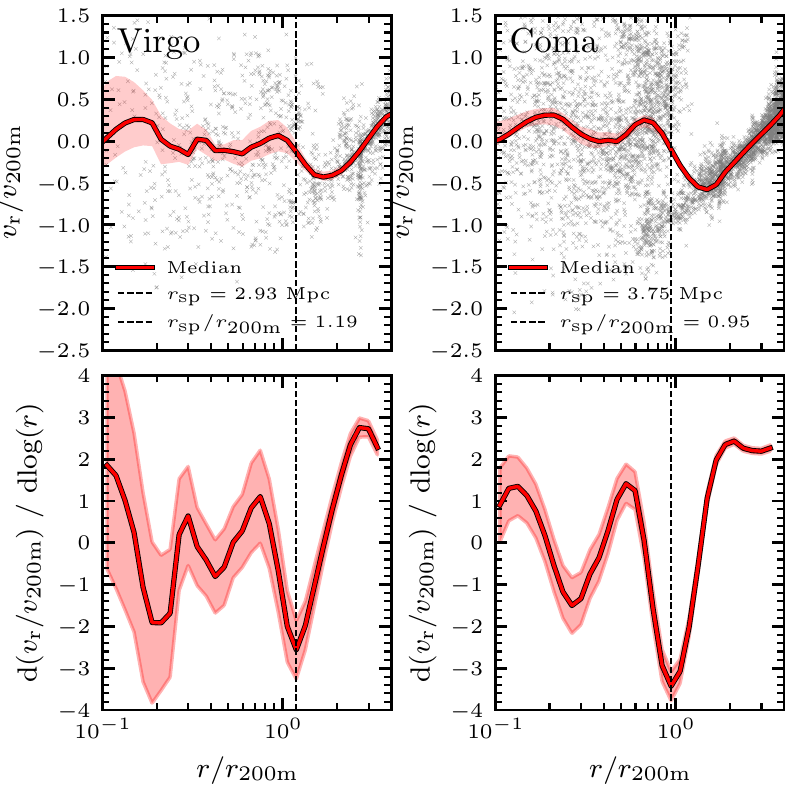}

\caption{A repeat of the analysis from the right hand panels of \cref{fig:caustics_virgo,fig:caustics_coma}, now using the clustercentric radial velocities and distances of the galaxies, rather than the dark matter particles. The grey points mark individual galaxies with stellar masses $M_{\mathrm{*}} \geq 10^{8}$~\Msol, and the solid lines represent the smoothed medians. We again find the primary caustic of the splashback feature, at $2.92 \pm 0.32$~Mpc and $3.75 \pm 0.41$~Mpc for the Virgo and Coma clusters, respectively. These values are $\approx 10$\% lower than the radii found when directly using the dark matter particles.}

\label{fig:caustics_subhaloes}

\end{figure}

\begin{figure} \includegraphics[width=\columnwidth]{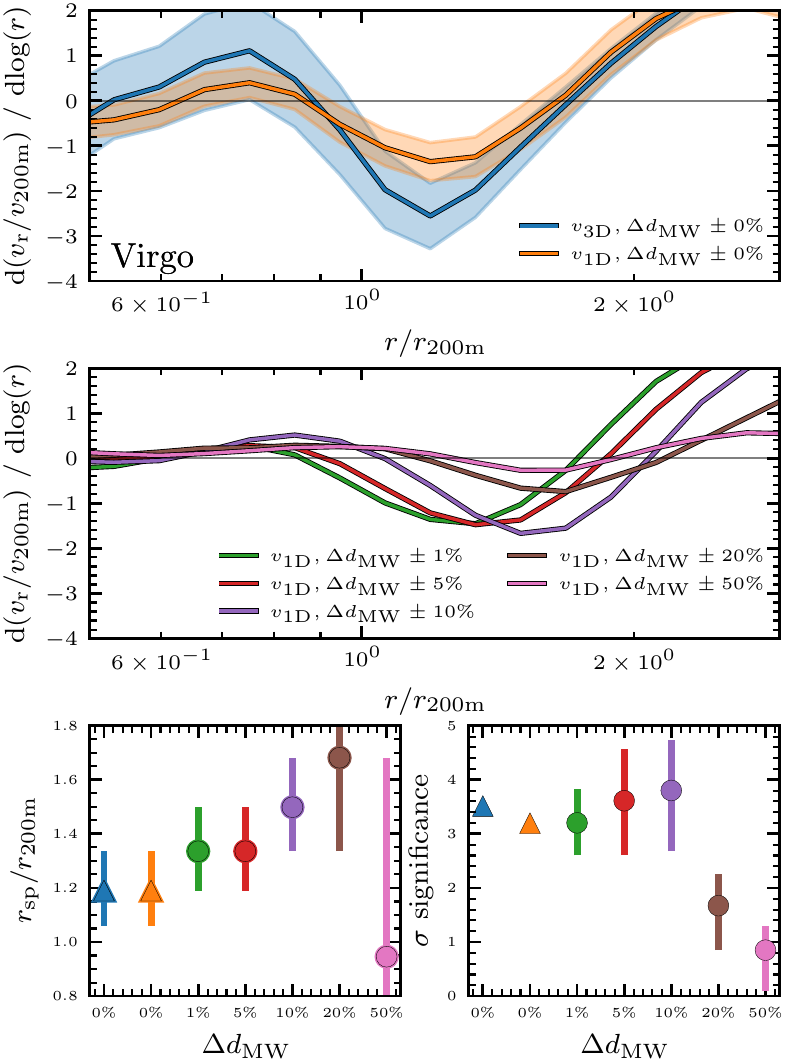}

\caption{\squotes{Observing} the splashback radius of the Virgo cluster using member galaxies above $M_* \geq 10^{8}$~\Msol. The top panel is a repeat of the lower left panel of \cref{fig:caustics_subhaloes}, i.e., the gradient of the clustercentric radial velocities, using 3D and 1D (line-of-sight velocity) velocity information. Simply going from 3D to 1D velocities already serves to reduce the prominence of the dip in the splashback feature. In the middle panel we now include random Milky Way distance errors to each galaxy sampled from a uniform distribution of the quoted magnitude. For each value of the distance error, we perform the experiment 100 times, with each line in the middle panel showing the median value of d($v_{\mathrm{r}}/v_{\mathrm{200m}}$)/dlog($r$) between the 100 measurements. No errors (shaded regions) are shown in this panel for clarity, yet they are comparable to those in the panel above. As the distance error increases, so to does the location of the splashback feature (shown in the lower left panel). Additionally, as the distance error increases above 10\%, the splashback feature is no longer significantly detected (lower right panel). We define the significance ($\sigma$) to be the ratio of the value of the splashback feature below zero divided by the error. The values in the lower panels are the median values between the 100 measurements, and the errors are the 10$^{\mathrm{th}}$ to 90$^{\mathrm{th}}$ percentile range (or for just the lower left panel, at least the bin width).}

\label{fig:caustics_subhaloes_observed}

\end{figure}

\begin{equation}
    \Gamma_{\mathrm{0.5}} = \frac{\Delta \mathrm{log} M_{\mathrm{200m}}}{\Delta \mathrm{log}(1+z)},
\end{equation}

\noindent computed in the range $z = 0.5$ to $z=0$ \citep{Diemer2014,More2015,Diemer2017}. In \sibeliusdark, the Coma cluster has a smaller value of $r_{\mathrm{sp}}/r_{\mathrm{200m}}$ both because it is more massive, and because it has a higher accretion rate ($\Gamma_{\mathrm{0.5}}[\mathrm{Coma}] = 5.4$ vs. $\Gamma_{\mathrm{0.5}}[\mathrm{Virgo}] = 1.5$). We note that the measured values for $r_{\mathrm{sp}}/r_{\mathrm{200m}}$ are consistent with the theoretical trends for the general population of dark matter haloes \citep{More2015,Diemer2017,Contigiani2020}. 

\begin{figure*} 
\includegraphics[width=\textwidth]{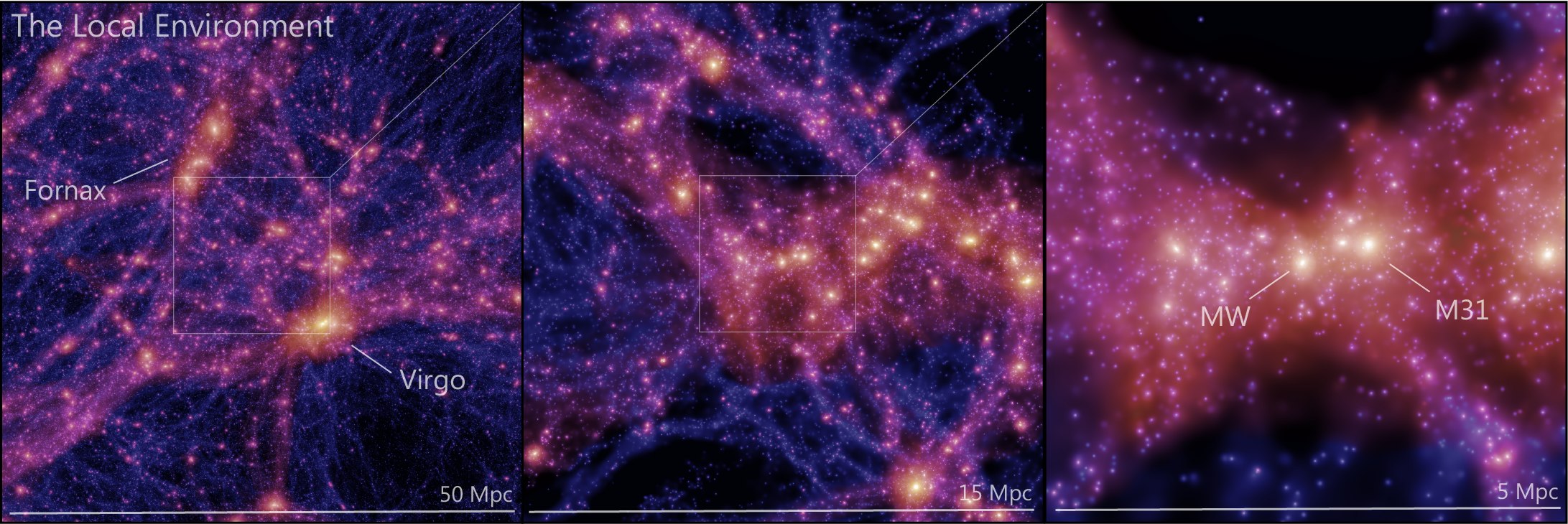}

\caption{The left panel shows the dark matter distribution in a 50x50x50~Mpc region centred on the Milky Way, coloured by the projected density and velocity dispersion of the particles. Our two most massive neighbours, the Virgo cluster and the Fornax/Eridanus groups, are highlighted. The middle and right panels zoom into a 15x15x15~Mpc and 5x5x5~Mpc region, respectively. The right panel highlights the location of the Milky Way and Andromeda (M31). Images are shown in $y$--$z$ equatorial coordinates, projected down the $x$ axis.}

\label{fig:local_environment}

\end{figure*}

Until now we have been considering the density and radial velocity profiles of Virgo and Coma using the abundance of dark matter particles. Yet, the location of the splashback feature can also be retrieved using the subhalo population of the cluster. \cref{fig:caustics_subhaloes} repeats the analysis of the right hand panels of  \cref{fig:caustics_virgo,fig:caustics_coma}, now using the clustercentric radial velocities and distances of the subhalo members with stellar masses $M_{\mathrm{*}} \geq 10^{8}$~\Msol. Due to the reduced numbers involved in this method, we no longer average the profiles over multiple azimuthal bins, and instead consider complete radial shells. The errors on the median are computed by bootstrap resampling the subhaloes within each radial bin, and the gradients in the lower panels are computed in the same way as for \cref{fig:caustics_virgo,fig:caustics_coma}. Once again the primary caustic of the splashback radius is robustly detected, yet now at values $\approx 10$\% lower than was found when directly using the dark matter particles. This is consistent with previous simulation studies, which argue that the increased dynamical friction felt by the subhalo members may be the cause for the smaller radii \citep{Xhakaj2020}.

We conclude this section with a brief observational experiment. The splashback feature has been detected observationally in galaxy clusters by stacking the data over many sources, either using the luminosity density profiles of the galaxy members \citep[e.g.,][]{More2016,Baxter2017,Nishizawa2018,Murata2020,Bianconi2021} or via weak gravitational lensing \citep{Contigiani2019}. The feature has also tentatively been observed in the intracluster light of individual clusters \citep[e.g.,][]{Gonzalez2021}. Here we consider a similar approach, investigating if the splashback radius of the \sibeliusdark Virgo cluster can be recovered using the line-of-sight velocities of the member galaxy population.

We select the galaxies with stellar masses greater than $M_* \geq 10^{8}$~\Msol within $3 \times r_{200{\mathrm{m}}}$ of the \sibeliusdark Virgo cluster. The clustercentric distances and clustercentric radial velocities of the Virgo members are computed \squotes{observationally}; using their distance from the Milky Way ($d_{\mathrm{MW}}$), their line-of-sight velocity ($v_{\mathrm{r}}$) and their angular position relative to the Virgo BCG using the transformations of \citet{Karachentsev2006} \citep[see also the Appendix of ][]{Sorce2021}. When considering distance errors to the Virgo member galaxies in the analysis below, we add a random scatter to the Milky Way distance ($\Delta d_{\mathrm{MW}}$) drawn from a uniform distribution of the quoted magnitude before computing the clustercentric properties. We assume no errors on the line-of-sight velocities of the cluster members or their positions on the sky, and we add no distance error to the Virgo BCG (i.e., our central reference point for the cluster).

We present the results in \cref{fig:caustics_subhaloes_observed}. The upper panel is a repeat of the lower left panel from \cref{fig:caustics_subhaloes}, i.e., the gradient of the slope of the clustercentric radial velocities for Virgo galaxies (blue line), computed using the full three-dimensional velocity information, with the characteristic dip at $r_{\mathrm{sp}}/r_{\mathrm{200}} = 1.19$ being the splashback radius as seen previously. In orange is the equivalent measurement, but now we have estimated the clustercentric radial velocities using the one-dimensional line-of-sight velocities as viewed from the Milky Way. In both cases we have not included an error on the galaxy distance from the Milky Way ($\Delta d_{\mathrm{MW}} = 0$). From this we can already see that the loss of velocity information (i.e., 3D $\rightarrow$ 1D) serves to reduce the prominence of the splashback feature, yet it does not change its location.

In the middle panel of \cref{fig:caustics_subhaloes_observed} we now include random errors on the distance from the Milky Way to the Virgo member galaxies ($\Delta d_{\mathrm{MW}}$), to investigate how this influences the location of the splashback radius. To do this we sample a random distance error for each galaxy from a uniform distribution of the quoted magnitude, compute d($v_{\mathrm{r}}/v_{\mathrm{200m}}$)/dlog($r$) for this new realisation, and retrieve the location of $r_{\mathrm{sp}} / r_{\mathrm{200m}}$. We then repeat this experiment 100 times for each magnitude of the distance error, with the lines in the middle panel showing the median values of d($v_{\mathrm{r}}/v_{\mathrm{200m}}$)/dlog($r$) between the 100 realisations. The median location of $r_{\mathrm{sp}} / r_{\mathrm{200m}}$ between the 100 realisations as a function of the distance error is shown in the lower left panel, including also the location of the splashback feature without any distance errors (i.e., the values from the upper panel). The errorbars represent the uncertainty, which is the maximum value between the bin width and the 10$^{\mathrm{th}}$ to 90$^{\mathrm{th}}$ percentile range between the 100 realisations (when applicable). We define the \squotes{significance} ($\sigma$) of the feature as the ratio between the absolute value of d($v_{\mathrm{r}}/v_{\mathrm{200m}}$)/dlog($r$) divided by the width of the error on d($v_{\mathrm{r}}/v_{\mathrm{200m}}$)/dlog($r$). The lower right panel shows the median value of $\sigma$ as a function of distance error between the 100 realisations (the errorbars represent the 10$^{\mathrm{th}}$ to 90$^{\mathrm{th}}$ percentile range).  

We find that as the distance error increases, the location of the splashback feature also increases to higher radii, from $r_{\mathrm{sp}}/r_{\mathrm{200}} \approx 1.3$ at $\Delta d_{\mathrm{MW}} = 1$\% to $r_{\mathrm{sp}}/r_{\mathrm{200}} \approx 1.5$ at $\Delta d_{\mathrm{MW}} = 10$\%. The significance of the dip is also dependent on the distance error, with distance errors below 10\% remaining detectable with a significance on average above $3\sigma$, yet when the distance errors increase above 10\% the feature is essentially lost. We then predict that only with a complete census of the Virgo cluster member galaxies above $M_* \geq 10^{8}$~\Msol, with distance errors $\Delta d_{\mathrm{MW}} \leq 10$\%, can the splashback feature be recovered. 

\subsection{The Local Group \& the Local Neighbourhood}
\label{sect:local_group}

\begin{figure} \includegraphics[width=\columnwidth]{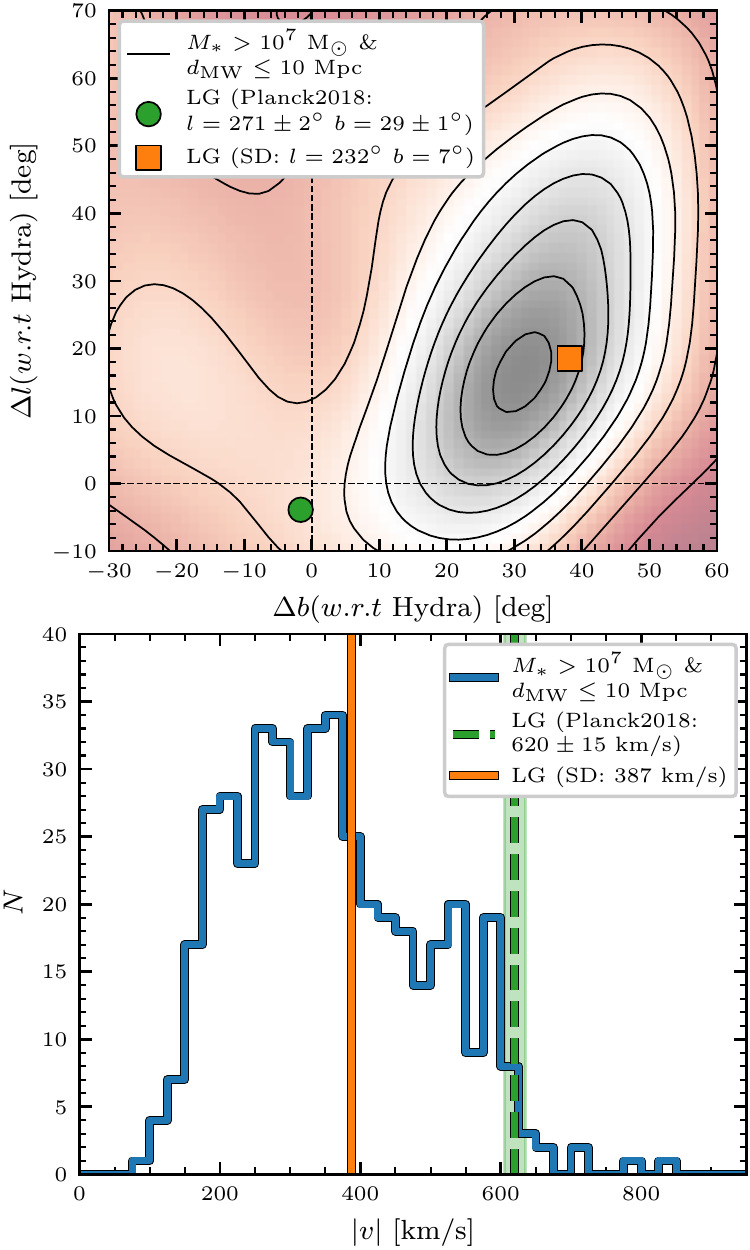}

\caption{\textit{Top}: direction of the velocity vectors in the CMB reference frame for central galaxies with mass $M_* \geq 10^{7}$~\Msol within a distance of 10~Mpc from the Milky Way. The velocity vectors are shown with respect to the location of the \sibeliusdark Hydra cluster on the sky, \emph{as viewed from each galaxy's reference frame}, using the galactic coordinate system ($l$, $b$). The observed direction of the Local Group's motion points very close to the Hydra cluster \citep[]{Planck2018}, whereas the \sibeliusdark Local Groups's motion, and that of the majority of galaxies surrounding the \sibeliusdark Local Group, points $\Delta l \approx +20^\circ$ and $\Delta b \approx +30^\circ$ relative to the Hydra cluster. \textit{Bottom}: magnitude of the velocity for each galaxy in the CMB reference frame. The \sibeliusdark Local Group has a peculiar velocity $\approx 225$~km/s less than the observed value \citep{Planck2018}.}

\label{fig:lg_velocity}

\end{figure}

At approximately the centre of the \sibeliusdark volume lies a dark-matter halo pair that matches closely the observed dynamics of our own Local Group's Milky Way and Andromeda haloes, a deliberate result of the embedment procedure that created the constrained initial condition set (see \cref{sect:constraints}). All the properties detailed below are also listed in \cref{table:lg}. The model Milky Way and Andromeda have halo masses $M_{\mathrm{200}} = 0.9 \times 10^{12}$~\Msol and $M_{\mathrm{200}} = 1.3 \times 10^{12}$~\Msol, respectively, matching well the current observational estimates \citep[$\sim 10^{12}$~\Msol, i.e.,][]{Watkins2010,Posti2019,Cautun2020}. Viewed from the model Milky Way, the model Andromeda galaxy is at a distance of 753~kpc, putting it well within the observed range \citep[$752 \pm 27$~kpc, ][]{Riess2012}; has an infalling radial velocity of -117~km/s, very close to the observed value \citep[$-109.3 \pm 4.4$~km/s,][]{vanderMarel2019} and has a low tangential velocity, 33 km/s, also well in line with observational constraints \citep[$56^{+35}_{-31}$~km/s from Gaia DR2 and HST data and $164.4 \pm 61.8$ km/s from satellite kinematics,][]{vanderMarel2019,Salomon2016}.

It is encouraging, yet unsurprising, that the halo masses and dynamics of the model Local Group match so well the observations, as they were deliberately designed to be so. As a reminder, the exploration runs of \citet{Sawala2021b} used to generate the eventual initial conditions of \sibeliusdark were dark-matter-only, and thus we were unable to consider the galaxy/baryonic properties of the Local Group candidates during the process. There was no guarantee therefore that the model properties of the Milky Way and Andromeda predicted by the semi-analytic model would turn out correct. Indeed, the total stellar mass for the model Milky Way, at $M_* = 5.43 \times 10^{9}$~\Msol, is approximately an order of magnitude too low compared to the estimated value of $M_* = (4$--$7) \times 10^{10}$~\Msol \citep[e.g.,][]{Licquia2015}, and the model Andromeda total stellar mass, at $M_* = 6.38 \times 10^{9}$~\Msol, is similarly an $\approx$order of magnitude too low compared to the estimated value of $M_* = (7$--$15) \times 10^{10}$~\Msol \citep[e.g.,][]{Tamm2012,Rahmani2016}. These are, of course, sensitive to the particular choice of semi-analytic galaxy formation model chosen, and also could change considerably if a full hydrodynamical simulation was performed on the same initial condition set. Additionally, the halo masses of the Milky Way and Andromeda lie at a critical juncture for galaxy formation physics, right at the onset of the AGN feedback regime \citep{Bower2012,McAlpine2018}, which results in a large spread of stellar masses for a given halo mass, particularly for the \galform model \citep[see Figure 4 of][]{Guo2016}. Regardless, for future generations of the \sibelius project we will consider also the baryonic properties during the exploration process, in order to further improve the Local Group analogues. 

\subsubsection{The \squotes{Bulk Flow} of the Local Neighbourhood}
\label{sect:bulk_flow}

We now investigate what immediately surrounds the model Local Group pair, what we dub the \squotes{Local Neighbourhood}. \cref{fig:local_environment} shows the projected dark matter distribution in a (50~Mpc)${^3}$, (15~Mpc)${^3}$ and (5~Mpc)${^3}$ volume centred on the model Milky Way. We find that the Local Group pair reside within a relatively isolated filament, and are eventually bracketed almost symmetrically by the Fornax and Virgo clusters at larger scales. 

An interesting observed feature relating to the nature of the Local Neighbourhood, and beyond, is the existence of the \squotes{bulk flow}; a coherent motion of galaxies in the vicinity of the Local Group towards a common location, against the direction of the Hubble flow. This common location is often referred to as the \dquotes{Great Attractor} \citep{Lynden-Bell1988}, and it is located in the approximate direction of the Hydra and Centaurus clusters. 

To investigate the motion of \sibeliusdark galaxies in the Local Neighbourhood we present \cref{fig:lg_velocity}. This shows, for all central galaxies within 10~Mpc of the Milky Way with a stellar mass greater than $M_* > 10^{7}$~\Msol, the direction of the peculiar velocity vector in the CMB reference frame with-respect-to the location of the \sibeliusdark Hydra cluster. That is, using an absolute set of coordinate axes (the conventional Galactic coordinate axes), we locate the observed position of the Hydra cluster on the sky \emph{from the reference frame of each galaxy} and compare that position vector to the CMB reference frame velocity vector for that galaxy. Or, put another way, we ask when viewed from each galaxy, \dquotes{how closely am I moving towards the simulated Hydra cluster?}.

From the upper panel of \cref{fig:lg_velocity}, we find that the majority of galaxies in the region of the Local Neighbourhood (including the Local Group) are flowing in a communal direction, approximately +30~degrees in Galactic longitude and +20~degrees in Galactic latitude with-respect-to the position of the Hydra cluster. From \cref{fig:sky_maps}, this points somewhere between the Virgo and Centaurus clusters in the northern hemisphere. Compared to the observed direction of the Local Group's motion, computed from the CMB dipole \citep{Planck2018}, this is off by approximately the same amount, as the observed vector of motion points almost exactly at the location of the \sibeliusdark Hydra cluster. In the lower panel we show the distribution of magnitudes for the velocity vectors of these galaxies. As with the direction, the Local Group is \squotes{typical} compared to the galaxies surrounding it, with an absolute velocity of 387~km/s. This is $\approx 225$~km/s below the observed value \citep[$620 \pm 15$~km/s][]{Planck2018}. It is extremely challenging to state where this discrepancy stems from, given how sensitive this result can be to the particular structural distribution surrounding the Local Group. One example could be the location of the \sibeliusdark Virgo cluster, which is $\approx 40$\% more distant than the observed Virgo cluster. Also, it has been argued that many objects contribute to the Local Group's bulk flow, including Hydra, Centaurus, Norma and even as far as Shapley, which all reside in a communal area on the sky at ever increasing distances. It is possible therefore that structures beyond our constraints ($d_{\mathrm{mw}} > 200$~Mpc) would provide the remaining contribution to the bulk flow \citep[e.,g][]{Turnbull2012,Boruah2020}, however it is perhaps unlikely that structures beyond 200~Mpc could make up for such a discrepancy. It would take a detailed study beyond the scope of this work to fully establish to origin of this discrepancy, however it will be an important consideration for improving future generations of \sibelius constrained initial conditions.

\section{Discussion \& Conclusions}
\label{sect:conclusion}

Whilst this study has largely been a presentation of the \sibeliusdark
simulation, in which here we have investigated the accuracy of the constrained
initial conditions at $z=0$ and provided examples of how to select specific
structures and associate them with observational counterparts, it has
also yielded interesting results relating to our local volume's
\squotes{unusual} nature. We discuss two aspects here: (1) the number
of haloes within the local volume with masses above $M_{\mathrm{200c}}
\geq 10^{15}h^{-1}$~\Msol and (2) the underdensity of the local volume
within $d_{\mathrm{MW}} \leq 200$~Mpc (or \dquotes{Local Hole}).  

\subsection{The number of haloes more massive than $M_{\mathrm{200c}} \geq 10^{15}h^{-1}$~\Msol in the local volume}

First, we investigate the prevalence of the most massive haloes within
the local volume, i.e., those with halo masses,
$M_{\mathrm{200c}} \geq 10^{15}h^{-1}$~\Msol. Given the theoretical
predicted exponential cutoff at the high-mass end of the halo mass
function, a few, or even a single unexpected halo at these masses can
pose a significant challenge to the underlying theory
\citep{Frenk1990}. Here, we compare the predictions of \sibeliusdark
to the recent study by \citet{Stopyra2021}, who attempt to quantify
the rarity of the local volume by comparing the number of observed
haloes above the $\geq 10^{15}h^{-1}$~\Msol threshold to that
predicted by a $\Lambda$CDM simulation. By analysing the halo mass
function from the Horizon Run 4 $\Lambda$CDM simulation
\citep{Kim2015}, \citet{Stopyra2021} predict an expectation value of
$\mathcal{O}(1)$ supermassive clusters per local volume. The likelihood
of finding $N$ clusters greater than
$M_{\mathrm{200c}} \geq 10^{15} h^{-1}$~\Msol in $\Lambda$CDM then
follows a Poisson distribution,

\begin{equation}
    \mathcal{L}(N|N_{\mathrm{exp}}) = \frac{N_{\mathrm{exp}}^{N} e^{-N_{\mathrm{exp}}}}{N!},
\end{equation}

\noindent where $N_{\mathrm{exp}} = 1$ as mentioned above.

Ultimately, \citet{Stopyra2021} are unable to conclusively estimate
the local volume's rarity in this respect, as the uncertainties in the
halo mass estimators from observations are simply too large, giving a
range of values of $N$ between 0 and 5. For \sibeliusdark, we find
$N=4$ halos more massive than
$M_{\mathrm{200c}} \geq 10^{15} h^{-1}$~\Msol: \textit{Perseus},
\textit{Hercules-A}, \textit{Hercules-B} and \textit{Norma} (see
\cref{table:cluster_masses}). \citet{Stopyra2021} theoretically
predict the likelihood of finding so many massive clusters in
something the size of the local volume (i,e., the volume of
\sibeliusdark) to be $1.3 \times 10^{-2}$, which could certainly be
classed as \squotes{rare}. We note, however, that this interpretation
is reliant on a reasonable estimate of the original expectation
value. Also, we accept that \sibeliusdark is a single realisation of
the \borg Markov chain, and, as the numbers here are so small,
investigating $N$ for multiple realisations would give a clearer
picture as to the predicted number of supermassive haloes from
\sibelius. Yet, from \sibeliusdark, we predict that there is
potentially an abnormally large number of supermassive clusters
occupying the local volume.

\subsection{The underdensity of the local volume, a \squotes{Local Hole}?}
\label{sect:local_hole}

Knowing if we reside in a particularly unusual region of the Universe,
say a large underdensity, can be crucial for interpreting the results
from local observations. One particular point of interest is in
relation to the measurement of $H_0$, where an unusually large
underdensity at the position of the observer would serve to increase
the inferred expansion rate, which, if true, would alleviate some of
the tension between the local \citep{Riess2016} and Cosmic Microwave
Background (CMB) measurements \citep{Planck2016} of $H_0$. Many
studies have argued for the existence of such a large-scale void,
potentially stretching as far out as 150--200$h^{-1}$~Mpc, quoting the
local volume to be underdense by up to values of $\approx 30$\%
\citep{Zehavi1998,Whitbourn2014,Whitbourn2016,Hoscheit2018,Bohringer2020}. Critiques
against many of these studies point to the assumption of isotropy when
extrapolating results from a subset of the sky to the entire
volume. However, an underdensity of 20\% has recently been proposed 
through a similar analysis using consistent all-sky data
\citep{Wong2021}. Whilst underdensities at  these levels could alter
the reported value of $H_0$ by as much as $\approx 5$\%, the scale of
such an underdensity in a volume so large is almost impossible in
$\Lambda$CDM \citep{Wu2017}.

\citet{Wong2021} find the $n(z)$ and $n(m)$ counts from the 2M++
survey to be $\approx 20$\% lower than the homogeneous model of
\citet{Whitbourn2014}, which would imply an underdensity at the same
level within $z<0.05$. In \cref{fig:K_band_number_counts} we
investigated the same $n(z)$ and $n(m)$ distributions for
\sibeliusdark galaxies with $K < 11.5$, and compared to the data of
the 2M++ survey. Given the agreement between \sibeliusdark and the
2M++ data, it is safe to assume that an analysis similar to
\citet{Wong2021} applied to the simulation data would yield a similar
underdensity estimate relative to the homogeneous model of
\citet{Whitbourn2014}. However, in the simulation we have the
advantage of being able to measure the density of the volume directly,
which was reported in \cref{fig:density_vs_rad}. We found the volume
to be 5\% underdense at the boundary of the constraints
($d_{\mathrm{MW}} = 200$~Mpc, or $z \approx 0.045$), a 2$\sigma$
deviation in $\Lambda$CDM \citep[a result that is unlikely to change
much for different choices of \borg Markov chain,
see][]{Jasche2019}. Therefore we would argue that an exceptional
deviation from $\Lambda$CDM, in terms of a large-scale underdensity,
is \emph{not} required to reproduce the observed number counts of
galaxies in the local volume.

\bigbreak

In this work we have presented \sibeliusdark, a dark-matter-only (DMO)
simulation paired with the semi-analytic model \galform that uses the
\sibelius constrained initial conditions to replicate the density and
velocity field of the local volume out to a distance,
$d_{\mathrm{MW}} \leq 200$~Mpc, from the Milky Way. Overall,
\sibeliusdark provides the most comprehensive constrained realisation
simulation to date. The local volume's large scale structure is
replicated \textit{en masse} with high fidelity \citep[see
\cref{fig:sky_maps} and the \borg results from
][]{Jasche2019}. Statistically, the galaxies that populate
\sibeliusdark reproduce well the observed stellar mass distribution,
luminosity distributions, central supermassive black hole population
(see \cref{fig:K_band_lum,fig:sdss_lum}) and number counts (see
\cref{fig:K_band_number_counts,fig:r_band_number_counts}) in the local
volume.

Specific structures are generally present in the correct location of
the sky and at the correct distance from the Milky Way (such as the
\textit{Perseus}, \textit{Hercules}, \textit{Norma}, \textit{Coma},
\textit{Leo}, \textit{Hydra}, \textit{Centaurus} and \textit{Virgo}
clusters; see \cref{fig:cluster_locations_1,fig:cluster_locations_2}
and reference \cref{table:cluster_masses} and
\cref{table:cluster_bcgs}), which provides a realistic spatial galaxy
distribution (see \cref{fig:cfa}). When these clusters are probed more
closely, they represent well their counterparts in the observations
(judged, for example, by the luminosity function of the cluster
members; see \cref{fig:virgo_lf}), which can then be used to make
predictions that are currently beyond our observational capabilities (for
example, the location of the \squotes{splashback radius} of 
local clusters, see \cref{fig:caustics_virgo}). Finally, at the centre
of the volume, \sibeliusdark contains a halo pair that closely
resembles the observed dynamics of the Local Group (see
\cref{sect:local_group}).

\sibeliusdark is the first production simulation of the \sibelius
project \citep{Sawala2021b}, demonstrating the potential of datasets from large-scale constrained realisation simulations using the \sibelius initial conditions. Some of the next stages of research for the \sibelius project include: (1) resimulating the
\sibelius volume with full hydrodynamics, (2) resimulating multiple realisations of the \borg constraints to estimate the variance in the formation histories and final-day properties of massive clusters within the local volume and (3) refining the reconstruction processes that produce the initial conditions, allowing
for more accurate and extensive constraints in the next generation of
\sibelius simulations.

With the publication of this study we publicly release the halo and
galaxy catalogues of \sibeliusdark at $z=0$. See
\cref{sect:public_data_release} for details of how to access the
catalogue and caveats of which to be aware when comparing to
observational data.

\section*{Acknowledgements}

The authors would like to thank the referee for their suggestions that have improved the quality of this work. We also thank Alastair Basden, Peter Draper and the
COSMA support team for their technical expertise that made this
simulation possible. SM thanks Tom Shanks and Jonathan Wong for their
helpful conversations, and particularly Ruari Mackenzie for his
assistance with many observational datasets.

The simulation in this
paper made use of the \swift open-source simulation code
\citep[\url{www.swiftsim.com},][]{Schaller2018} version 0.9.0.

This work was supported by the Academy of Finland: grant numbers
314238 and 335607. PHJ acknowledges the support by the European
Research Council via ERC Consolidator Grant KETJU (no. 818930). MS is
supported by the Netherlands Organisation for Scientific Research
(NWO) through VENI grant 639.041.749. CSF acknowledges support by the
European Research Council (ERC) through Advanced Investigator DMIDAS
(GA 786910). GL acknowledges financial support by the French National
Research Agency for the project BIG4, under reference
ANR-16-CE23-0002, and MMUniverse, under reference ANR-19-CE31-0020.

This work used the DiRAC Data Centric system at Durham University,
operated by the Institute for Computational Cosmology on behalf of the
STFC DiRAC HPC Facility (www.dirac.ac.uk). This equipment was funded
by BIS National E-infrastructure capital grant ST/K00042X/1, STFC
capital grants ST/H008519/1 and ST/K00087X/1, STFC DiRAC Operations
grant ST/K003267/1 and Durham University. DiRAC is part of the
National E-Infrastructure.

This work has been done within the Aquila
Consortium (\url{https://www.aquila-consortium.org}).

\section*{Data Availability}

We make public the halo and galaxy catalogues of the \sibeliusdark simulation at $z=0$ (see \cref{sect:public_data_release}). The particle data for the simulation is available upon reasonable request.
\bibliographystyle{mnras}
\bibliography{mybibfile}

\begin{thebibliography}{}
\makeatletter
\relax
\def\mn@urlcharsother{\let\do\@makeother \do\$\do\&\do\#\do\^\do\_\do\%\do\~}
\def\mn@doi{\begingroup\mn@urlcharsother \@ifnextchar [ {\mn@doi@}
  {\mn@doi@[]}}
\def\mn@doi@[#1]#2{\def\@tempa{#1}\ifx\@tempa\@empty \href
  {http://dx.doi.org/#2} {doi:#2}\else \href {http://dx.doi.org/#2} {#1}\fi
  \endgroup}
\def\mn@eprint#1#2{\mn@eprint@#1:#2::\@nil}
\def\mn@eprint@arXiv#1{\href {http://arxiv.org/abs/#1} {{\tt arXiv:#1}}}
\def\mn@eprint@dblp#1{\href {http://dblp.uni-trier.de/rec/bibtex/#1.xml}
  {dblp:#1}}
\def\mn@eprint@#1:#2:#3:#4\@nil{\def\@tempa {#1}\def\@tempb {#2}\def\@tempc
  {#3}\ifx \@tempc \@empty \let \@tempc \@tempb \let \@tempb \@tempa \fi \ifx
  \@tempb \@empty \def\@tempb {arXiv}\fi \@ifundefined
  {mn@eprint@\@tempb}{\@tempb:\@tempc}{\expandafter \expandafter \csname
  mn@eprint@\@tempb\endcsname \expandafter{\@tempc}}}

\bibitem[\protect\citeauthoryear{{Abazajian} et~al.,}{{Abazajian}
  et~al.}{2009}]{Abazajian2009}
{Abazajian} K.~N.,  et~al., 2009, \mn@doi [\apjs]
  {10.1088/0067-0049/182/2/543}, \href
  {https://ui.adsabs.harvard.edu/abs/2009ApJS..182..543A} {182, 543}

\bibitem[\protect\citeauthoryear{{Abell}}{{Abell}}{1958}]{Abell1958}
{Abell} G.~O.,  1958, \mn@doi [\apjs] {10.1086/190036}, \href
  {https://ui.adsabs.harvard.edu/abs/1958ApJS....3..211A} {3, 211}

\bibitem[\protect\citeauthoryear{{Abell}, {Corwin}  \& {Olowin}}{{Abell}
  et~al.}{1989}]{Abell1989}
{Abell} G.~O.,  {Corwin} Harold~G. J.,   {Olowin} R.~P.,  1989, \mn@doi [\apjs]
  {10.1086/191333}, \href
  {https://ui.adsabs.harvard.edu/abs/1989ApJS...70....1A} {70, 1}

\bibitem[\protect\citeauthoryear{{Adhikari}, {Dalal}  \&
  {Chamberlain}}{{Adhikari} et~al.}{2014}]{Adhikari2014}
{Adhikari} S.,  {Dalal} N.,   {Chamberlain} R.~T.,  2014, \mn@doi [\jcap]
  {10.1088/1475-7516/2014/11/019}, \href
  {https://ui.adsabs.harvard.edu/abs/2014JCAP...11..019A} {2014, 019}

\bibitem[\protect\citeauthoryear{{Ahumada} et~al.,}{{Ahumada}
  et~al.}{2020}]{Ahumada2020}
{Ahumada} R.,  et~al., 2020, \mn@doi [\apjs] {10.3847/1538-4365/ab929e}, \href
  {https://ui.adsabs.harvard.edu/abs/2020ApJS..249....3A} {249, 3}

\bibitem[\protect\citeauthoryear{{Baldry} et~al.,}{{Baldry}
  et~al.}{2018}]{Baldry2018}
{Baldry} I.~K.,  et~al., 2018, \mn@doi [\mnras] {10.1093/mnras/stx3042}, \href
  {https://ui.adsabs.harvard.edu/abs/2018MNRAS.474.3875B} {474, 3875}

\bibitem[\protect\citeauthoryear{{Baugh} et~al.,}{{Baugh}
  et~al.}{2019}]{Baugh2019}
{Baugh} C.~M.,  et~al., 2019, \mn@doi [\mnras] {10.1093/mnras/sty3427}, \href
  {https://ui.adsabs.harvard.edu/abs/2019MNRAS.483.4922B} {483, 4922}

\bibitem[\protect\citeauthoryear{{Baxter} et~al.,}{{Baxter}
  et~al.}{2017}]{Baxter2017}
{Baxter} E.,  et~al., 2017, \mn@doi [\apj] {10.3847/1538-4357/aa6ff0}, \href
  {https://ui.adsabs.harvard.edu/abs/2017ApJ...841...18B} {841, 18}

\bibitem[\protect\citeauthoryear{{Benson}, {Borgani}, {De Lucia},
  {Boylan-Kolchin}  \& {Monaco}}{{Benson} et~al.}{2012}]{Benson2012}
{Benson} A.~J.,  {Borgani} S.,  {De Lucia} G.,  {Boylan-Kolchin} M.,   {Monaco}
  P.,  2012, \mn@doi [\mnras] {10.1111/j.1365-2966.2011.20002.x}, \href
  {https://ui.adsabs.harvard.edu/abs/2012MNRAS.419.3590B} {419, 3590}

\bibitem[\protect\citeauthoryear{{Bertschinger}}{{Bertschinger}}{1987}]{Bertschinger1987}
{Bertschinger} E.,  1987, \mn@doi [\apjl] {10.1086/185066}, \href
  {https://ui.adsabs.harvard.edu/abs/1987ApJ...323L.103B} {323, L103}

\bibitem[\protect\citeauthoryear{{Bianconi}, {Buscicchio}, {Smith}, {McGee},
  {Haines}, {Finoguenov}  \& {Babul}}{{Bianconi} et~al.}{2021}]{Bianconi2021}
{Bianconi} M.,  {Buscicchio} R.,  {Smith} G.~P.,  {McGee} S.~L.,  {Haines}
  C.~P.,  {Finoguenov} A.,   {Babul} A.,  2021, \mn@doi [\apj]
  {10.3847/1538-4357/abebd7}, \href
  {https://ui.adsabs.harvard.edu/abs/2021ApJ...911..136B} {911, 136}

\bibitem[\protect\citeauthoryear{{Binggeli}, {Sandage}  \&
  {Tammann}}{{Binggeli} et~al.}{1985}]{Binggeli1985}
{Binggeli} B.,  {Sandage} A.,   {Tammann} G.~A.,  1985, \mn@doi [\aj]
  {10.1086/113874}, \href
  {https://ui.adsabs.harvard.edu/abs/1985AJ.....90.1681B} {90, 1681}

\bibitem[\protect\citeauthoryear{{Binggeli}, {Tammann}  \&
  {Sandage}}{{Binggeli} et~al.}{1987}]{Binggeli1987}
{Binggeli} B.,  {Tammann} G.~A.,   {Sandage} A.,  1987, \mn@doi [\aj]
  {10.1086/114467}, \href
  {https://ui.adsabs.harvard.edu/abs/1987AJ.....94..251B} {94, 251}

\bibitem[\protect\citeauthoryear{{Blakeslee} et~al.,}{{Blakeslee}
  et~al.}{2009}]{Blakeslee2009}
{Blakeslee} J.~P.,  et~al., 2009, \mn@doi [\apj] {10.1088/0004-637X/694/1/556},
  \href {https://ui.adsabs.harvard.edu/abs/2009ApJ...694..556B} {694, 556}

\bibitem[\protect\citeauthoryear{{Blanton} et~al.,}{{Blanton}
  et~al.}{2003}]{Blanton2003}
{Blanton} M.~R.,  et~al., 2003, \mn@doi [\apj] {10.1086/375776}, \href
  {https://ui.adsabs.harvard.edu/abs/2003ApJ...592..819B} {592, 819}

\bibitem[\protect\citeauthoryear{{B{\"o}hringer}, {Chon}  \&
  {Collins}}{{B{\"o}hringer} et~al.}{2020}]{Bohringer2020}
{B{\"o}hringer} H.,  {Chon} G.,   {Collins} C.~A.,  2020, \mn@doi [\aap]
  {10.1051/0004-6361/201936400}, \href
  {https://ui.adsabs.harvard.edu/abs/2020A&A...633A..19B} {633, A19}

\bibitem[\protect\citeauthoryear{{Boruah}, {Hudson}  \& {Lavaux}}{{Boruah}
  et~al.}{2020}]{Boruah2020}
{Boruah} S.~S.,  {Hudson} M.~J.,   {Lavaux} G.,  2020, \mn@doi [\mnras]
  {10.1093/mnras/staa2485}, \href
  {https://ui.adsabs.harvard.edu/abs/2020MNRAS.498.2703B} {498, 2703}

\bibitem[\protect\citeauthoryear{{Bower}, {Benson}  \& {Crain}}{{Bower}
  et~al.}{2012}]{Bower2012}
{Bower} R.~G.,  {Benson} A.~J.,   {Crain} R.~A.,  2012, \mn@doi [\mnras]
  {10.1111/j.1365-2966.2012.20516.x}, \href
  {https://ui.adsabs.harvard.edu/abs/2012MNRAS.422.2816B} {422, 2816}

\bibitem[\protect\citeauthoryear{{Boylan-Kolchin}, {Springel}, {White},
  {Jenkins}  \& {Lemson}}{{Boylan-Kolchin} et~al.}{2009}]{BoylanKolchin2009}
{Boylan-Kolchin} M.,  {Springel} V.,  {White} S. D.~M.,  {Jenkins} A.,
  {Lemson} G.,  2009, \mn@doi [\mnras] {10.1111/j.1365-2966.2009.15191.x},
  \href {https://ui.adsabs.harvard.edu/abs/2009MNRAS.398.1150B} {398, 1150}

\bibitem[\protect\citeauthoryear{{Boylan-Kolchin}, {Bullock}  \&
  {Kaplinghat}}{{Boylan-Kolchin} et~al.}{2011}]{Boylan-Kolchin2011}
{Boylan-Kolchin} M.,  {Bullock} J.~S.,   {Kaplinghat} M.,  2011, \mn@doi
  [\mnras] {10.1111/j.1745-3933.2011.01074.x}, \href
  {https://ui.adsabs.harvard.edu/abs/2011MNRAS.415L..40B} {415, L40}

\bibitem[\protect\citeauthoryear{{Bullock} \& {Boylan-Kolchin}}{{Bullock} \&
  {Boylan-Kolchin}}{2017}]{Bullock2017}
{Bullock} J.~S.,  {Boylan-Kolchin} M.,  2017, \mn@doi [\araa]
  {10.1146/annurev-astro-091916-055313}, \href
  {https://ui.adsabs.harvard.edu/abs/2017ARA&A..55..343B} {55, 343}

\bibitem[\protect\citeauthoryear{{Carlesi} et~al.,}{{Carlesi}
  et~al.}{2016}]{Carlesi2016}
{Carlesi} E.,  et~al., 2016, \mn@doi [\mnras] {10.1093/mnras/stw357}, \href
  {https://ui.adsabs.harvard.edu/abs/2016MNRAS.458..900C} {458, 900}

\bibitem[\protect\citeauthoryear{{Carlesi}, {Hoffman}, {Gottl{\"o}ber},
  {Libeskind}, {Knebe}, {Yepes}  \& {Pilipenko}}{{Carlesi}
  et~al.}{2020}]{Carlesi2020}
{Carlesi} E.,  {Hoffman} Y.,  {Gottl{\"o}ber} S.,  {Libeskind} N.~I.,  {Knebe}
  A.,  {Yepes} G.,   {Pilipenko} S.~V.,  2020, \mn@doi [\mnras]
  {10.1093/mnras/stz3089}, \href
  {https://ui.adsabs.harvard.edu/abs/2020MNRAS.491.1531C} {491, 1531}

\bibitem[\protect\citeauthoryear{{Cautun} et~al.,}{{Cautun}
  et~al.}{2020}]{Cautun2020}
{Cautun} M.,  et~al., 2020, \mn@doi [\mnras] {10.1093/mnras/staa1017}, \href
  {https://ui.adsabs.harvard.edu/abs/2020MNRAS.494.4291C} {494, 4291}

\bibitem[\protect\citeauthoryear{Cheng, Greengard  \& Rokhlin}{Cheng
  et~al.}{1999}]{Cheng1999}
Cheng H.,  Greengard L.,   Rokhlin V.,  1999, \mn@doi [Journal of Computational
  Physics] {http://dx.doi.org/10.1006/jcph.1999.6355}, 155, 468

\bibitem[\protect\citeauthoryear{{Contigiani}, {Hoekstra}  \&
  {Bah{\'e}}}{{Contigiani} et~al.}{2019}]{Contigiani2019}
{Contigiani} O.,  {Hoekstra} H.,   {Bah{\'e}} Y.~M.,  2019, \mn@doi [\mnras]
  {10.1093/mnras/stz404}, \href
  {https://ui.adsabs.harvard.edu/abs/2019MNRAS.485..408C} {485, 408}

\bibitem[\protect\citeauthoryear{{Contigiani}, {Bah{\'e}}  \&
  {Hoekstra}}{{Contigiani} et~al.}{2020}]{Contigiani2020}
{Contigiani} O.,  {Bah{\'e}} Y.~M.,   {Hoekstra} H.,  2020, arXiv e-prints,
  \href {https://ui.adsabs.harvard.edu/abs/2020arXiv201201336C} {p.
  arXiv:2012.01336}

\bibitem[\protect\citeauthoryear{{Cortese}, {Gavazzi}, {Boselli},
  {Iglesias-Paramo}  \& {Carrasco}}{{Cortese} et~al.}{2004}]{Cortese2004}
{Cortese} L.,  {Gavazzi} G.,  {Boselli} A.,  {Iglesias-Paramo} J.,   {Carrasco}
  L.,  2004, \mn@doi [\aap] {10.1051/0004-6361:20040381}, \href
  {https://ui.adsabs.harvard.edu/abs/2004A&A...425..429C} {425, 429}

\bibitem[\protect\citeauthoryear{{Cowley}, {Lacey}, {Baugh}  \&
  {Cole}}{{Cowley} et~al.}{2015}]{Cowley2015}
{Cowley} W.~I.,  {Lacey} C.~G.,  {Baugh} C.~M.,   {Cole} S.,  2015, \mn@doi
  [\mnras] {10.1093/mnras/stu2179}, \href
  {https://ui.adsabs.harvard.edu/abs/2015MNRAS.446.1784C} {446, 1784}

\bibitem[\protect\citeauthoryear{{Crain} et~al.,}{{Crain}
  et~al.}{2015}]{Crain2015}
{Crain} R.~A.,  et~al., 2015, \mn@doi [\mnras] {10.1093/mnras/stv725}, \href
  {https://ui.adsabs.harvard.edu/abs/2015MNRAS.450.1937C} {450, 1937}

\bibitem[\protect\citeauthoryear{{Davis}, {Efstathiou}, {Frenk}  \&
  {White}}{{Davis} et~al.}{1985}]{Davis1985}
{Davis} M.,  {Efstathiou} G.,  {Frenk} C.~S.,   {White} S.~D.~M.,  1985,
  \mn@doi [\apj] {10.1086/163168}, \href
  {https://ui.adsabs.harvard.edu/abs/1985ApJ...292..371D} {292, 371}

\bibitem[\protect\citeauthoryear{{Deason}, {Fattahi}, {Frenk}, {Grand}, {Oman},
  {Garrison-Kimmel}, {Simpson}  \& {Navarro}}{{Deason}
  et~al.}{2020}]{Deason2020}
{Deason} A.~J.,  {Fattahi} A.,  {Frenk} C.~S.,  {Grand} R. J.~J.,  {Oman}
  K.~A.,  {Garrison-Kimmel} S.,  {Simpson} C.~M.,   {Navarro} J.~F.,  2020,
  \mn@doi [\mnras] {10.1093/mnras/staa1711}, \href
  {https://ui.adsabs.harvard.edu/abs/2020MNRAS.496.3929D} {496, 3929}

\bibitem[\protect\citeauthoryear{{Dehnen}}{{Dehnen}}{2014}]{Dehnen2014}
{Dehnen} W.,  2014, \mn@doi [Computational Astrophysics and Cosmology]
  {10.1186/s40668-014-0001-7}, \href
  {https://ui.adsabs.harvard.edu/abs/2014ComAC...1....1D} {1, 1}

\bibitem[\protect\citeauthoryear{{Diaferio} \& {Geller}}{{Diaferio} \&
  {Geller}}{1997}]{Diaferio1997}
{Diaferio} A.,  {Geller} M.~J.,  1997, \mn@doi [\apj] {10.1086/304075}, \href
  {https://ui.adsabs.harvard.edu/abs/1997ApJ...481..633D} {481, 633}

\bibitem[\protect\citeauthoryear{{Diemer} \& {Kravtsov}}{{Diemer} \&
  {Kravtsov}}{2014}]{Diemer2014}
{Diemer} B.,  {Kravtsov} A.~V.,  2014, \mn@doi [\apj]
  {10.1088/0004-637X/789/1/1}, \href
  {https://ui.adsabs.harvard.edu/abs/2014ApJ...789....1D} {789, 1}

\bibitem[\protect\citeauthoryear{{Diemer}, {More}  \& {Kravtsov}}{{Diemer}
  et~al.}{2013}]{Diemer2013}
{Diemer} B.,  {More} S.,   {Kravtsov} A.~V.,  2013, \mn@doi [\apj]
  {10.1088/0004-637X/766/1/25}, \href
  {https://ui.adsabs.harvard.edu/abs/2013ApJ...766...25D} {766, 25}

\bibitem[\protect\citeauthoryear{{Diemer}, {Mansfield}, {Kravtsov}  \&
  {More}}{{Diemer} et~al.}{2017}]{Diemer2017}
{Diemer} B.,  {Mansfield} P.,  {Kravtsov} A.~V.,   {More} S.,  2017, \mn@doi
  [\apj] {10.3847/1538-4357/aa79ab}, \href
  {https://ui.adsabs.harvard.edu/abs/2017ApJ...843..140D} {843, 140}

\bibitem[\protect\citeauthoryear{{Driver} et~al.,}{{Driver}
  et~al.}{2012}]{Driver2012}
{Driver} S.~P.,  et~al., 2012, \mn@doi [\mnras]
  {10.1111/j.1365-2966.2012.22036.x}, \href
  {https://ui.adsabs.harvard.edu/abs/2012MNRAS.427.3244D} {427, 3244}

\bibitem[\protect\citeauthoryear{{Dubois} et~al.,}{{Dubois}
  et~al.}{2014}]{Dubois2014}
{Dubois} Y.,  et~al., 2014, \mn@doi [\mnras] {10.1093/mnras/stu1227}, \href
  {https://ui.adsabs.harvard.edu/abs/2014MNRAS.444.1453D} {444, 1453}

\bibitem[\protect\citeauthoryear{{En{\ss}lin}, {Frommert}  \&
  {Kitaura}}{{En{\ss}lin} et~al.}{2009}]{Ensslin2009}
{En{\ss}lin} T.~A.,  {Frommert} M.,   {Kitaura} F.~S.,  2009, \mn@doi [\prd]
  {10.1103/PhysRevD.80.105005}, \href
  {https://ui.adsabs.harvard.edu/abs/2009PhRvD..80j5005E} {80, 105005}

\bibitem[\protect\citeauthoryear{{Erdo{\v{g}}du} et~al.,}{{Erdo{\v{g}}du}
  et~al.}{2006}]{Erdogdu2006}
{Erdo{\v{g}}du} P.,  et~al., 2006, \mn@doi [\mnras]
  {10.1111/j.1365-2966.2006.10243.x}, \href
  {https://ui.adsabs.harvard.edu/abs/2006MNRAS.368.1515E} {368, 1515}

\bibitem[\protect\citeauthoryear{{Ferrarese} et~al.,}{{Ferrarese}
  et~al.}{2012}]{Ferrarese2012}
{Ferrarese} L.,  et~al., 2012, \mn@doi [\apjs] {10.1088/0067-0049/200/1/4},
  \href {https://ui.adsabs.harvard.edu/abs/2012ApJS..200....4F} {200, 4}

\bibitem[\protect\citeauthoryear{{Ferrarese} et~al.,}{{Ferrarese}
  et~al.}{2016}]{Ferrarese2016}
{Ferrarese} L.,  et~al., 2016, \mn@doi [\apj] {10.3847/0004-637X/824/1/10},
  \href {https://ui.adsabs.harvard.edu/abs/2016ApJ...824...10F} {824, 10}

\bibitem[\protect\citeauthoryear{{Flores} \& {Primack}}{{Flores} \&
  {Primack}}{1994}]{Flores1994}
{Flores} R.~A.,  {Primack} J.~R.,  1994, \mn@doi [\apjl] {10.1086/187350},
  \href {https://ui.adsabs.harvard.edu/abs/1994ApJ...427L...1F} {427, L1}

\bibitem[\protect\citeauthoryear{{Frenk} \& {White}}{{Frenk} \&
  {White}}{2012}]{Frenk2012}
{Frenk} C.~S.,  {White} S.~D.~M.,  2012, \mn@doi [Annalen der Physik]
  {10.1002/andp.201200212}, \href
  {https://ui.adsabs.harvard.edu/abs/2012AnP...524..507F} {524, 507}

\bibitem[\protect\citeauthoryear{{Frenk}, {White}, {Efstathiou}  \&
  {Davis}}{{Frenk} et~al.}{1990}]{Frenk1990}
{Frenk} C.~S.,  {White} S. D.~M.,  {Efstathiou} G.,   {Davis} M.,  1990,
  \mn@doi [\apj] {10.1086/168439}, \href
  {https://ui.adsabs.harvard.edu/abs/1990ApJ...351...10F} {351, 10}

\bibitem[\protect\citeauthoryear{{Gao}, {Navarro}, {Frenk}, {Jenkins},
  {Springel}  \& {White}}{{Gao} et~al.}{2012}]{Gao2012}
{Gao} L.,  {Navarro} J.~F.,  {Frenk} C.~S.,  {Jenkins} A.,  {Springel} V.,
  {White} S.~D.~M.,  2012, \mn@doi [\mnras] {10.1111/j.1365-2966.2012.21564.x},
  \href {https://ui.adsabs.harvard.edu/abs/2012MNRAS.425.2169G} {425, 2169}

\bibitem[\protect\citeauthoryear{{Garrison-Kimmel}, {Boylan-Kolchin}, {Bullock}
   \& {Lee}}{{Garrison-Kimmel} et~al.}{2014}]{GarrisonKimmel2014}
{Garrison-Kimmel} S.,  {Boylan-Kolchin} M.,  {Bullock} J.~S.,   {Lee} K.,
  2014, \mn@doi [\mnras] {10.1093/mnras/stt2377}, \href
  {https://ui.adsabs.harvard.edu/abs/2014MNRAS.438.2578G} {438, 2578}

\bibitem[\protect\citeauthoryear{{Garrison-Kimmel} et~al.,}{{Garrison-Kimmel}
  et~al.}{2019}]{GarrisonKimmel2019}
{Garrison-Kimmel} S.,  et~al., 2019, \mn@doi [\mnras] {10.1093/mnras/stz1317},
  \href {https://ui.adsabs.harvard.edu/abs/2019MNRAS.487.1380G} {487, 1380}

\bibitem[\protect\citeauthoryear{{Gonzalez}, {George}, {Connor}, {Deason},
  {Donahue}, {Montes}, {Zabludoff}  \& {Zaritsky}}{{Gonzalez}
  et~al.}{2021}]{Gonzalez2021}
{Gonzalez} A.~H.,  {George} T.,  {Connor} T.,  {Deason} A.,  {Donahue} M.,
  {Montes} M.,  {Zabludoff} A.~I.,   {Zaritsky} D.,  2021, arXiv e-prints,
  \href {https://ui.adsabs.harvard.edu/abs/2021arXiv210404306G} {p.
  arXiv:2104.04306}

\bibitem[\protect\citeauthoryear{{Guo}, {White}, {Angulo}, {Henriques},
  {Lemson}, {Boylan-Kolchin}, {Thomas}  \& {Short}}{{Guo}
  et~al.}{2013}]{Guo2013}
{Guo} Q.,  {White} S.,  {Angulo} R.~E.,  {Henriques} B.,  {Lemson} G.,
  {Boylan-Kolchin} M.,  {Thomas} P.,   {Short} C.,  2013, \mn@doi [\mnras]
  {10.1093/mnras/sts115}, \href
  {https://ui.adsabs.harvard.edu/abs/2013MNRAS.428.1351G} {428, 1351}

\bibitem[\protect\citeauthoryear{{Guo} et~al.,}{{Guo} et~al.}{2016}]{Guo2016}
{Guo} Q.,  et~al., 2016, \mn@doi [\mnras] {10.1093/mnras/stw1525}, \href
  {https://ui.adsabs.harvard.edu/abs/2016MNRAS.461.3457G} {461, 3457}

\bibitem[\protect\citeauthoryear{{Han}, {Jing}, {Wang}  \& {Wang}}{{Han}
  et~al.}{2012}]{Han2012}
{Han} J.,  {Jing} Y.~P.,  {Wang} H.,   {Wang} W.,  2012, \mn@doi [\mnras]
  {10.1111/j.1365-2966.2012.22111.x}, \href
  {https://ui.adsabs.harvard.edu/abs/2012MNRAS.427.2437H} {427, 2437}

\bibitem[\protect\citeauthoryear{{Han}, {Cole}, {Frenk}, {Benitez-Llambay}  \&
  {Helly}}{{Han} et~al.}{2018}]{Han2018}
{Han} J.,  {Cole} S.,  {Frenk} C.~S.,  {Benitez-Llambay} A.,   {Helly} J.,
  2018, \mn@doi [\mnras] {10.1093/mnras/stx2792}, \href
  {https://ui.adsabs.harvard.edu/abs/2018MNRAS.474..604H} {474, 604}

\bibitem[\protect\citeauthoryear{{Hirschmann}, {Dolag}, {Saro}, {Bachmann},
  {Borgani}  \& {Burkert}}{{Hirschmann} et~al.}{2014}]{Hirschmann2014}
{Hirschmann} M.,  {Dolag} K.,  {Saro} A.,  {Bachmann} L.,  {Borgani} S.,
  {Burkert} A.,  2014, \mn@doi [\mnras] {10.1093/mnras/stu1023}, \href
  {https://ui.adsabs.harvard.edu/abs/2014MNRAS.442.2304H} {442, 2304}

\bibitem[\protect\citeauthoryear{{Hoffman} \& {Ribak}}{{Hoffman} \&
  {Ribak}}{1991}]{Hoffman1991}
{Hoffman} Y.,  {Ribak} E.,  1991, \mn@doi [\apjl] {10.1086/186160}, \href
  {https://ui.adsabs.harvard.edu/abs/1991ApJ...380L...5H} {380, L5}

\bibitem[\protect\citeauthoryear{{Hoscheit} \& {Barger}}{{Hoscheit} \&
  {Barger}}{2018}]{Hoscheit2018}
{Hoscheit} B.~L.,  {Barger} A.~J.,  2018, \mn@doi [\apj]
  {10.3847/1538-4357/aaa59b}, \href
  {https://ui.adsabs.harvard.edu/abs/2018ApJ...854...46H} {854, 46}

\bibitem[\protect\citeauthoryear{{Hou}, {Lacey}  \& {Frenk}}{{Hou}
  et~al.}{2018}]{Hou2018}
{Hou} J.,  {Lacey} C.~G.,   {Frenk} C.~S.,  2018, \mn@doi [\mnras]
  {10.1093/mnras/stx3218}, \href
  {https://ui.adsabs.harvard.edu/abs/2018MNRAS.475..543H} {475, 543}

\bibitem[\protect\citeauthoryear{{Hou}, {Lacey}  \& {Frenk}}{{Hou}
  et~al.}{2019}]{Hou2019}
{Hou} J.,  {Lacey} C.~G.,   {Frenk} C.~S.,  2019, \mn@doi [\mnras]
  {10.1093/mnras/stz730}, \href
  {https://ui.adsabs.harvard.edu/abs/2019MNRAS.486.1691H} {486, 1691}

\bibitem[\protect\citeauthoryear{{Huchra}, {Davis}, {Latham}  \&
  {Tonry}}{{Huchra} et~al.}{1983}]{Huchra1983}
{Huchra} J.,  {Davis} M.,  {Latham} D.,   {Tonry} J.,  1983, \mn@doi [\apjs]
  {10.1086/190860}, \href
  {https://ui.adsabs.harvard.edu/abs/1983ApJS...52...89H} {52, 89}

\bibitem[\protect\citeauthoryear{{Huchra} et~al.,}{{Huchra}
  et~al.}{2012}]{Huchra2012}
{Huchra} J.~P.,  et~al., 2012, \mn@doi [\apjs] {10.1088/0067-0049/199/2/26},
  \href {https://ui.adsabs.harvard.edu/abs/2012ApJS..199...26H} {199, 26}

\bibitem[\protect\citeauthoryear{{Impey}, {Bothun}  \& {Malin}}{{Impey}
  et~al.}{1988}]{Impey1998}
{Impey} C.,  {Bothun} G.,   {Malin} D.,  1988, \mn@doi [\apj] {10.1086/166500},
  \href {https://ui.adsabs.harvard.edu/abs/1988ApJ...330..634I} {330, 634}

\bibitem[\protect\citeauthoryear{{Jasche} \& {Lavaux}}{{Jasche} \&
  {Lavaux}}{2019}]{Jasche2019}
{Jasche} J.,  {Lavaux} G.,  2019, \mn@doi [\aap] {10.1051/0004-6361/201833710},
  \href {https://ui.adsabs.harvard.edu/abs/2019A&A...625A..64J} {625, A64}

\bibitem[\protect\citeauthoryear{{Jasche} \& {Wandelt}}{{Jasche} \&
  {Wandelt}}{2013}]{Jasche2013}
{Jasche} J.,  {Wandelt} B.~D.,  2013, \mn@doi [\mnras] {10.1093/mnras/stt449},
  \href {https://ui.adsabs.harvard.edu/abs/2013MNRAS.432..894J} {432, 894}

\bibitem[\protect\citeauthoryear{{Jenkins}}{{Jenkins}}{2010}]{Jenkins2010}
{Jenkins} A.,  2010, \mn@doi [\mnras] {10.1111/j.1365-2966.2010.16259.x}, \href
  {https://ui.adsabs.harvard.edu/abs/2010MNRAS.403.1859J} {403, 1859}

\bibitem[\protect\citeauthoryear{{Jenkins}}{{Jenkins}}{2013}]{Jenkins2013}
{Jenkins} A.,  2013, \mn@doi [\mnras] {10.1093/mnras/stt1154}, \href
  {https://ui.adsabs.harvard.edu/abs/2013MNRAS.434.2094J} {434, 2094}

\bibitem[\protect\citeauthoryear{{Jiang}, {Helly}, {Cole}  \& {Frenk}}{{Jiang}
  et~al.}{2014}]{Jiang2014}
{Jiang} L.,  {Helly} J.~C.,  {Cole} S.,   {Frenk} C.~S.,  2014, \mn@doi
  [\mnras] {10.1093/mnras/stu390}, \href
  {https://ui.adsabs.harvard.edu/abs/2014MNRAS.440.2115J} {440, 2115}

\bibitem[\protect\citeauthoryear{{Jones} et~al.,}{{Jones}
  et~al.}{2009}]{Jones2009}
{Jones} D.~H.,  et~al., 2009, \mn@doi [\mnras]
  {10.1111/j.1365-2966.2009.15338.x}, \href
  {https://ui.adsabs.harvard.edu/abs/2009MNRAS.399..683J} {399, 683}

\bibitem[\protect\citeauthoryear{{Karachentsev} \& {Kashibadze}}{{Karachentsev}
  \& {Kashibadze}}{2006}]{Karachentsev2006}
{Karachentsev} I.~D.,  {Kashibadze} O.~G.,  2006, \mn@doi [Astrophysics]
  {10.1007/s10511-006-0002-6}, \href
  {https://ui.adsabs.harvard.edu/abs/2006Ap.....49....3K} {49, 3}

\bibitem[\protect\citeauthoryear{{Kashibadze}, {Karachentsev}  \&
  {Karachentseva}}{{Kashibadze} et~al.}{2020}]{Kashibadze2020}
{Kashibadze} O.~G.,  {Karachentsev} I.~D.,   {Karachentseva} V.~E.,  2020,
  \mn@doi [\aap] {10.1051/0004-6361/201936172}, \href
  {https://ui.adsabs.harvard.edu/abs/2020A&A...635A.135K} {635, A135}

\bibitem[\protect\citeauthoryear{{Kim}, {Park}, {L'Huillier}  \& {Hong}}{{Kim}
  et~al.}{2015}]{Kim2015}
{Kim} J.,  {Park} C.,  {L'Huillier} B.,   {Hong} S.~E.,  2015, \mn@doi [Journal
  of Korean Astronomical Society] {10.5303/JKAS.2015.48.4.213}, \href
  {https://ui.adsabs.harvard.edu/abs/2015JKAS...48..213K} {48, 213}

\bibitem[\protect\citeauthoryear{{Kitaura} \& {En{\ss}lin}}{{Kitaura} \&
  {En{\ss}lin}}{2008}]{Kitaura2008}
{Kitaura} F.~S.,  {En{\ss}lin} T.~A.,  2008, \mn@doi [\mnras]
  {10.1111/j.1365-2966.2008.13341.x}, \href
  {https://ui.adsabs.harvard.edu/abs/2008MNRAS.389..497K} {389, 497}

\bibitem[\protect\citeauthoryear{{Klypin}, {Kravtsov}, {Valenzuela}  \&
  {Prada}}{{Klypin} et~al.}{1999}]{Klypin1999}
{Klypin} A.,  {Kravtsov} A.~V.,  {Valenzuela} O.,   {Prada} F.,  1999, \mn@doi
  [\apj] {10.1086/307643}, \href
  {https://ui.adsabs.harvard.edu/abs/1999ApJ...522...82K} {522, 82}

\bibitem[\protect\citeauthoryear{{Klypin}, {Hoffman}, {Kravtsov}  \&
  {Gottl{\"o}ber}}{{Klypin} et~al.}{2003}]{Klypin2003}
{Klypin} A.,  {Hoffman} Y.,  {Kravtsov} A.~V.,   {Gottl{\"o}ber} S.,  2003,
  \mn@doi [\apj] {10.1086/377574}, \href
  {https://ui.adsabs.harvard.edu/abs/2003ApJ...596...19K} {596, 19}

\bibitem[\protect\citeauthoryear{{Kochanek} et~al.,}{{Kochanek}
  et~al.}{2001}]{Kochanek2001}
{Kochanek} C.~S.,  et~al., 2001, \mn@doi [\apj] {10.1086/322488}, \href
  {https://ui.adsabs.harvard.edu/abs/2001ApJ...560..566K} {560, 566}

\bibitem[\protect\citeauthoryear{{Kopylova} \& {Kopylov}}{{Kopylova} \&
  {Kopylov}}{2013}]{Kopylova2013}
{Kopylova} F.~G.,  {Kopylov} A.~I.,  2013, \mn@doi [Astronomy Letters]
  {10.1134/S1063773712120043}, \href
  {https://ui.adsabs.harvard.edu/abs/2013AstL...39....1K} {39, 1}

\bibitem[\protect\citeauthoryear{{Krempe{\'c}-Krygier}, {Krygier}  \&
  {Krywult}}{{Krempe{\'c}-Krygier} et~al.}{2002}]{Kremp2002}
{Krempe{\'c}-Krygier} J.,  {Krygier} B.,   {Krywult} J.,  2002, Baltic
  Astronomy, \href {https://ui.adsabs.harvard.edu/abs/2002BaltA..11..269K} {11,
  269}

\bibitem[\protect\citeauthoryear{{Kubo} et~al.,}{{Kubo}
  et~al.}{2009}]{Kubo2009}
{Kubo} J.~M.,  et~al., 2009, \mn@doi [\apjl] {10.1088/0004-637X/702/2/L110},
  \href {https://ui.adsabs.harvard.edu/abs/2009ApJ...702L.110K} {702, L110}

\bibitem[\protect\citeauthoryear{{Lacey}, {Baugh}, {Frenk}  \&
  {Benson}}{{Lacey} et~al.}{2011}]{Lacey2011}
{Lacey} C.~G.,  {Baugh} C.~M.,  {Frenk} C.~S.,   {Benson} A.~J.,  2011, \mn@doi
  [\mnras] {10.1111/j.1365-2966.2010.18021.x}, \href
  {https://ui.adsabs.harvard.edu/abs/2011MNRAS.412.1828L} {412, 1828}

\bibitem[\protect\citeauthoryear{{Lacey} et~al.,}{{Lacey}
  et~al.}{2016}]{Lacey2016}
{Lacey} C.~G.,  et~al., 2016, \mn@doi [\mnras] {10.1093/mnras/stw1888}, \href
  {https://ui.adsabs.harvard.edu/abs/2016MNRAS.462.3854L} {462, 3854}

\bibitem[\protect\citeauthoryear{{Lavaux} \& {Hudson}}{{Lavaux} \&
  {Hudson}}{2011}]{Lavaux2011}
{Lavaux} G.,  {Hudson} M.~J.,  2011, \mn@doi [\mnras]
  {10.1111/j.1365-2966.2011.19233.x}, \href
  {https://ui.adsabs.harvard.edu/abs/2011MNRAS.416.2840L} {416, 2840}

\bibitem[\protect\citeauthoryear{{Lemson} \& {Virgo Consortium}}{{Lemson} \&
  {Virgo Consortium}}{2006}]{Lemson2006}
{Lemson} G.,  {Virgo Consortium} t.,  2006, arXiv e-prints, \href
  {https://ui.adsabs.harvard.edu/abs/2006astro.ph..8019L} {pp
  astro--ph/0608019}

\bibitem[\protect\citeauthoryear{{Li} \& {White}}{{Li} \&
  {White}}{2009}]{Li2009}
{Li} C.,  {White} S. D.~M.,  2009, \mn@doi [\mnras]
  {10.1111/j.1365-2966.2009.15268.x}, \href
  {https://ui.adsabs.harvard.edu/abs/2009MNRAS.398.2177L} {398, 2177}

\bibitem[\protect\citeauthoryear{{Libeskind}, {Knebe}, {Hoffman},
  {Gottl{\"o}ber}, {Yepes}  \& {Steinmetz}}{{Libeskind}
  et~al.}{2011}]{Libeskind2011}
{Libeskind} N.~I.,  {Knebe} A.,  {Hoffman} Y.,  {Gottl{\"o}ber} S.,  {Yepes}
  G.,   {Steinmetz} M.,  2011, \mn@doi [\mnras]
  {10.1111/j.1365-2966.2010.17786.x}, \href
  {https://ui.adsabs.harvard.edu/abs/2011MNRAS.411.1525L} {411, 1525}

\bibitem[\protect\citeauthoryear{{Libeskind} et~al.,}{{Libeskind}
  et~al.}{2020}]{Libeskind2020}
{Libeskind} N.~I.,  et~al., 2020, \mn@doi [\mnras] {10.1093/mnras/staa2541},
  \href {https://ui.adsabs.harvard.edu/abs/2020MNRAS.498.2968L} {498, 2968}

\bibitem[\protect\citeauthoryear{{Licquia} \& {Newman}}{{Licquia} \&
  {Newman}}{2015}]{Licquia2015}
{Licquia} T.~C.,  {Newman} J.~A.,  2015, \mn@doi [\apj]
  {10.1088/0004-637X/806/1/96}, \href
  {https://ui.adsabs.harvard.edu/abs/2015ApJ...806...96L} {806, 96}

\bibitem[\protect\citeauthoryear{{Lopes}, {Trevisan}, {Lagan{\'a}}, {Durret},
  {Ribeiro}  \& {Rembold}}{{Lopes} et~al.}{2018}]{Lopes2018}
{Lopes} P. A.~A.,  {Trevisan} M.,  {Lagan{\'a}} T.~F.,  {Durret} F.,  {Ribeiro}
  A.~L.~B.,   {Rembold} S.~B.,  2018, \mn@doi [\mnras] {10.1093/mnras/sty1374},
  \href {https://ui.adsabs.harvard.edu/abs/2018MNRAS.478.5473L} {478, 5473}

\bibitem[\protect\citeauthoryear{{Ludlow}, {Schaye}, {Schaller}  \&
  {Bower}}{{Ludlow} et~al.}{2020}]{Ludlow2020}
{Ludlow} A.~D.,  {Schaye} J.,  {Schaller} M.,   {Bower} R.,  2020, \mn@doi
  [\mnras] {10.1093/mnras/staa316}, \href
  {https://ui.adsabs.harvard.edu/abs/2020MNRAS.493.2926L} {493, 2926}

\bibitem[\protect\citeauthoryear{{Lynden-Bell}, {Faber}, {Burstein}, {Davies},
  {Dressler}, {Terlevich}  \& {Wegner}}{{Lynden-Bell}
  et~al.}{1988}]{Lynden-Bell1988}
{Lynden-Bell} D.,  {Faber} S.~M.,  {Burstein} D.,  {Davies} R.~L.,  {Dressler}
  A.,  {Terlevich} R.~J.,   {Wegner} G.,  1988, \mn@doi [\apj]
  {10.1086/166066}, \href
  {https://ui.adsabs.harvard.edu/abs/1988ApJ...326...19L} {326, 19}

\bibitem[\protect\citeauthoryear{{Mansfield}, {Kravtsov}  \&
  {Diemer}}{{Mansfield} et~al.}{2017}]{Mansfield2017}
{Mansfield} P.,  {Kravtsov} A.~V.,   {Diemer} B.,  2017, \mn@doi [\apj]
  {10.3847/1538-4357/aa7047}, \href
  {https://ui.adsabs.harvard.edu/abs/2017ApJ...841...34M} {841, 34}

\bibitem[\protect\citeauthoryear{{Mathis}, {Lemson}, {Springel}, {Kauffmann},
  {White}, {Eldar}  \& {Dekel}}{{Mathis} et~al.}{2002}]{Mathis2002}
{Mathis} H.,  {Lemson} G.,  {Springel} V.,  {Kauffmann} G.,  {White} S.~D.~M.,
  {Eldar} A.,   {Dekel} A.,  2002, \mn@doi [\mnras]
  {10.1046/j.1365-8711.2002.05447.x}, \href
  {https://ui.adsabs.harvard.edu/abs/2002MNRAS.333..739M} {333, 739}

\bibitem[\protect\citeauthoryear{{McAlpine} et~al.,}{{McAlpine}
  et~al.}{2016}]{McAlpine2016}
{McAlpine} S.,  et~al., 2016, \mn@doi [Astronomy and Computing]
  {10.1016/j.ascom.2016.02.004}, \href
  {https://ui.adsabs.harvard.edu/abs/2016A&C....15...72M} {15, 72}

\bibitem[\protect\citeauthoryear{{McAlpine}, {Bower}, {Rosario}, {Crain},
  {Schaye}  \& {Theuns}}{{McAlpine} et~al.}{2018}]{McAlpine2018}
{McAlpine} S.,  {Bower} R.~G.,  {Rosario} D.~J.,  {Crain} R.~A.,  {Schaye} J.,
   {Theuns} T.,  2018, \mn@doi [\mnras] {10.1093/mnras/sty2489}, \href
  {https://ui.adsabs.harvard.edu/abs/2018MNRAS.481.3118M} {481, 3118}

\bibitem[\protect\citeauthoryear{{McConnell}, {Ma}, {Gebhardt}, {Wright},
  {Murphy}, {Lauer}, {Graham}  \& {Richstone}}{{McConnell}
  et~al.}{2011}]{McConnell2011}
{McConnell} N.~J.,  {Ma} C.-P.,  {Gebhardt} K.,  {Wright} S.~A.,  {Murphy}
  J.~D.,  {Lauer} T.~R.,  {Graham} J.~R.,   {Richstone} D.~O.,  2011, \mn@doi
  [\nat] {10.1038/nature10636}, \href
  {https://ui.adsabs.harvard.edu/abs/2011Natur.480..215M} {480, 215}

\bibitem[\protect\citeauthoryear{{Meusinger}, {Rudolf}, {Stecklum}, {Hoeft},
  {Mauersberger}  \& {Apai}}{{Meusinger} et~al.}{2020}]{Meusinger2020}
{Meusinger} H.,  {Rudolf} C.,  {Stecklum} B.,  {Hoeft} M.,  {Mauersberger} R.,
   {Apai} D.,  2020, \mn@doi [\aap] {10.1051/0004-6361/202037574}, \href
  {https://ui.adsabs.harvard.edu/abs/2020A&A...640A..30M} {640, A30}

\bibitem[\protect\citeauthoryear{{Mitchell}, {Lacey}, {Baugh}  \&
  {Cole}}{{Mitchell} et~al.}{2013}]{Mitchell2013}
{Mitchell} P.~D.,  {Lacey} C.~G.,  {Baugh} C.~M.,   {Cole} S.,  2013, \mn@doi
  [\mnras] {10.1093/mnras/stt1280}, \href
  {https://ui.adsabs.harvard.edu/abs/2013MNRAS.435...87M} {435, 87}

\bibitem[\protect\citeauthoryear{{Moore}}{{Moore}}{1994}]{Moore1994}
{Moore} B.,  1994, \mn@doi [\nat] {10.1038/370629a0}, \href
  {https://ui.adsabs.harvard.edu/abs/1994Natur.370..629M} {370, 629}

\bibitem[\protect\citeauthoryear{{Moore}, {Quinn}, {Governato}, {Stadel}  \&
  {Lake}}{{Moore} et~al.}{1999}]{Moore1999}
{Moore} B.,  {Quinn} T.,  {Governato} F.,  {Stadel} J.,   {Lake} G.,  1999,
  \mn@doi [\mnras] {10.1046/j.1365-8711.1999.03039.x}, \href
  {https://ui.adsabs.harvard.edu/abs/1999MNRAS.310.1147M} {310, 1147}

\bibitem[\protect\citeauthoryear{{More}, {Diemer}  \& {Kravtsov}}{{More}
  et~al.}{2015}]{More2015}
{More} S.,  {Diemer} B.,   {Kravtsov} A.~V.,  2015, \mn@doi [\apj]
  {10.1088/0004-637X/810/1/36}, \href
  {https://ui.adsabs.harvard.edu/abs/2015ApJ...810...36M} {810, 36}

\bibitem[\protect\citeauthoryear{{More} et~al.,}{{More}
  et~al.}{2016}]{More2016}
{More} S.,  et~al., 2016, \mn@doi [\apj] {10.3847/0004-637X/825/1/39}, \href
  {https://ui.adsabs.harvard.edu/abs/2016ApJ...825...39M} {825, 39}

\bibitem[\protect\citeauthoryear{{Murata}, {Sunayama}, {Oguri}, {More},
  {Nishizawa}, {Nishimichi}  \& {Osato}}{{Murata} et~al.}{2020}]{Murata2020}
{Murata} R.,  {Sunayama} T.,  {Oguri} M.,  {More} S.,  {Nishizawa} A.~J.,
  {Nishimichi} T.,   {Osato} K.,  2020, \mn@doi [\pasj] {10.1093/pasj/psaa041},
  \href {https://ui.adsabs.harvard.edu/abs/2020PASJ...72...64M} {72, 64}

\bibitem[\protect\citeauthoryear{{Navarro}, {Frenk}  \& {White}}{{Navarro}
  et~al.}{1996}]{Navarro1996}
{Navarro} J.~F.,  {Frenk} C.~S.,   {White} S. D.~M.,  1996, \mn@doi [\apj]
  {10.1086/177173}, \href
  {https://ui.adsabs.harvard.edu/abs/1996ApJ...462..563N} {462, 563}

\bibitem[\protect\citeauthoryear{{Navarro}, {Frenk}  \& {White}}{{Navarro}
  et~al.}{1997}]{Navarro1997}
{Navarro} J.~F.,  {Frenk} C.~S.,   {White} S. D.~M.,  1997, \mn@doi [\apj]
  {10.1086/304888}, \href
  {https://ui.adsabs.harvard.edu/abs/1997ApJ...490..493N} {490, 493}

\bibitem[\protect\citeauthoryear{{Navarro} et~al.,}{{Navarro}
  et~al.}{2010}]{Navarro2010}
{Navarro} J.~F.,  et~al., 2010, \mn@doi [\mnras]
  {10.1111/j.1365-2966.2009.15878.x}, \href
  {https://ui.adsabs.harvard.edu/abs/2010MNRAS.402...21N} {402, 21}

\bibitem[\protect\citeauthoryear{{Nishizawa} et~al.,}{{Nishizawa}
  et~al.}{2018}]{Nishizawa2018}
{Nishizawa} A.~J.,  et~al., 2018, \mn@doi [\pasj] {10.1093/pasj/psx106}, \href
  {https://ui.adsabs.harvard.edu/abs/2018PASJ...70S..24N} {70, S24}

\bibitem[\protect\citeauthoryear{{O'Neil}, {Barnes}, {Vogelsberger}  \&
  {Diemer}}{{O'Neil} et~al.}{2020}]{ONeil2020}
{O'Neil} S.,  {Barnes} D.~J.,  {Vogelsberger} M.,   {Diemer} B.,  2020, arXiv
  e-prints, \href {https://ui.adsabs.harvard.edu/abs/2020arXiv201200025O} {p.
  arXiv:2012.00025}

\bibitem[\protect\citeauthoryear{{Piffaretti}, {Arnaud}, {Pratt},
  {Pointecouteau}  \& {Melin}}{{Piffaretti} et~al.}{2011}]{Piffaretti2011}
{Piffaretti} R.,  {Arnaud} M.,  {Pratt} G.~W.,  {Pointecouteau} E.,   {Melin}
  J.~B.,  2011, \mn@doi [\aap] {10.1051/0004-6361/201015377}, \href
  {https://ui.adsabs.harvard.edu/abs/2011A&A...534A.109P} {534, A109}

\bibitem[\protect\citeauthoryear{{Pillepich} et~al.,}{{Pillepich}
  et~al.}{2018}]{Pillepich2018}
{Pillepich} A.,  et~al., 2018, \mn@doi [\mnras] {10.1093/mnras/stx2656}, \href
  {https://ui.adsabs.harvard.edu/abs/2018MNRAS.473.4077P} {473, 4077}

\bibitem[\protect\citeauthoryear{{Planck Collaboration} et~al.,}{{Planck
  Collaboration} et~al.}{2014}]{Planck2013}
{Planck Collaboration} et~al., 2014, \mn@doi [Astronomy and Astrophysics]
  {10.1051/0004-6361/201321529}, \href
  {http://adsabs.harvard.edu/abs/2014A%26A...571A...1P} {571, A1}

\bibitem[\protect\citeauthoryear{{Planck Collaboration} et~al.,}{{Planck
  Collaboration} et~al.}{2016a}]{Planck2016}
{Planck Collaboration} et~al., 2016a, \mn@doi [\aap]
  {10.1051/0004-6361/201525830}, \href
  {https://ui.adsabs.harvard.edu/abs/2016A&A...594A..13P} {594, A13}

\bibitem[\protect\citeauthoryear{{Planck Collaboration} et~al.,}{{Planck
  Collaboration} et~al.}{2016b}]{Planck2016b}
{Planck Collaboration} et~al., 2016b, \mn@doi [\aap]
  {10.1051/0004-6361/201525823}, \href
  {https://ui.adsabs.harvard.edu/abs/2016A&A...594A..27P} {594, A27}

\bibitem[\protect\citeauthoryear{{Planck Collaboration} et~al.,}{{Planck
  Collaboration} et~al.}{2020}]{Planck2018}
{Planck Collaboration} et~al., 2020, \mn@doi [\aap]
  {10.1051/0004-6361/201833880}, \href
  {https://ui.adsabs.harvard.edu/abs/2020A&A...641A...1P} {641, A1}

\bibitem[\protect\citeauthoryear{{Posti} \& {Helmi}}{{Posti} \&
  {Helmi}}{2019}]{Posti2019}
{Posti} L.,  {Helmi} A.,  2019, \mn@doi [\aap] {10.1051/0004-6361/201833355},
  \href {https://ui.adsabs.harvard.edu/abs/2019A&A...621A..56P} {621, A56}

\bibitem[\protect\citeauthoryear{{Rahmani}, {Lianou}  \& {Barmby}}{{Rahmani}
  et~al.}{2016}]{Rahmani2016}
{Rahmani} S.,  {Lianou} S.,   {Barmby} P.,  2016, \mn@doi [\mnras]
  {10.1093/mnras/stv2951}, \href
  {https://ui.adsabs.harvard.edu/abs/2016MNRAS.456.4128R} {456, 4128}

\bibitem[\protect\citeauthoryear{{Raychaudhury}}{{Raychaudhury}}{1989}]{Raychaudhury1989}
{Raychaudhury} S.,  1989, \mn@doi [\nat] {10.1038/342251a0}, \href
  {https://ui.adsabs.harvard.edu/abs/1989Natur.342..251R} {342, 251}

\bibitem[\protect\citeauthoryear{{Richtler}, {Salinas}, {Misgeld}, {Hilker},
  {Hau}, {Romanowsky}, {Schuberth}  \& {Spolaor}}{{Richtler}
  et~al.}{2011}]{Richtler2011}
{Richtler} T.,  {Salinas} R.,  {Misgeld} I.,  {Hilker} M.,  {Hau} G.~K.~T.,
  {Romanowsky} A.~J.,  {Schuberth} Y.,   {Spolaor} M.,  2011, \mn@doi [\aap]
  {10.1051/0004-6361/201015948}, \href
  {https://ui.adsabs.harvard.edu/abs/2011A&A...531A.119R} {531, A119}

\bibitem[\protect\citeauthoryear{{Riess}, {Fliri}  \& {Valls-Gabaud}}{{Riess}
  et~al.}{2012}]{Riess2012}
{Riess} A.~G.,  {Fliri} J.,   {Valls-Gabaud} D.,  2012, \mn@doi [\apj]
  {10.1088/0004-637X/745/2/156}, \href
  {https://ui.adsabs.harvard.edu/abs/2012ApJ...745..156R} {745, 156}

\bibitem[\protect\citeauthoryear{{Riess} et~al.,}{{Riess}
  et~al.}{2016}]{Riess2016}
{Riess} A.~G.,  et~al., 2016, \mn@doi [\apj] {10.3847/0004-637X/826/1/56},
  \href {https://ui.adsabs.harvard.edu/abs/2016ApJ...826...56R} {826, 56}

\bibitem[\protect\citeauthoryear{{Rines} \& {Geller}}{{Rines} \&
  {Geller}}{2008}]{Rines2008}
{Rines} K.,  {Geller} M.~J.,  2008, \mn@doi [\aj]
  {10.1088/0004-6256/135/5/1837}, \href
  {https://ui.adsabs.harvard.edu/abs/2008AJ....135.1837R} {135, 1837}

\bibitem[\protect\citeauthoryear{{Rines}, {Geller}, {Kurtz}, {Diaferio},
  {Jarrett}  \& {Huchra}}{{Rines} et~al.}{2001}]{Rines2001}
{Rines} K.,  {Geller} M.~J.,  {Kurtz} M.~J.,  {Diaferio} A.,  {Jarrett} T.~H.,
   {Huchra} J.~P.,  2001, \mn@doi [\apjl] {10.1086/324457}, \href
  {https://ui.adsabs.harvard.edu/abs/2001ApJ...561L..41R} {561, L41}

\bibitem[\protect\citeauthoryear{{Rines}, {Geller}, {Kurtz}  \&
  {Diaferio}}{{Rines} et~al.}{2003}]{Rines2003}
{Rines} K.,  {Geller} M.~J.,  {Kurtz} M.~J.,   {Diaferio} A.,  2003, \mn@doi
  [\aj] {10.1086/378599}, \href
  {https://ui.adsabs.harvard.edu/abs/2003AJ....126.2152R} {126, 2152}

\bibitem[\protect\citeauthoryear{{Sahu}, {Graham}  \& {Davis}}{{Sahu}
  et~al.}{2019}]{Sahu2019}
{Sahu} N.,  {Graham} A.~W.,   {Davis} B.~L.,  2019, \mn@doi [\apj]
  {10.3847/1538-4357/ab0f32}, \href
  {https://ui.adsabs.harvard.edu/abs/2019ApJ...876..155S} {876, 155}

\bibitem[\protect\citeauthoryear{{Salomon}, {Ibata}, {Famaey}, {Martin}  \&
  {Lewis}}{{Salomon} et~al.}{2016}]{Salomon2016}
{Salomon} J.~B.,  {Ibata} R.~A.,  {Famaey} B.,  {Martin} N.~F.,   {Lewis}
  G.~F.,  2016, \mn@doi [\mnras] {10.1093/mnras/stv2865}, \href
  {https://ui.adsabs.harvard.edu/abs/2016MNRAS.456.4432S} {456, 4432}

\bibitem[\protect\citeauthoryear{{Sandage}, {Binggeli}  \& {Tammann}}{{Sandage}
  et~al.}{1985}]{Sandage1985}
{Sandage} A.,  {Binggeli} B.,   {Tammann} G.~A.,  1985, \mn@doi [\aj]
  {10.1086/113875}, \href
  {https://ui.adsabs.harvard.edu/abs/1985AJ.....90.1759S} {90, 1759}

\bibitem[\protect\citeauthoryear{{Savitzky} \& {Golay}}{{Savitzky} \&
  {Golay}}{1964}]{Savitzky1964}
{Savitzky} A.,  {Golay} M.~J.~E.,  1964, Analytical Chemistry, \href
  {https://ui.adsabs.harvard.edu/abs/1964AnaCh..36.1627S} {36, 1627}

\bibitem[\protect\citeauthoryear{{Sawala} et~al.,}{{Sawala}
  et~al.}{2016}]{Sawala2016}
{Sawala} T.,  et~al., 2016, \mn@doi [\mnras] {10.1093/mnras/stw145}, \href
  {https://ui.adsabs.harvard.edu/abs/2016MNRAS.457.1931S} {457, 1931}

\bibitem[\protect\citeauthoryear{{Sawala}, {McAlpine}, {Jasche}, {Lavaux},
  {Jenkins}, {Johansson}  \& {Frenk}}{{Sawala} et~al.}{2021a}]{Sawala2021b}
{Sawala} T.,  {McAlpine} S.,  {Jasche} J.,  {Lavaux} G.,  {Jenkins} A.,
  {Johansson} P.~H.,   {Frenk} C.~S.,  2021a, arXiv e-prints, \href
  {https://ui.adsabs.harvard.edu/abs/2021arXiv210312073S} {p. arXiv:2103.12073}

\bibitem[\protect\citeauthoryear{{Sawala}, {Jenkins}, {McAlpine}, {Jasche},
  {Lavaux}, {Johansson}  \& {Frenk}}{{Sawala} et~al.}{2021b}]{Sawala2021a}
{Sawala} T.,  {Jenkins} A.,  {McAlpine} S.,  {Jasche} J.,  {Lavaux} G.,
  {Johansson} P.~H.,   {Frenk} C.~S.,  2021b, \mn@doi [\mnras]
  {10.1093/mnras/staa3568}, \href
  {https://ui.adsabs.harvard.edu/abs/2021MNRAS.501.4759S} {501, 4759}

\bibitem[\protect\citeauthoryear{{Schaller}, {Gonnet}, {Chalk}  \&
  {Draper}}{{Schaller} et~al.}{2016}]{Schaller2016}
{Schaller} M.,  {Gonnet} P.,  {Chalk} A. B.~G.,   {Draper} P.~W.,  2016, arXiv
  e-prints, \href {https://ui.adsabs.harvard.edu/abs/2016arXiv160602738S} {p.
  arXiv:1606.02738}

\bibitem[\protect\citeauthoryear{{Schaller}, {Gonnet}, {Draper}, {Chalk},
  {Bower}, {Willis}  \& {Hausammann}}{{Schaller} et~al.}{2018}]{Schaller2018}
{Schaller} M.,  {Gonnet} P.,  {Draper} P.~W.,  {Chalk} A. B.~G.,  {Bower}
  R.~G.,  {Willis} J.,   {Hausammann} L.,  2018, {SWIFT: SPH With
  Inter-dependent Fine-grained Tasking} (\mn@eprint {ascl} {1805.020})

\bibitem[\protect\citeauthoryear{{Schaye} et~al.,}{{Schaye}
  et~al.}{2015}]{Schaye2015}
{Schaye} J.,  et~al., 2015, \mn@doi [\mnras] {10.1093/mnras/stu2058}, \href
  {https://ui.adsabs.harvard.edu/abs/2015MNRAS.446..521S} {446, 521}

\bibitem[\protect\citeauthoryear{{Seth} \& {Raychaudhury}}{{Seth} \&
  {Raychaudhury}}{2020}]{Seth2020}
{Seth} R.,  {Raychaudhury} S.,  2020, \mn@doi [\mnras]
  {10.1093/mnras/staa1779}, \href
  {https://ui.adsabs.harvard.edu/abs/2020MNRAS.497..466S} {497, 466}

\bibitem[\protect\citeauthoryear{{Shaya}, {Tully}, {Hoffman}  \&
  {Pomar{\`e}de}}{{Shaya} et~al.}{2017}]{Shaya2017}
{Shaya} E.~J.,  {Tully} R.~B.,  {Hoffman} Y.,   {Pomar{\`e}de} D.,  2017,
  \mn@doi [\apj] {10.3847/1538-4357/aa9525}, \href
  {https://ui.adsabs.harvard.edu/abs/2017ApJ...850..207S} {850, 207}

\bibitem[\protect\citeauthoryear{{Simionescu} et~al.,}{{Simionescu}
  et~al.}{2011}]{Simionescu2011}
{Simionescu} A.,  et~al., 2011, \mn@doi [Science] {10.1126/science.1200331},
  \href {https://ui.adsabs.harvard.edu/abs/2011Sci...331.1576S} {331, 1576}

\bibitem[\protect\citeauthoryear{{Skrutskie} et~al.,}{{Skrutskie}
  et~al.}{2006}]{Skrutskie2006}
{Skrutskie} M.~F.,  et~al., 2006, \mn@doi [\aj] {10.1086/498708}, \href
  {https://ui.adsabs.harvard.edu/abs/2006AJ....131.1163S} {131, 1163}

\bibitem[\protect\citeauthoryear{{Sohn}, {Geller}, {Zahid}, {Fabricant},
  {Diaferio}  \& {Rines}}{{Sohn} et~al.}{2017}]{Sohn2017}
{Sohn} J.,  {Geller} M.~J.,  {Zahid} H.~J.,  {Fabricant} D.~G.,  {Diaferio} A.,
    {Rines} K.~J.,  2017, \mn@doi [\apjs] {10.3847/1538-4365/aa653e}, \href
  {https://ui.adsabs.harvard.edu/abs/2017ApJS..229...20S} {229, 20}

\bibitem[\protect\citeauthoryear{{Sorce} et~al.,}{{Sorce}
  et~al.}{2016a}]{Sorce2016a}
{Sorce} J.~G.,  et~al., 2016a, \mn@doi [\mnras] {10.1093/mnras/stv2407}, \href
  {https://ui.adsabs.harvard.edu/abs/2016MNRAS.455.2078S} {455, 2078}

\bibitem[\protect\citeauthoryear{{Sorce}, {Gottl{\"o}ber}, {Hoffman}  \&
  {Yepes}}{{Sorce} et~al.}{2016b}]{Sorce2016b}
{Sorce} J.~G.,  {Gottl{\"o}ber} S.,  {Hoffman} Y.,   {Yepes} G.,  2016b,
  \mn@doi [\mnras] {10.1093/mnras/stw1085}, \href
  {https://ui.adsabs.harvard.edu/abs/2016MNRAS.460.2015S} {460, 2015}

\bibitem[\protect\citeauthoryear{{Sorce}, {Dubois}, {Blaizot}, {McGee}, {Yepes}
   \& {Knebe}}{{Sorce} et~al.}{2021}]{Sorce2021}
{Sorce} J.~G.,  {Dubois} Y.,  {Blaizot} J.,  {McGee} S.~L.,  {Yepes} G.,
  {Knebe} A.,  2021, \mn@doi [\mnras] {10.1093/mnras/stab1021}, \href
  {https://ui.adsabs.harvard.edu/abs/2021MNRAS.504.2998S} {504, 2998}

\bibitem[\protect\citeauthoryear{{Stopyra}, {Peiris}, {Pontzen}, {Jasche}  \&
  {Natarajan}}{{Stopyra} et~al.}{2021}]{Stopyra2021}
{Stopyra} S.,  {Peiris} H.~V.,  {Pontzen} A.,  {Jasche} J.,   {Natarajan} P.,
  2021, arXiv e-prints, \href
  {https://ui.adsabs.harvard.edu/abs/2021arXiv210706903S} {p. arXiv:2107.06903}

\bibitem[\protect\citeauthoryear{{Strauss} et~al.,}{{Strauss}
  et~al.}{2002}]{Strauss2002}
{Strauss} M.~A.,  et~al., 2002, \mn@doi [\aj] {10.1086/342343}, \href
  {https://ui.adsabs.harvard.edu/abs/2002AJ....124.1810S} {124, 1810}

\bibitem[\protect\citeauthoryear{{Tamm}, {Tempel}, {Tenjes}, {Tihhonova}  \&
  {Tuvikene}}{{Tamm} et~al.}{2012}]{Tamm2012}
{Tamm} A.,  {Tempel} E.,  {Tenjes} P.,  {Tihhonova} O.,   {Tuvikene} T.,  2012,
  \mn@doi [\aap] {10.1051/0004-6361/201220065}, \href
  {https://ui.adsabs.harvard.edu/abs/2012A&A...546A...4T} {546, A4}

\bibitem[\protect\citeauthoryear{{Turnbull}, {Hudson}, {Feldman}, {Hicken},
  {Kirshner}  \& {Watkins}}{{Turnbull} et~al.}{2012}]{Turnbull2012}
{Turnbull} S.~J.,  {Hudson} M.~J.,  {Feldman} H.~A.,  {Hicken} M.,  {Kirshner}
  R.~P.,   {Watkins} R.,  2012, \mn@doi [\mnras]
  {10.1111/j.1365-2966.2011.20050.x}, \href
  {https://ui.adsabs.harvard.edu/abs/2012MNRAS.420..447T} {420, 447}

\bibitem[\protect\citeauthoryear{Virtanen et~al.,}{Virtanen
  et~al.}{2020}]{SciPy2020}
Virtanen P.,  et~al., 2020, \mn@doi [Nature Methods]
  {10.1038/s41592-019-0686-2}, \href {https://rdcu.be/b08Wh} {17, 261}

\bibitem[\protect\citeauthoryear{{Vogelsberger}, {Marinacci}, {Torrey}  \&
  {Puchwein}}{{Vogelsberger} et~al.}{2020}]{Vogelsberger2020}
{Vogelsberger} M.,  {Marinacci} F.,  {Torrey} P.,   {Puchwein} E.,  2020,
  \mn@doi [Nature Reviews Physics] {10.1038/s42254-019-0127-2}, \href
  {https://ui.adsabs.harvard.edu/abs/2020NatRP...2...42V} {2, 42}

\bibitem[\protect\citeauthoryear{{Wang}, {Mo}, {Yang}, {Jing}  \& {Lin}}{{Wang}
  et~al.}{2014}]{Wang2014}
{Wang} H.,  {Mo} H.~J.,  {Yang} X.,  {Jing} Y.~P.,   {Lin} W.~P.,  2014,
  \mn@doi [\apj] {10.1088/0004-637X/794/1/94}, \href
  {https://ui.adsabs.harvard.edu/abs/2014ApJ...794...94W} {794, 94}

\bibitem[\protect\citeauthoryear{{Wang} et~al.,}{{Wang}
  et~al.}{2016}]{Wang2016}
{Wang} H.,  et~al., 2016, \mn@doi [\apj] {10.3847/0004-637X/831/2/164}, \href
  {https://ui.adsabs.harvard.edu/abs/2016ApJ...831..164W} {831, 164}

\bibitem[\protect\citeauthoryear{{Watkins}, {Evans}  \& {An}}{{Watkins}
  et~al.}{2010}]{Watkins2010}
{Watkins} L.~L.,  {Evans} N.~W.,   {An} J.~H.,  2010, \mn@doi [\mnras]
  {10.1111/j.1365-2966.2010.16708.x}, \href
  {https://ui.adsabs.harvard.edu/abs/2010MNRAS.406..264W} {406, 264}

\bibitem[\protect\citeauthoryear{{Whitbourn} \& {Shanks}}{{Whitbourn} \&
  {Shanks}}{2014}]{Whitbourn2014}
{Whitbourn} J.~R.,  {Shanks} T.,  2014, \mn@doi [\mnras]
  {10.1093/mnras/stt2024}, \href
  {https://ui.adsabs.harvard.edu/abs/2014MNRAS.437.2146W} {437, 2146}

\bibitem[\protect\citeauthoryear{{Whitbourn} \& {Shanks}}{{Whitbourn} \&
  {Shanks}}{2016}]{Whitbourn2016}
{Whitbourn} J.~R.,  {Shanks} T.,  2016, \mn@doi [\mnras]
  {10.1093/mnras/stw555}, \href
  {https://ui.adsabs.harvard.edu/abs/2016MNRAS.459..496W} {459, 496}

\bibitem[\protect\citeauthoryear{{Williams} \& {Kerr}}{{Williams} \&
  {Kerr}}{1981}]{Williams1981}
{Williams} B.~A.,  {Kerr} F.~J.,  1981, \mn@doi [\aj] {10.1086/112972}, \href
  {https://ui.adsabs.harvard.edu/abs/1981AJ.....86..953W} {86, 953}

\bibitem[\protect\citeauthoryear{{Wong}, {Shanks}  \& {Metcalfe}}{{Wong}
  et~al.}{2021}]{Wong2021}
{Wong} J. H.~W.,  {Shanks} T.,   {Metcalfe} N.,  2021, arXiv e-prints, \href
  {https://ui.adsabs.harvard.edu/abs/2021arXiv210708505W} {p. arXiv:2107.08505}

\bibitem[\protect\citeauthoryear{{Woudt}, {Kraan-Korteweg}, {Lucey}, {Fairall}
  \& {Moore}}{{Woudt} et~al.}{2008}]{Woudt2008}
{Woudt} P.~A.,  {Kraan-Korteweg} R.~C.,  {Lucey} J.,  {Fairall} A.~P.,
  {Moore} S.~A.~W.,  2008, \mn@doi [\mnras] {10.1111/j.1365-2966.2007.12571.x},
  \href {https://ui.adsabs.harvard.edu/abs/2008MNRAS.383..445W} {383, 445}

\bibitem[\protect\citeauthoryear{{Wright} et~al.,}{{Wright}
  et~al.}{2017}]{Wright2017}
{Wright} A.~H.,  et~al., 2017, \mn@doi [\mnras] {10.1093/mnras/stx1149}, \href
  {https://ui.adsabs.harvard.edu/abs/2017MNRAS.470..283W} {470, 283}

\bibitem[\protect\citeauthoryear{{Wu} \& {Huterer}}{{Wu} \&
  {Huterer}}{2017}]{Wu2017}
{Wu} H.-Y.,  {Huterer} D.,  2017, \mn@doi [\mnras] {10.1093/mnras/stx1967},
  \href {https://ui.adsabs.harvard.edu/abs/2017MNRAS.471.4946W} {471, 4946}

\bibitem[\protect\citeauthoryear{{Xhakaj}, {Diemer}, {Leauthaud}, {Wasserman},
  {Huang}, {Luo}, {Adhikari}  \& {Singh}}{{Xhakaj} et~al.}{2020}]{Xhakaj2020}
{Xhakaj} E.,  {Diemer} B.,  {Leauthaud} A.,  {Wasserman} A.,  {Huang} S.,
  {Luo} Y.,  {Adhikari} S.,   {Singh} S.,  2020, \mn@doi [\mnras]
  {10.1093/mnras/staa3046}, \href
  {https://ui.adsabs.harvard.edu/abs/2020MNRAS.499.3534X} {499, 3534}

\bibitem[\protect\citeauthoryear{{Yepes}, {Gottl{\"o}ber}  \&
  {Hoffman}}{{Yepes} et~al.}{2014}]{Yepes2014}
{Yepes} G.,  {Gottl{\"o}ber} S.,   {Hoffman} Y.,  2014, \mn@doi [\nar]
  {10.1016/j.newar.2013.11.001}, \href
  {https://ui.adsabs.harvard.edu/abs/2014NewAR..58....1Y} {58, 1}

\bibitem[\protect\citeauthoryear{{Zehavi}, {Riess}, {Kirshner}  \&
  {Dekel}}{{Zehavi} et~al.}{1998}]{Zehavi1998}
{Zehavi} I.,  {Riess} A.~G.,  {Kirshner} R.~P.,   {Dekel} A.,  1998, \mn@doi
  [\apj] {10.1086/306015}, \href
  {https://ui.adsabs.harvard.edu/abs/1998ApJ...503..483Z} {503, 483}

\bibitem[\protect\citeauthoryear{{van der Marel}, {Fardal}, {Sohn}, {Patel},
  {Besla}, {del Pino}, {Sahlmann}  \& {Watkins}}{{van der Marel}
  et~al.}{2019}]{vanderMarel2019}
{van der Marel} R.~P.,  {Fardal} M.~A.,  {Sohn} S.~T.,  {Patel} E.,  {Besla}
  G.,  {del Pino} A.,  {Sahlmann} J.,   {Watkins} L.~L.,  2019, \mn@doi [\apj]
  {10.3847/1538-4357/ab001b}, \href
  {https://ui.adsabs.harvard.edu/abs/2019ApJ...872...24V} {872, 24}

\makeatother
\end{thebibliography}
\appendix

\section{Public data release}
\label{sect:public_data_release}

With the publication of this study comes the public data release of the \sibeliusdark galaxy and halo properties at $z=0$. As with previous \textit{Virgo Consortium} projects, the data will be made available through an {\sc sql} database. Users familiar with the \textit{Millennium} database \citep{Lemson2006} and the \eagle database \citep{McAlpine2016} will recognise the main features of the interface and should be able to adapt their scripts easily to the \sibeliusdark database. Therefore here we do not provide an in-depth description of the database layout, but we do describe the unique considerations when using the \sibeliusdark data, along with some examples. In the future we plan to append the \sibeliusdark database with additional simulation outputs, which will allow users to track the evolution of particular objects through cosmic time.

The \sibeliusdark data can be accessed through the VirgoDB portal (\href{https://virgodb.dur.ac.uk/}{https://virgodb.dur.ac.uk/}), either via the \textit{Millennium} or \eagle interface. The database containing the \sibeliusdark data is named \squotes{McAlpine2022a}, and is divided between two tables: a \textit{Halo} table (properties listed in \cref{table:fof_properties}), which stores the mass and size for each dark matter halo, and a \textit{Galaxy} table (properties listed in \cref{table:subhalo1}), which stores various properties for each galaxy/subhalo that populate these haloes. The objects from the \textit{Galaxy} table can be linked to the \textit{Halo} table using the {\tt hosthaloid} (see the examples). 

There are always unique considerations to be aware of when using any particular dataset. When interpreting the results from \sibeliusdark data, users should consider the following:

\begin{itemize}
    \item \textbf{When comparing object-by-object between \sibeliusdark and observational data}: as was mentioned in \cref{sect:clusters_and_groups}, the large-scale structure constraints are designed \emph{en masse}, giving no guarantee that any particular object within the \sibeliusdark volume will be located at exactly the same location at exactly the right distance relative to the data. Therefore when using \sibeliusdark to compare to a particular structure/cluster/void in the data, we recommend doing an association approach similar to that in \cref{fig:cluster_locations_1,fig:cluster_locations_2}, by plotting the \sibeliusdark galaxies and the data together to first establish the layout of the region of interest, and associate the analogue structures to the data appropriately.
    \item \textbf{Coordinates and velocities frame}: the coordinates ({\tt [x,y,z]}), right ascension's ({\tt ra}) and declination's ({\tt dec}) of \sibeliusdark galaxies are in the equatorial frame (i.e., the Milky Way galaxy is located at [0,0,0] Mpc). The velocities ({\tt v\_[x,y,z]}) of \sibeliusdark galaxies are in the CMB reference frame. There exist no galaxies beyond a distance of {\tt dist} = 200~Mpc from the Milky Way analogue, as this defines the edge of the constrained region.
    \item \textbf{The data is not a lightcone}: the public data is all galaxies and haloes from the $z=0$ simulation snapshot. Thus we have assumed a negligible evolution in the positions and properties of the \sibeliusdark galaxies between $z=0.045$ (the edge of the constrained region) and $z=0$. We define the redshift of a galaxy as $z=v_{\mathrm{r}} / c$, where $v_{\mathrm{r}}$ is the radial velocity (which includes the Hubble flow) and $c$ is the speed of light, and we define the apparent magnitude as $m = M + 5 \mathrm{log}_{\mathrm{10}}(d/10)$, where $d$ is the distance to the simulated Milky Way in pc and $M$ is the absolute magnitude.
    \item \textbf{Simulation resolution}: as with all simulations there is a  minimum halo mass we can resolve. The minimum number of bound particles to constitute a substructure is {\tt nbound}=20, which, given the dark matter particle mass of $1.15 \times 10^{7}$~\Msol, equates to $2.3 \times 10^{8}$~\Msol. For this study we applied a conservative mass cut of $10^{9}$~\Msol, or $\approx 90$ particles.
    
    Substructures that once existed but have later been stripped below the {\tt nbound}=20 limit have their most bound particle tracked for the remainder of the simulation. These are sometimes referred to in semi-analytic models as \squotes{orphan} galaxies, and they can be identified in the \textit{Galaxy} table as those with {\tt nbound}=0 or 1 or {\tt type} = 2.
    
    We would recommend users generally select/cut their samples on total  mass ({\tt mstars\_bulge} + {\tt mstars\_disk}) or luminosity, for which there is no lower value limit. 
    
    \item \textbf{The Local Group pair}: due to their close proximity, the \sibeliusdark Milky Way and Andromeda subhaloes/galaxies have come to occupy the same Friends-of-Friends halo by $z=0$ (even although they remain distinctly separated by eye, see \cref{fig:local_environment}). By traditional conventions, this classifies the Milky Way as a \squotes{satellite} of the Andromeda in the simulation, as the more massive Andromeda subhalo/galaxy is assigned to be the \squotes{central}. This is why the Milky Way has no reported halo mass in \cref{table:lg} (as the structure finder, {\sc hbt+}, only computes overdensities for central subhaloes), and why it shares the same {\tt hosthaloid}/halo mass as Andromeda in the public database.
    
    \item \textbf{The data is not perfect:} whilst we believe \sibeliusdark to be the most comprehensive constrained realisation simulation to date, we understand it is not a perfect representation of our Local Volume. We would therefore urge caution as to not overinterperate the results from the simulation, and instead use them as plausible predictions for how structures in the Local Volume may have come to pass.
\end{itemize}

\subsection{{\sc sql} examples}

Here we list three example {\sc sql} queries that return the necessary data to reproduce \cref{fig:mass_function,fig:r_band_number_counts,fig:caustics_subhaloes}. Barring \cref{fig:sky_maps,fig:density_vs_rad,fig:caustics_virgo,fig:caustics_coma,fig:local_environment}, which make use of the raw particle data, all the figures of this paper are reproducible with similar queries.\\

\noindent \textbf{\cref{fig:mass_function}: the halo mass function in spherical apertures}. This example returns the masses and distances of all \sibeliusdark haloes with a mass in excess of $M_{\mathrm{200c}} \geq 10^{9}$~\Msol. We must link the \textit{Halo} table to the \textit{Galaxy} table via the {\tt hosthaloid} in order to retrieve the distance information. The distance to a halo is assumed as the distance to the haloes most massive subhalo/galaxy, i.e., the one with {\tt rank}=0.

\sqlstyle
\footnotesize
\begin{lstlisting}[numbers = none]
-- Select the quantities we want
SELECT      
  -- Halo mass
  halo.m200crit,
  -- Distance to the central galaxy      
  gal.dist
-- Sibelius-DARK tables
FROM      
  mcalpine2022a..halo as halo,
  mcalpine2022a..galaxy as gal
-- Apply the conditions        
WHERE
  -- Link tables via hosthaloid
  gal.hosthaloid = halo.hosthaloid
  -- Only interested in the central galaxy
  AND gal.rank = 0
  -- Those above the mass limit
  AND halo.m200crit >= 1e9
\end{lstlisting}
\normalsize

\noindent \textbf{\cref{fig:r_band_number_counts}: $r$-band counts in the North Galactic Cap}. This example returns the redshift and apparent $r$-band magnitude of \sibeliusdark galaxies within the North Galactic Cap region of the SDSS survey.  

\sqlstyle
\footnotesize
\begin{lstlisting}[numbers = none]
-- Select the quantities we want
SELECT      
  redshift,
  -- Apparent magnitude in the r-band      
  mag_r_ext + 5*log10(dist*1e6/10) as r_band
-- Sibelius-DARK table
FROM      
  mcalpine2022a..galaxy
-- Apply the conditions        
WHERE
  -- Area on sky of North Galactic Cap
  ra BETWEEN 120 AND 240
  AND dec BETWEEN 0 AND 60
  -- Redshift (volume) limit
  AND redshift <= 0.045
\end{lstlisting}
\normalsize

\noindent \textbf{\cref{fig:caustics_subhaloes}: the Virgo cluster's member galaxies}. This example returns all galaxies with a stellar mass $M_* \geq 10^{8}$~\Msol within 2~Mpc of the \sibeliusdark Virgo cluster. This is done in two steps: first, we retrieve the details of the model Virgo cluster knowing the unique {\tt galaxyid} (listed in \cref{table:cluster_bcgs}). We then use the retrieved position information to select the galaxies around the model BCG position. This gives us all the information needed to reproduce the left hand panels of \cref{fig:caustics_subhaloes}.  

\sqlstyle
\footnotesize
\begin{lstlisting}[numbers = none]
-- Select the quantities we want
SELECT      
  -- Virgo halo size and halo mass.
  halo.m200mean, halo.r200mean,
  -- Virgo position and velocity.
  gal.x, gal.y, gal.z,
  gal.v_x, gal.v_y, gal.v_z 
-- Sibelius-DARK table
FROM      
  mcalpine2022a..halo as halo
  mcalpine2022a..galaxy as gal
-- Apply the conditions        
WHERE
  -- Unique galaxyid of Virgo
  gal.galaxyid = 58233
  -- Link to halo table
  AND gal.hosthaloid = halo.hosthaloid
\end{lstlisting}
\normalsize

\sqlstyle
\footnotesize
\begin{lstlisting}[numbers = none]
-- Select the quantities we want
SELECT      
  -- Positions of Virgo member galaxies
  x, y, z
  -- Velocities of Virgo member galaxies
  v_x, v_y, v_z
-- Sibelius-DARK table
FROM      
  mcalpine2022a..galaxy
-- Apply the conditions        
WHERE
  -- Within 2 Mpc of Virgo centre
  x BETWEEN <VIRGO_X>-2 AND <VIRGO_X>+2
  AND y <VIRGO_Y>-2 AND <VIRGO_Y>+2
  AND z <VIRGO_Z>-2 AND <VIRGO_Z>+2
  -- Above stellar mass limit.
  AND mstars_bulge + mstars_disk > 1e8
\end{lstlisting}
\normalsize

\noindent where {\tt VIRGO\_[X,Y,Z]} are the positions of the Virgo cluster found by the first query. 

\begin{table*}
\caption{Full listing of the \textit{Halo} properties table and
  description of the columns. Haloes can be linked to their member galaxies in the \textit{Galaxy} table via the {\tt hosthaloid}. The centre of the halo for computing the overdensities is taken as the position of the most massive subhalo member (i.e., {\tt rank}=0). There are no $h$-factors in any of the units.}
\label{table:fof_properties}
\begin{center}
\footnotesize
\renewcommand{\arraystretch}{1.5}
\begin{tabular}{ >{\ttfamily}p{2.5cm}p{1.5cm}p{12.5cm}}
{\large \bf Halo} & & \\
\hline
\normalfont Field & Units & Description \\
\hline\hline

hosthaloid &
- &
Unique identifier for a halo, used to link to the member galaxies in the \textit{Galaxy} table \\

m[200/500]\_crit &
\Msol &
Total halo mass contained within {\tt r[200/500]crit}. \\
 
m[200/500]\_mean &
\Msol &
Total halo mass contained within {\tt r[200/500]mean}. \\

r[200/500]\_crit &
Mpc &
The radius at which the enclosed density is equal to [200/500] times the critical density ($\rho_{\mathrm{crit}}$). \\
 
r[200/500]\_mean &
Mpc &
The radius at which the enclosed density is equal to [200/500] times the mean density ($\Omega_{\mathrm{m}}\rho_{\mathrm{crit}}$).\\

\hline
 
\end{tabular}
\end{center}
\end{table*}

\begin{table*}
\caption{Full listing of the \textit{Galaxy} properties table and
  description of the columns. Galaxies can be linked to their host halo properties in the \textit{Halo} table via the {\tt hosthaloid}. Two magnitudes are available for each band; including the effects of dust extinction (e.g., {\tt mag\_K\_ext}) and not including the effects of dust extinction (e.g., {\tt mag\_K}). There are no $h$-factors in any of the units.}
\label{table:subhalo1}
\begin{center}
\footnotesize
\renewcommand{\arraystretch}{1.5}
\begin{tabular}{ >{\ttfamily}p{2.5cm}p{1.5cm}p{12.5cm}}
{\large \bf Galaxy} & & \\
\hline
\normalfont Field & Units & Description \\
\hline\hline

galaxyid &
- &
Unique identifier of a galaxy. \\

hosthaloid & 
- &
Unique identifier of a galaxy's host halo (used to link to the \textit{Halo} table).
\\
\hline

mag\_H[\_ext] & 
Mag (Vega) & 
Absolute $H$-band magnitude (2MASS). \\

mag\_J[\_ext] & 
Mag (Vega) & 
Absolute $J$-band magnitude (2MASS). \\

mag\_K[\_ext] & 
Mag (Vega) & 
Absolute $K$-band magnitude (2MASS). \\

mag\_u[\_ext] & 
Mag (AB) & 
Absolute $u$-band magnitude (SDSS). \\

mag\_g[\_ext] & 
Mag (AB) & 
Absolute $g$-band magnitude (SDSS). \\

mag\_r[\_ext] & 
Mag (AB) & 
Absolute $r$-band magnitude (SDSS). \\

mag\_i[\_ext] & 
Mag (AB) & 
Absolute $i$-band magnitude (SDSS). \\

mag\_z[\_ext] & 
Mag (AB) & 
Absolute $z$-band magnitude (SDSS). \\

\hline

x & Mpc & x-coordinate of the galaxy's centre of potential (equatorial). \\

y & Mpc & y-coordinate of the galaxy's centre of potential (equatorial). \\

z & Mpc & z-coordinate of the galaxy's centre of potential (equatorial). \\

v\_x & Mpc & x-coordinate of the galaxy's velocity (CMB rest frame). \\

v\_y & Mpc & y-coordinate of the galaxy's velocity (CMB rest frame). \\

v\_z & Mpc & z-coordinate of the galaxy's velocity (CMB rest frame). \\

\hline

ra & degrees & Right Ascension (equatorial). \\

dec & degrees & Declination (equatorial). \\

v\_r & km/s & Radial velocity of galaxy relative to Milky Way. This is the radial component of the galaxy's peculiar velocity ({\tt v}) plus the Hubble flow ($H \times$ {\tt dist}, where $H$ is the Hubble constant). \\

dist & Mpc & Distance from Milky Way. \\

redshift & - & Redshift of galaxy ($z=v_{\mathrm{r}} / c$). \\
\hline
rank & - & Rank order of galaxy in host halo ({\tt rank} $=0$ is the central galaxy, {\tt rank} $>0$ are satellite galaxies) \\
type & - & \galform galaxy type: 0: central galaxy, 1: satellite galaxy and 2: \squotes{orphan} galaxy.\\
\hline

mbound & \Msol & Total mass bound to the subhalo. We refer to this as the subhalo mass, or $M_{\mathrm{sub}}$. \\

mstars\_bulge & \Msol & Stellar bulge mass. \\

mstars\_disk & \Msol & Stellar disk mass. \\

mbh & \Msol & Central supermassive black hole mass. \\

\hline
 
\end{tabular}
\end{center}
\end{table*}

\section{Cluster and galaxy properties}

Listed in \cref{table:cluster_masses,table:cluster_bcgs} are the halo and brightest-cluster-galaxy properties for the twelve haloes discussed in \cref{sect:clusters_and_groups}. \cref{table:lg} shows the properties for the Milky Way and Andromeda analogues discussed in \cref{sect:local_group}.

\begin{table*}

\caption{The properties of the nine most massive haloes in the \sibeliusdark volume, including also the properties of three well known lower-mass haloes. Four overdensity masses are listed, along with their enclosed mass radius, where $c$ denotes the critical density and $m$ denotes the mean density. $M_{\mathrm{rank}}$ is the mass ranking of the halo in units of $M_{\mathrm{200c}}$. The final column shows the observational counterpart cluster for the \sibeliusdark halo, identified using \cref{fig:cluster_locations_1,fig:cluster_locations_2}. The properties of each haloes most massive subhalo/galaxy (i.e., the BCG) are listed in \cref{table:cluster_bcgs}.}

\begin{tabular}{lrrrrrrrrrc} \hline

Object & $M_{\mathrm{200c}}$ & $M_{\mathrm{200m}}$ & $M_{\mathrm{500c}}$ & $M_{\mathrm{500m}}$ & $r_{\mathrm{200c}}$ & $r_{\mathrm{200m}}$ & $r_{\mathrm{500c}}$ & $r_{\mathrm{500m}}$ & $M_{\mathrm{rank}}$ & Obs. Counterpart \\
 & [$10^{15}$~\Msol] & [$10^{15}$~\Msol] & [$10^{15}$~\Msol] & [$10^{15}$~\Msol] & [Mpc] & [Mpc] & [Mpc] & [Mpc] & & Cat. ID \\

\hline\hline

Perseus & 2.72 & 4.05 &  1.87 & 2.97 &  2.94 & 4.99 &  1.91 & 3.31 &  1 & Abell 426 \\
Hercules-A & 1.89 & 2.93 &  1.13 & 2.08 &  2.60 & 4.48 &  1.62 & 2.94 &  2 & Abell 2199 \\
Hercules-B & 1.78 & 2.46 &  0.70 & 2.00 &  2.56 & 4.22 &  1.38 & 2.91 &  3 & Abell 2151 \\
Norma & 1.72 & 2.19 &  1.19 & 1.87 &  2.53 & 4.06 &  1.65 & 2.84 &  4 & Abell 3627 \\
Coma & 1.27 & 2.04 &  0.87 & 1.47 &  2.28 & 3.97 &  1.48 & 2.62 &  5 & Abell 1656 \\
Leo & 1.17 & 1.42 &  0.84 & 1.25 &  2.22 & 3.52 &  1.47 & 2.48 &  6 & Abell 1367 \\
SD-C7 & 1.15 & 1.34 &  0.88 & 1.22 &  2.21 & 3.45 &  1.49 & 2.46 &  7 & - \\
SD-C8 & 0.90 & 1.31 &  0.61 & 1.00 &  2.03 & 3.43 &  1.31 & 2.30 &  8 & - \\
SD-C9 & 0.83 & 1.16 &  0.53 & 0.93 &  1.98 & 3.29 &  1.26 & 2.25 &  9 & Abell 1185 \\
Hydra & 0.49 & 0.73 &  0.35 & 0.56 &  1.66 & 2.81 &  1.10 & 1.90 &  32 & Abell 1060 \\
Centaurus & 0.41 & 0.57 &  0.30 & 0.44 &  1.57 & 2.59 &  1.04 & 1.76 &  44 & Abell 3526 \\
Virgo & 0.35 & 0.48 &  0.27 & 0.38 &  1.49 & 2.46 &  1.00 & 1.67 &  56 & - \\

\hline
\end{tabular}
\label{table:cluster_masses}
\end{table*}

\begin{table*}

\caption{The properties of the most massive subhalo/galaxy (i.e., the BCG) hosted by the twelve haloes listed in \cref{table:cluster_masses}. From left to right, the common name for the halo/cluster, the \sibeliusdark public database ID ({\tt galaxyid}), the subhalo mass (i.e., total bound mass of the subhalo), the distance from the Milky Way, the recession velocity relative to the Milky Way, the galaxy's right ascension and declination on the \sibeliusdark sky, the stellar mass of the galaxy and the central supermassive black hole mass.}

\begin{tabular}{llrrrrrrr} \hline

Object & {\tt galaxyid} & $M_{\mathrm{sub}}$ & $r$ & $v_r$ & RA & DEC & $M_{\mathrm{*}}$ & $M_{\mathrm{BH}}$ \\
 & & [$10^{15}$~\Msol] & [Mpc] & [km~s$^{-1}$] & [deg] & [deg] & [$10^{11}$~\Msol] & [$10^{9}$~\Msol] \\

\hline\hline

Perseus & 207645 & 2.81 & 77.2 & 5187 & 44.31 & 41.87 & 11.95 & 34.93 \\
Hercules-A & 6720 & 1.79 & 141.5 & 10386 & 250.11 & 40.35 & 10.44 & 15.19 \\
Hercules-B & 2017405 & 0.66 & 156.4 & 11259 & 241.94 & 16.99 & 4.02 & 4.26 \\
Norma & 220176 & 1.45 & 73.9 & 4898 & 250.91 & -59.80 & 11.81 & 10.78 \\
Coma & 337006 & 1.22 & 108.2 & 7410 & 196.76 & 30.13& 8.58 & 5.31 \\
Leo & 84853 & 0.58 & 100.6 & 6629 & 177.23 & 22.25& 7.93 & 5.83 \\
SD-C7 & 118936 & 0.68 & 180.4 & 12347 & 29.53 & -0.63& 6.28 & 4.79 \\
SD-C8 & 8636 & 0.98 & 126.5 & 8690 & 108.30 & -36.79& 5.57 & 11.97 \\
SD-C9 & 346825 & 0.63 & 157.4 & 10059 & 167.84 & 30.27& 4.27 & 6.67 \\
Hydra & 37323 & 0.54 & 57.5 & 3516 & 159.27 & -28.50 & 6.85 & 6.51 \\
Centaurus & 624232 & 0.45 & 49.0 & 3535 & 197.79 & -41.98 & 4.74 & 5.38 \\
Virgo & 58233 & 0.35 & 21.1 & 1536 & 196.90 & 15.96 & 3.65 & 3.25 \\

\hline
\end{tabular}
\label{table:cluster_bcgs}
\end{table*}

\begin{table*}

\caption{Properties of the two primary Local Group members at $z=0$. From left to right, the object name, the \sibeliusdark public database ID ({\tt galaxyid}), the subhalo mass (i.e., total bound mass of the subhalo), the distance from the Milky Way, the recession and tangential velocity relative to the Milky Way, the galaxy's right ascension and declination on the Sibelius-DARK sky, the stellar mass of the galaxy, the central supermassive black hole mass and the halo mass. The Milky Way has no reported halo mass (and shares the same {\tt hosthaloid} in the public database as the Andromeda galaxy), because the Milky Way and Andromeda are associated with the same Friends-of-Friends group by $z=0$ in the simulation, which makes the Milky Way technically a satellite of Andromeda.}

\begin{tabular}{llrrrrrrrrr} \hline

Object & {\tt galaxyid} & $M_{\mathrm{sub}}$ & $r$ & $v_r$ & $v_t$ & RA & DEC & $M_{\mathrm{*}}$ & $M_{\mathrm{BH}}$ & $M_{\mathrm{200c}}$ \\
 & & [\Msol] & [Mpc] & [km~s$^{-1}$] & [km~s$^{-1}$] & [deg] & [deg] & [\Msol] & [\Msol] & [\Msol] \\

\hline\hline

Milky Way & 17791952 & $9.42 \times 10^{11}$ & - & - & - & - & - & $5.43 \times 10^{9}$ & $2.62 \times 10^{6}$ & - \\
Andromeda & 5098129 & $1.48 \times 10^{12}$ & 0.753 & -117 & 33 & 25.8 & 57.5 & $6.38 \times 10^{9}$ & $4.40 \times 10^{6}$ & $1.30 \times 10^{12}$\\

\hline
\end{tabular}
\label{table:lg}
\end{table*}

\bsp	
\label{lastpage}

\end{document}